\documentclass{article}

 \usepackage[preprint]{neurips_2026}

\usepackage[utf8]{inputenc} 
\usepackage[T1]{fontenc}    
\usepackage[table]{xcolor}
\definecolor{dark-blue}{RGB}{0,0,191}
\usepackage[hidelinks,colorlinks=true,citecolor=dark-blue,linkcolor=dark-blue,urlcolor=dark-blue, backref=page]{hyperref}
\usepackage{url}            
\usepackage{booktabs}       
\usepackage{amsfonts}       
\usepackage{nicefrac}       
\usepackage{microtype}      
\usepackage{xcolor}         
\usepackage{tabularx}
\usepackage{booktabs}
\usepackage{graphicx}
\usepackage[table]{xcolor}
\usepackage{seqsplit}

\usepackage{listings}
\lstdefinestyle{rawouttxt}{
  basicstyle=\ttfamily\footnotesize,
  breaklines=true,
  breakatwhitespace=false,
  columns=fullflexible,
  keepspaces=true,
  showstringspaces=false,
  frame=single,
  numbers=none
}

\title{AstroAlertBench: Evaluating the Accuracy, Reasoning, and Honesty of Multimodal LLMs in Astronomical Classification}

\usepackage{pifont}

\usepackage{amsmath}
\usepackage{amssymb}
\usepackage{mathtools}
\usepackage{amsthm}
\usepackage{color}
\mathtoolsset{showonlyrefs}

\usepackage{algorithm}
\usepackage{algpseudocode}

\usepackage{enumitem}

\usepackage[capitalize,noabbrev]{cleveref}

\usepackage{xurl} 
\usepackage{listings}
\usepackage{xcolor}
\usepackage{thmtools}

\lstdefinestyle{prompttxt}{
  basicstyle=\ttfamily\footnotesize,
  breaklines=true,
  breakatwhitespace=false,
  columns=fullflexible,
  keepspaces=true,
  frame=none,
}

\usepackage{pifont}
\theoremstyle{plain}

\theoremstyle{definition}

\theoremstyle{remark}

\usepackage{algorithm}
\usepackage{algorithmicx}
\usepackage{algpseudocode}

\usepackage[textsize=tiny]{todonotes}

\author{
  \textbf{Claire Chen}$^{1*}$ \quad  \textbf{Jiabao Sean Xiao}$^{1*}$  \quad \textbf{Shuze Daniel Liu}$^{2,3}\thanks{Equal contribution.}$  \quad
  \textbf{Facundo Perez Paolino}$^{1}$\\
  \textbf{Luke Handley}$^{1}$\thanks{Equal contribution; listed alphabetically.}\quad
  \textbf{Theophile Jegou du Laz}$^{1\dagger}$ \quad \textbf{Ricky Nilsson}$^{1\dagger}$  \quad  \textbf{Alice Zou}$^{1\dagger}$ \\
  \textbf{Matthew Graham}$^{1}$\thanks{Equal senior supervision.}\quad\textbf{Ashish Mahabal}$^{1\ddagger}$
  \\
$^{1}$California Institute of Technology \\
  $^{2}$Massachusetts Institute of Technology  \\
  $^{3}$Purdue University \\
}
\usepackage{xcolor}
\definecolor{hlred}{HTML}{FF0000}
\definecolor{hlyellow}{HTML}{DAA520}
\definecolor{hlgreen}{HTML}{008000}
\usepackage{listings}

\lstdefinestyle{rawouttxt}{
  basicstyle=\ttfamily\footnotesize,
  breaklines=true,
  breakatwhitespace=false,
  columns=fullflexible,
  keepspaces=true,
  showstringspaces=false,
  frame=none,
  escapeinside={(*@}{@*)},
}

\usepackage{footnote}
\usepackage[most]{tcolorbox}
\tcbuselibrary{breakable,listings}

\lstdefinestyle{prompttxt}{
  basicstyle=\ttfamily\footnotesize,
  breaklines=true,
  breakatwhitespace=false,
  columns=fullflexible,
  keepspaces=true,
  frame=none,
}

\tcbset{
  chatturn/.style={
    boxrule=0.4pt,
    arc=2pt,
    left=6pt,
    right=6pt,
    top=2pt,
    bottom=4pt,
    titlerule=0pt,
    toptitle=0.5mm,
    fonttitle=\sffamily\bfseries\small,
  },
  chatblue/.style={
    colback=blue!4!white,
    colframe=blue!50!black,
    colbacktitle=blue!50!black,
    coltitle=white,
  },
  chatgreen/.style={
    colback=green!4!white,
    colframe=green!45!black,
    colbacktitle=green!40!black,
    coltitle=white,
  },
  chatgrey/.style={
    colback=gray!8,
    colframe=gray!55,
    colbacktitle=gray!55,
    coltitle=white,
  },
}

\NewTColorBox{chatmsg}{O{blue} m O{}}{
  chatturn,
  chat#1,
  title={#2},
  #3
}
\makesavenoteenv{chatmsg}

\begin{document}

\maketitle
\begin{abstract}
Modern astronomical observatories generate a massive volume of multimodal data, creating a critical bottleneck for expert human review. While multimodal Large Language Models (LLMs) have shown promise in interpreting complex visual and textual inputs, their ability to perform specialized scientific classification while providing interpretable reasoning remains understudied. We introduce AstroAlertBench, a comprehensive multimodal benchmark designed to evaluate LLMs' performance in astronomical event review along a three-stage logical chain: metadata grounding, scientific reasoning, and hierarchical classification over five categories. We utilize a pilot sample of 1,500 real-world alerts from the Zwicky Transient Facility (ZTF), a wide-field survey that scans the northern sky to detect transient astronomical events. On this dataset, we benchmark 13 frontier closed-source and open-weight LLMs supporting visual input.
Our results reveal that high accuracy does not always align with model ‘honesty'—the ability to self-evaluate its reasoning—impacting its reliability as a real-world assistant. 
We further initialize a human-in-the-loop evaluation protocol as a precursor to future community-scale participation. Together, \textit{AstroAlertBench} provides a framework for developing calibrated and interpretable astronomical assistants.
\end{abstract}

\section{Introduction}

The landscape of artificial intelligence has been fundamentally altered by the emergence of Large Language Models (LLMs) based on the Transformer architecture \citep{vaswani2017attention}. These models, including proprietary series such as GPT \citep{brown2020language, achiam2023gpt} and high-performance open-weight families like Qwen \citep{bai2023qwen, yang2025qwen3}, have demonstrated proficiency across a broad spectrum of zero-shot reasoning tasks \citep{wei2022chain}. The scaling of these architectures has enabled capabilities in multi-step reasoning and the integration of information from diverse sources, typically facilitated by extensive pre-training on diverse corpora followed by instruction tuning and alignment \citep{bai2023qwen}. Furthermore, the integration of visual encoders has birthed a new generation of Multimodal LLMs (MLLMs) capable of processing interleaved visual and textual data \citep{team2024gemini}. By mapping disparate modalities into a shared feature space, these models can address complex, open-ended problems that require the simultaneous interpretation of both semantic and structural visual cues \citep{radford2021learning, alayrac2022flamingo, li2023blip, liu2023visual}.

This progress has prompted a transition toward utilizing LLMs as specialized scientific assistants \citep{taylor2022galactica, romera2024mathematical}. Research is increasingly moving beyond static information retrieval toward the development of autonomous LLM agents designed to execute end-to-end research workflows \citep{boiko2023emergent, bran2023chemcrow}. However, as scientific inquiries increase in complexity, general-purpose models encounter a dual bottleneck. First, high costs for commercial, closed-source APIs often prohibit large-scale deployment on high-volume scientific data streams \citep{pan2024astromlab}. Second, general open-weight models frequently struggle with the long-tail domain knowledge and reasoning required for specialized research tasks \citep{pan2024astromlab, guo2024can}. These limitations create a critical need for comprehensive frameworks to evaluate whether these models possess the underlying scientific logic necessary for reliable deployment.

Addressing these challenges requires a shift from outcome-based metrics toward a staged evaluation logic \citep{guo2024can}. Scientific problem-solving is not merely a task of pattern recognition; it requires a verifiable logical chain that connects perception to reasoning and final application. For LLMs to be integrated into expert discovery pipelines, they must provide more than correct classification labels \citep{stoppa2025textual, lightman2023let}. They must offer interpretable reasoning and accurate self-evaluations regarding their internal logic to ensure reliability in scientific environments \citep{guo2017calibration, kadavath2022language, lightman2023let}.

Astronomy provides a representative environment for evaluating these capabilities. Time-domain surveys, such as the Zwicky Transient Facility (ZTF) \citep{bellm2019zwicky, graham2019zwicky, masci2019zwicky}, generate millions of nightly ``alerts'' signaling potential astrophysical discoveries \citep{masci2019zwicky}. To process this data, the astronomy community relies on machine learning pipelines for automated categorization \citep{duev2019real, mahabal2019machine, forster2021automatic}. However, as noted by \citet{stoppa2025textual}, these models primarily function as black boxes, lacking the human-readable rationales necessary for experts to prioritize follow-up resources. While Multimodal LLMs offer a path toward interpretable review, they face a significant domain gap in specialized astronomical logic, exacerbated by a lack of domain-specific benchmarks \citep{pan2024astromlab}.

\begin{figure}[t]
    \centering
    \includegraphics[width=0.85\linewidth]{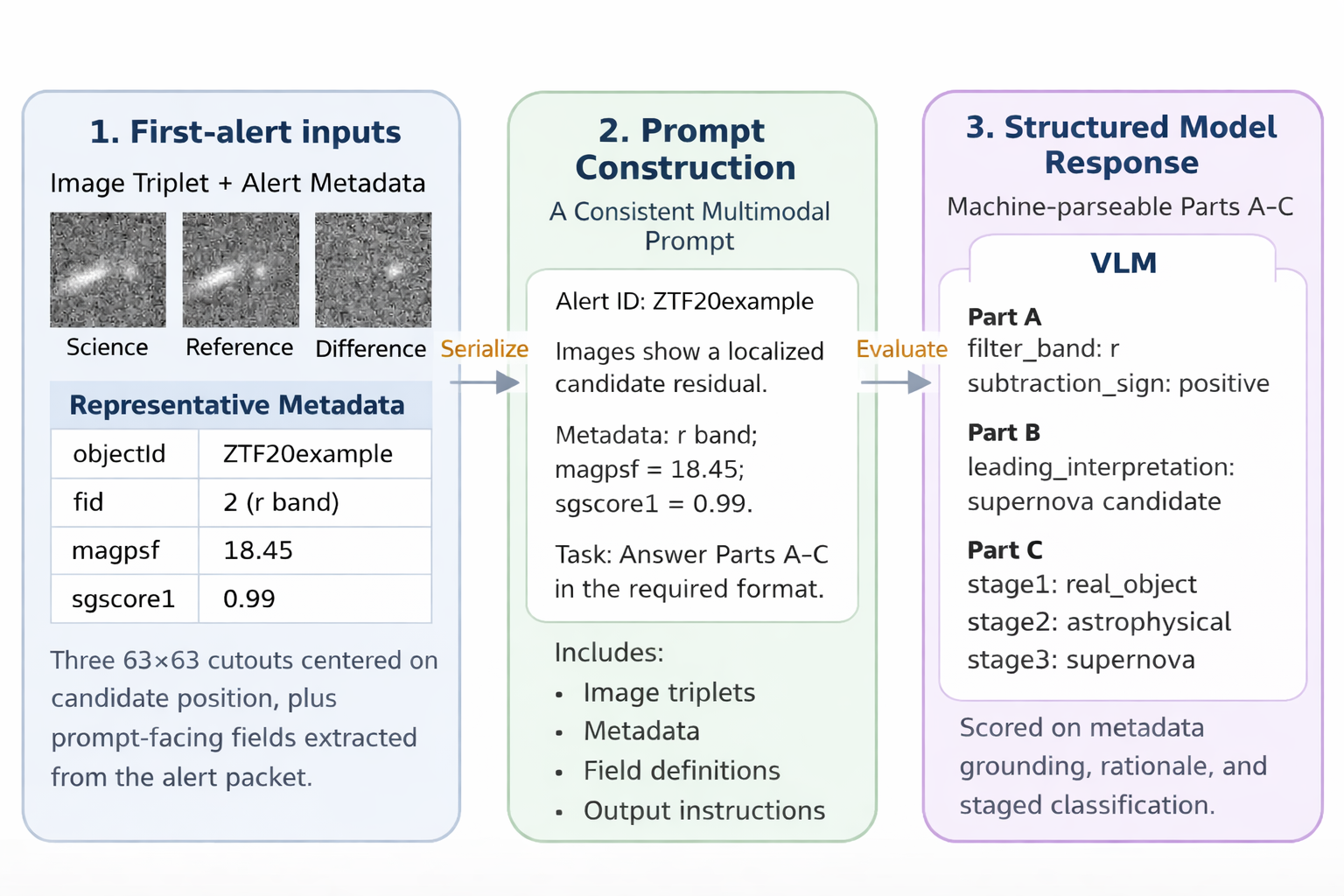}
  \caption{Overview of the AstroAlertBench pipeline. Starting from a first-detection alert, we construct each benchmark example from the ZTF science, reference, and difference cutouts together with selected alert metadata. These inputs are serialized into a standardized multimodal prompt, which is then given to a multimodal language model. The model returns a machine-parseable response containing Part~A metadata grounding, Part~B scientific rationale, and Part~C staged classification decisions, which are evaluated under the AstroAlertBench protocol.}
\label{fig:astroalertbench-pipeline}
\end{figure}

In this work, we introduce \textit{AstroAlertBench}\footnote{The project website, dataset, and code are available at \url{https://astroalertbench.com}, \url{https://huggingface.co/datasets/AnonymousUser16384/AstroAlertBench}, and \url{https://github.com/LLM-for-Astronomy/AstroAlertBench} respectively.}, a multimodal benchmark designed to evaluate LLMs in astronomical event review. We design the evaluation as a three-stage logical chain: (1) \textbf{Metadata Grounding}, assessing the model's ability to accurately read and restate verifiable observational parameters from the alert metadata; (2) \textbf{Scientific Rationale}, assessing the ability to propose astrophysical hypotheses and provide grounded reasoning; and (3) \textbf{Staged Classification}, testing a hierarchical decision process covering artifact rejection, physical origin, and final astrophysical categorization across five operationally meaningful categories. This staged approach allows for the isolation of failures in perception from errors in high-level reasoning \citep{guo2024can}.

A central component of this benchmark is the evaluation of model \textbf{Honesty}---the alignment between a system's internal self-evaluation and its actual performance. We evaluate this through three lenses: (i) whether models that perform better on average are more self-confident, (ii) whether the confidence score assigned to an individual alert reliably signals its accuracy, and (iii) a behavioral ``second-rollout'' test of whether models can successfully fix mistakes on alerts they originally flagged as uncertain. Across these dimensions, we benchmark 13 model configurations from the GPT, Claude, Gemini, Qwen, and Kimi families, using human baselines as a performance reference. Together, AstroAlertBench provides a framework for evaluating the scientific reasoning and reliability of multimodal LLMs as  astronomical assistants.

\section{Related Work}

\subsection{Astronomical Alert Triage and Broker Systems}
The operational demands of modern time-domain surveys, most notably the Zwicky Transient Facility (ZTF) \citep{bellm2019zwicky, graham2019zwicky, masci2019zwicky}, have necessitated the development of high-throughput alert distribution and filtering systems \citep{patterson2019zwicky}. These surveys generate millions of real-time alerts that must be triaged to separate instrumental or subtraction artifacts from genuine astrophysical phenomena \citep{duev2019real, mahabal2019machine}. Community broker systems, such as the Automatic Learning for the Rapid Classification of Events (ALeRCE), utilize specialized machine-learning pipelines to discriminate between classes such as supernovae, variable stars, and active galactic nuclei (AGN) \citep{carrasco2021alert, forster2021automatic}. Similarly, the Bright Transient Survey (BTS) utilizes deep-learning frameworks like BTSbot to automate the identification and follow-up of bright transients \citep{rehemtulla2024zwicky}. While these traditional models achieve high classification accuracy, they predominantly function as ``black boxes,'' providing scores or labels without the human-readable scientific rationales required for expert review and resource prioritization.

\subsection{Multimodal Reasoning and Scientific Benchmarks}
The evaluation of Vision-Language Models (VLMs) has shifted from basic object recognition toward assessing complex, multi-discipline reasoning capabilities \citep{radford2021learning, alayrac2022flamingo, li2023blip, liu2023visual}. Large-scale benchmarks such as \citet{yue2024mmmu} and \citet{lu2023mathvista} have established baseline performance for frontier models like GPT-4 \citep{achiam2023gpt}, Gemini \citep{team2024gemini}, and Qwen3 \citep{yang2025qwen3} across college-level academic tasks. However, as scientific inquiries move into specialized domains, general-purpose benchmarks often fail to capture the long-tail knowledge and cascading complexity required for research-grade problem solving \citep{taylor2022galactica, romera2024mathematical}. To address this, recent work has introduced domain-specific benchmarks in fields like chemistry and laboratory automation \citep{boiko2023emergent, bran2023chemcrow}. Specifically, \citet{guo2024can} developed staged reasoning protocols for molecular structure elucidation, while recent efforts in scientific discovery benchmarking have begun to decompose tractable tasks into hierarchical sub-goals, such as inspiration retrieval and hypothesis composition \citep{liu2025researchbench}.

Within the astronomical domain, early efforts to evaluate Large Language Models have remained largely unimodal and restricted in their classification scope. For instance, \citet{pan2024astromlab} introduced a suite to assess textual domain expertise and fact-retrieval through multiple-choice questions, while \citet{stoppa2025textual} utilized foundation models for the textual interpretation of binary real-versus-bogus image classifications. These approaches, however, are fundamentally limited by their reliance on single-modality inputs and simplified decision tasks—either focusing exclusively on textual scientific knowledge retrieval \citep{pan2024astromlab} or providing binary interpretations based solely on visual stamps without integrating the accompanying numerical alert metadata \citep{stoppa2025textual}. \textbf{In this work, we present \textit{AstroAlertBench} as an advancement that bridges and expands upon these efforts} by introducing a multimodal, three-stage evaluation framework that addresses the complex requirements of multi-class alert triage. By requiring models to ground high-level astrophysical rationales in both raw image triplets and serialized alert parameters across a five-category taxonomy, our benchmark establishes a more comprehensive standard for evaluating the reliability, scientific logic, and internal calibration of LLM agents in time-domain astronomy.
\vspace{-0.5em}
\section{The AstroAlertBench Dataset}

\begin{figure}[h!]
    \centering
    \includegraphics[width=0.65\linewidth]{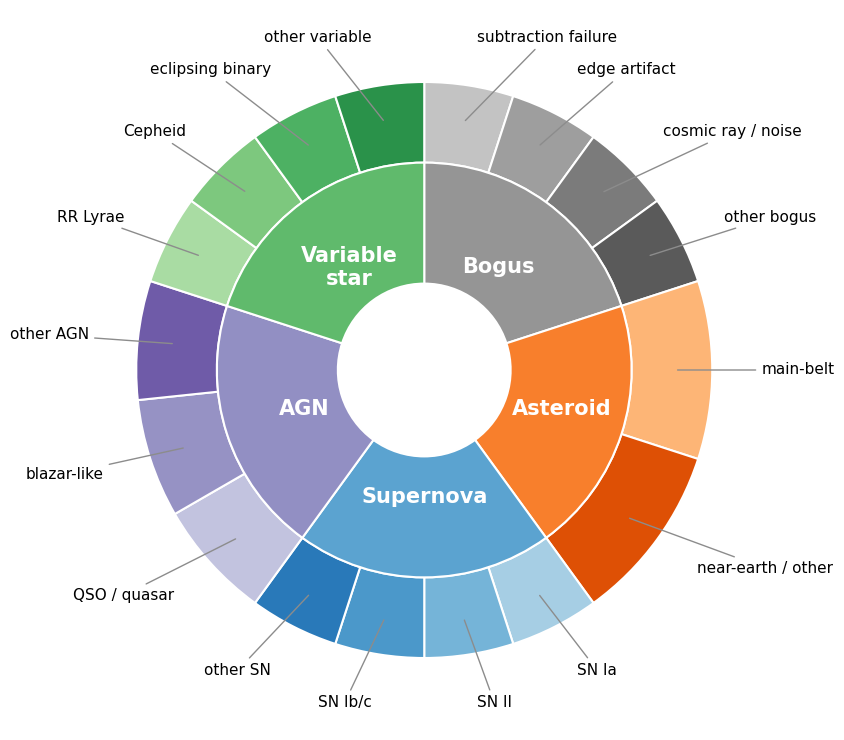}
    \caption{\textbf{AstroAlertBench taxonomy.} The inner ring shows the five benchmark classes with equal weight: \textit{bogus}, \textit{asteroid}, \textit{supernova}, \textit{AGN}, and \textit{variable star}. The outer ring illustrates representative subtype categories; this is not an exhaustive classification tree. Wedge sizes are schematic and do not encode subtype counts.}
    \label{fig:astroalertbench-taxonomy}
\end{figure}
\vspace{-0.25em}
This section details the construction and composition of \textbf{AstroAlertBench}, focusing on the transition from raw astronomical data to a structured multimodal benchmark. We decompose the dataset generation into data acquisition, automated visual preprocessing, and taxonomic design.

\subsection{Data Acquisition and Multimodal Content}
We utilize the fully public ALeRCE real-time stamp classifier dataset \citep{carrasco2021alert} as the authoritative source for ground-truth labels and alert-level inputs. This dataset is derived from the Zwicky Transient Facility (ZTF) alert stream \citep{patterson2019zwicky}, which issues nightly notifications of transient, variable, and moving objects. Each benchmark example is constructed from the \textbf{first-detection alert} of a unique astrophysical object to evaluate rapid triage capabilities without reliance on multi-epoch light curves.

Each datapoint consists of two primary modalities:
\begin{itemize}
    \item \textbf{Image Triplet:} Three $63 \times 63$ pixel cutouts centered on the candidate position: the \textit{Science} image (current observation), the \textit{Reference} image (historical baseline), and the \textit{Difference} image (Science minus Reference).
   \item \textbf{Alert Metadata:} A subset of 19 prompt-facing features extracted from the ZTF AVRO packet, including photometric attributes (e.g., PSF magnitude), spatial and morphological context (e.g., distance to the nearest reference source), and historical detection counts.
\end{itemize}

\subsection{Automated Image Preprocessing and Serialization}
To ensure 32-bit floating-point astronomical data is compatible with standard 8-bit Vision-Language Model encoders, we implement an automated visual pipeline. 

\paragraph{Visual Normalization.} 


Our image input for one alert is a combination image of three corresponding images from ZTF: science, reference, and difference of 63×63 pixels placed side by side, with text labels above each panel (Science, Reference, Difference) so panel identity and order are unambiguous. The reference image is the historical baseline at the same sky position; the difference image is the science-minus-reference subtraction. The transient is taken to lie at the geometric center of each image, so localization is well defined despite the compact field of view. 

\paragraph{Multimodal Prompt Construction.} As illustrated in Figure~\ref{fig:astroalertbench-pipeline}, the preprocessed image cutouts are tiled horizontally into a single image strip. 
Metadata is serialized into a compact human-readable format, providing the model with observational context alongside explicit field definitions (see Appendix \ref{app:main prompt} for the full prompt schema).

\subsection{Benchmark Taxonomy and Evaluation Set Construction}

Mirroring the real-time filtering standards used by major astronomical data brokers \citep{carrasco2021alert, rehemtulla2024zwicky}, AstroAlertBench assesses LLMs across five primary categories: \textit{bogus}, \textit{asteroid}, \textit{supernova}, \textit{AGN}, and \textit{variable star} (Figure~\ref{fig:astroalertbench-taxonomy}). This structure evaluates a model's ability to reject artifacts and classify astronomical objects. We construct a balanced evaluation set of 1,500 examples (300 per class) to ensure that benchmark scores reflect multimodal reasoning rather than majority-class priors, as the natural ZTF alert stream is highly imbalanced \citep{rehemtulla2024zwicky}. Details in dataset construction are in Appendix~\ref{app: data}.


\section{The AstroAlertBench Discovery Chain and Results}
\label{sec:results}

\subsection{Protocol Overview}
\label{sec:protocol-overview}

AstroAlertBench evaluates model performance through a three-stage discovery chain designed to isolate failures in data perception from errors in high-level reasoning: \textbf{Metadata Grounding} (Perception), \textbf{Scientific Rationale} (Reasoning), and \textbf{Staged Classification} (Decision). We benchmark 13 models across various architectures and reasoning modes; specific model configurations and hyperparameters are detailed in Appendix~\ref{app:model-configs}. To provide context for these results, we establish a human baseline on a representative subset of the benchmark
(Appendix~\ref{app:human-baseline}) and initialize a Zooniverse project\footnote{\url{https://www.zooniverse.org/projects/AstroAlert}}
as a precursor to future community-scale human-in-the-loop participation.
\subsection{Part A: Metadata Grounding}
\label{sec:results-parta}


Part~A evaluates metadata perception to ensure that reasoning failures are not caused by grounding errors. Models must map raw numerical parameters to their scientific definitions using a mapping key provided in the prompt. Full details of these definitions and the prompt template are provided in Appendix~\ref{app:main prompt}.
For example, given the raw metadata \texttt{fid: 1}, the model must cross-reference the provided information table to correctly translate the numerical filter ID (\texttt{fid}) into its scientific name, mapping \texttt{1} to the \textit{g}-band filter \citep{patterson2019zwicky}. We evaluate this interpretive mapping across six objective fields: \texttt{fid} (filter ID), \texttt{isdiffpos} (difference-residual sign), \texttt{magpsf} (PSF magnitude), \texttt{sigmapsf} (magnitude error), \texttt{ndethist} (historical detections), and \texttt{ncovhist} (historical coverages). Success is measured by a match with the ground-truth alert packet. \textbf{All 13 evaluated models achieved 100\% accuracy on Part~A across all six metadata fields}, confirming they correctly receive, parse, and relate the multimodal input. This result effectively isolates failures in the subsequent Reasoning (Part~B) and Decision (Part~C) stages to scientific logic rather than cross-modal serialization or input-parsing errors. The full per-field accuracy results for each model are reported in Appendix~\ref{app:metric-definitions}.

\subsection{Part B: Scientific Rationale}
\label{sec:results-partb}

In Part B, models must produce a structured scientific justification across three dimensions: \textbf{key evidence}, \textbf{leading interpretation}, and \textbf{alternative analysis}. Following each rationale, the model assigns itself a quality score (0--5) for each dimension based on the rubric in \citet{stoppa2025textual}. We summarize these responses using the \textit{mean self-reasoning score} and the \textit{self pass rate} (the proportion of alerts averaging $\ge 4$ on the rubric). 

Detailed performance metrics for all 13 runs are provided in Appendix~\ref{app: part b additional}. In summary, self-scores clustered near the top of the rubric for most models, with 11 of 13 runs exceeding a 90\% self pass rate. \texttt{Claude Opus 4.7} established the floor for self-evaluation at $3.99$, while \texttt{Gemini 2.5 Pro} reported the highest mean confidence at $4.89$. Because absolute self-scores exhibit high ceilings and low variance across most families, they carry limited discriminative power as raw metrics; their primary value lies in the calibration and honesty analysis developed in Section~\ref{sec:honesty}.




\subsection{Part C: Staged Classification}
\label{sec:results-partc}
\begin{figure}[t]
    \centering
    \includegraphics[width=0.8\linewidth]{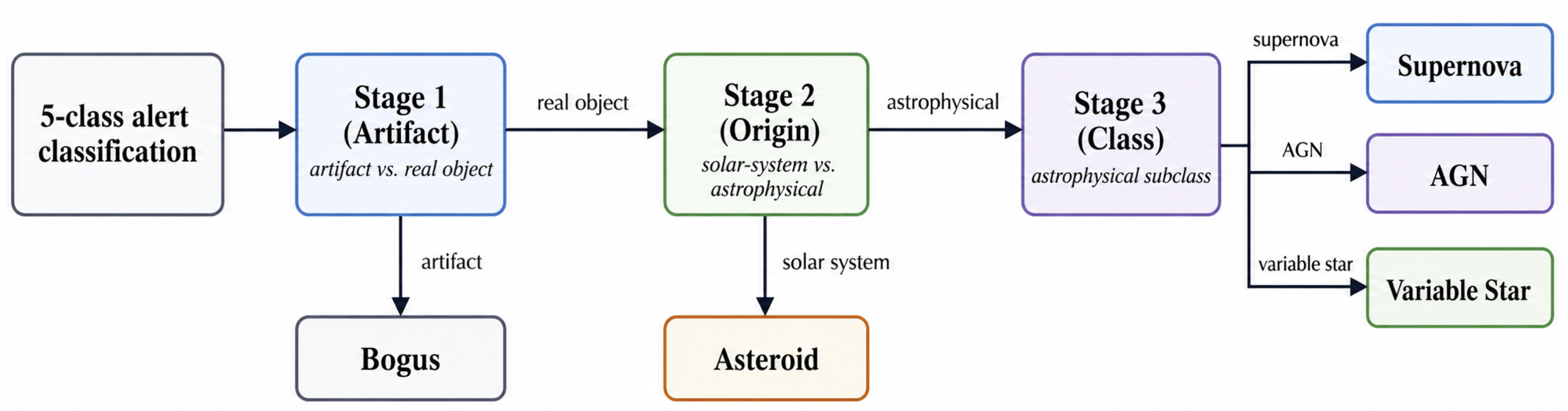}
    \caption{\textbf{Hierarchical classification cascade.} The five benchmark categories are mapped through three sequential decision stages: artifact rejection (Stage-1), physical origin (Stage-2), and astrophysical subclass assignment (Stage-3).}
    \label{fig:classification-hierarchy}
\end{figure}
\begin{figure}[t]
    \centering
    \IfFileExists{figure/perclass_heatmap.png}{\includegraphics[width=0.75\linewidth]{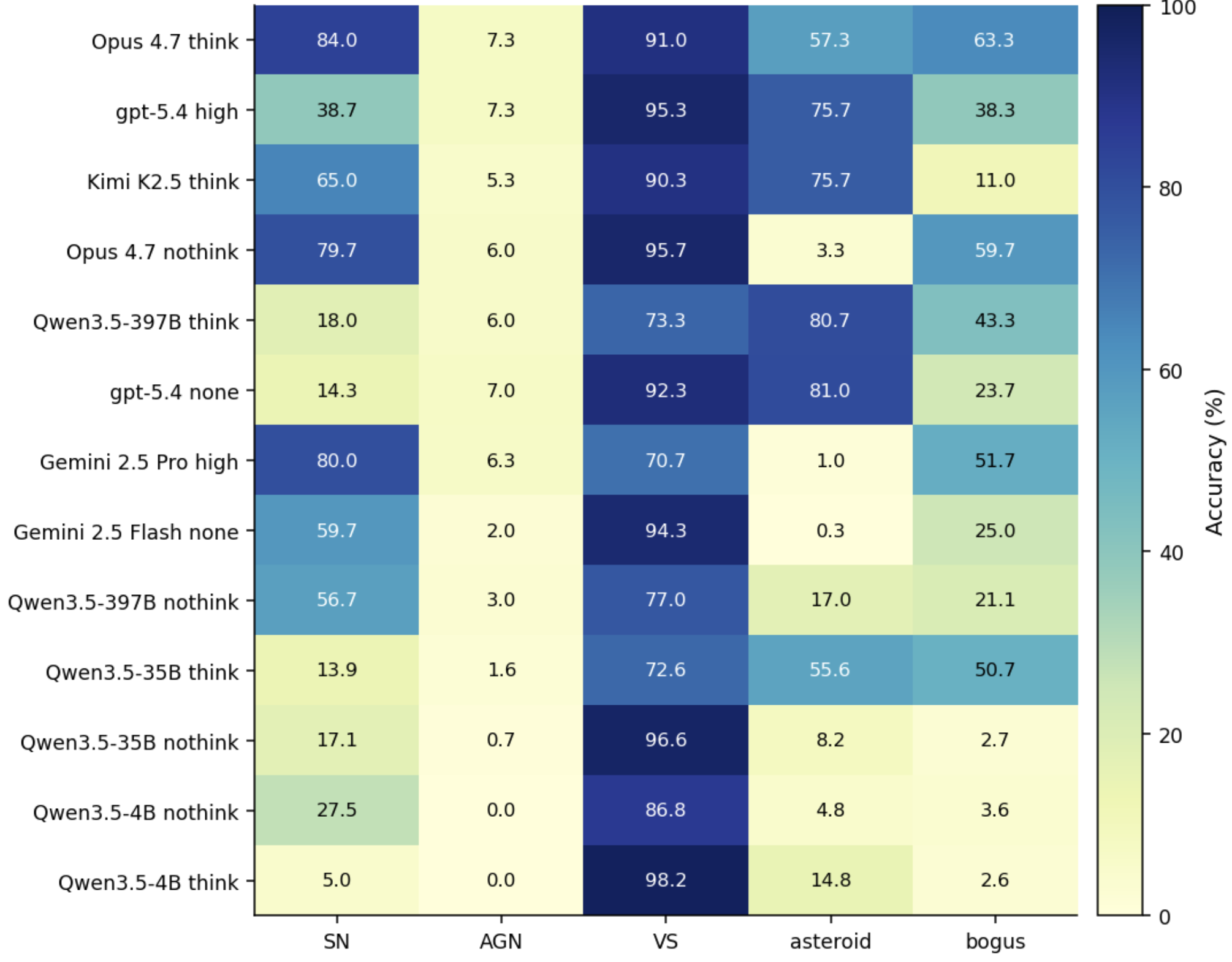}}{\fbox{\parbox[c][5cm][c]{0.85\linewidth}{\centering\itshape [Figure placeholder: figure/perclass\_heatmap.png \textemdash{} per-class accuracy heatmap, 13 runs $\times$ 5 classes]}}}
    \caption{\textbf{Per-class accuracy heatmap across 13 runs.} Rows are the 13 evaluated models, sorted top-to-bottom by absolute 5-class accuracy; columns are the five benchmark classes (\textit{supernova}, \textit{AGN}, \textit{variable star}, \textit{asteroid}, \textit{bogus}).
    }
    \label{fig:perclass-heatmap}
\end{figure}

Part~C decomposes the five-class decision into a hierarchical cascade: \textbf{Stage-1} distinguishes instrumental \textit{artifacts} from \textit{real\_object}s, \textbf{Stage-2} routes real objects to \textit{solar\_system} or \textit{astrophysical}, and \textbf{Stage-3} assigns the specific astrophysical subclass among \textit{supernova}, \textit{variable\_star}, and \textit{AGN}. Figure~\ref{fig:classification-hierarchy} maps the five benchmark classes to this cascade. We measure per-stage accuracy, conditional accuracy at each decision point (restricted to where the previous stages were correct), and end-to-end 5-class accuracy. Table~\ref{tab:partc-cascade} shows the stage-wise results; Figure~\ref{fig:perclass-heatmap} shows the per-class accuracy heatmap.


\begin{table}[H]
\centering
\footnotesize
\setlength{\tabcolsep}{4pt}
\resizebox{\linewidth}{!}{%
\begin{tabular}{clcccccc}
\toprule
\textbf{\#} & \textbf{Run} & \textbf{Stage-1} & \textbf{Stage-2} & \textbf{Stage-3} & \textbf{Stage-3 cond.} & \textbf{5-class (E2E)} & \textbf{S3 macro-F1} \\
\midrule
\rowcolor{gray!15}
--- & Ensemble Human Expert\textsuperscript{$\ddagger$} & --- & --- & --- & --- & $30.67 \pm 5.32$ & --- \\
\midrule
1  & Claude Opus 4.7 think       & \underline{$86.60 \pm 0.88$} & $\mathbf{81.60 \pm 1.00}$ & $\mathbf{66.80 \pm 1.22} $& $\mathbf{60.78 \pm 1.26}$& $\mathbf{60.60 \pm 1.26}$ & $\mathbf{0.558} $\\
2  & GPT-5.4 high-think          & $83.93 \pm 0.95$ & \underline{$80.00 \pm 1.03$} & $59.40 \pm 1.27$ & $47.11 \pm 1.29$ & \underline{$51.07 \pm 1.29$} & $0.435$ \\
3  & Kimi K2.5 think             & $81.36 \pm 1.01$ & $71.94 \pm 1.16$ & \underline{$60.19 \pm 1.27$} & $53.45 \pm 1.29$ & $49.34 \pm 1.29$ & $0.500$ \\
4  & Claude Opus 4.7 nothink     & $85.20 \pm 0.92$ & $70.93 \pm 1.17$ & $54.47 \pm 1.29$ & \underline{$60.44 \pm 1.26$} & $48.87 \pm 1.29$ & \underline{$0.548$} \\
5  & Qwen3.5-397B-A17B think     & $80.87 \pm 1.02$ & $72.80 \pm 1.15$ & $52.80 \pm 1.29$ & $32.44 \pm 1.21$ & $44.27 \pm 1.28$ & $0.310$ \\
6  & GPT-5.4 no-think            & $77.47 \pm 1.08$ & $71.53 \pm 1.17$ & $57.00 \pm 1.28$ & $37.89 \pm 1.25$ & $43.67 \pm 1.28$ & $0.329$ \\
7  & Gemini 2.5 Pro high-think   & $82.27 \pm 0.99$ & $63.60 \pm 1.24$ & $45.53 \pm 1.29$ & $52.33 \pm 1.29$ & $41.93 \pm 1.27$ & $0.509$ \\
8  & Gemini 2.5 Flash no-think   & $83.47 \pm 0.96$ & $63.73 \pm 1.24$ & $37.80 \pm 1.25$ & $52.00 \pm 1.29$ & $36.27 \pm 1.24$ & $0.456$ \\
9  & Qwen3.5-397B-A17B nothink   & $79.17 \pm 1.05$ & $60.95 \pm 1.26$ & $40.39 \pm 1.27$ & $45.61 \pm 1.29$ & $34.98 \pm 1.23$ & $0.430$ \\
10 & Qwen3.5-35B-A3B think       & $77.15 \pm 1.35$ & $62.98 \pm 1.55$ & $50.98 \pm 1.61$ & $32.12 \pm 1.50$ & $26.73 \pm 1.14$ & $0.271$ \\
11 & Qwen3.5-35B-A3B nothink     & $80.40 \pm 1.03$ & $60.88 \pm 1.27$ & $31.65 \pm 1.21$ & $37.51 \pm 1.26$ & $25.50 \pm 1.13$ & $0.273$ \\
12 & Qwen3.5-4B nothink          & $79.30 \pm 1.10$ & $54.21 \pm 1.35$ & $28.68 \pm 1.22$ & $38.67 \pm 1.32$ & $22.32 \pm 1.08$ & $0.307$ \\
13 & Qwen3.5-4B think            & $\mathbf{87.07 \pm 1.88}$ & $78.86 \pm 2.29$ & $42.59 \pm 2.78$ & $45.60 \pm 2.80$ & $8.41 \pm 0.72$ & $0.241$ \\
\bottomrule
\end{tabular}%
}
\caption{\textbf{Part~C staged classification performance.} Metrics include artifact rejection (Stage-1), origin identification (Stage-2), subclass assignment (Stage-3), conditional subclass accuracy (Stage-3 cond.), joint end-to-end accuracy (5-class), and macro-F1 for astrophysical subclasses (S3 macro-F1). \textbf{Bold} and \underline{underline} denote the first and second best performance in each column, respectively. Results are reported over $1,500$ alerts; $\ddagger$ denotes ensemble human performance (see Appendix~\ref{app:human-baseline} for details).}
\label{tab:partc-cascade}
\end{table}

\paragraph{Five-class classification performance.}
As shown in Table~\ref{tab:partc-cascade}, \texttt{Claude Opus~4.7 think} leads the benchmark with $60.60 \pm 1.26\%$ end-to-end 5-class accuracy, representing a significant $9.53$ percentage point lead over the second-ranked \texttt{GPT-5.4 high-think}. The three highest-ranking models---\texttt{Opus 4.7 think}, \texttt{GPT-5.4 high-think}, and \texttt{Kimi K2.5 think}---all utilize reasoning-enabled modes. Notably, \texttt{Kimi K2.5 think} achieves the highest accuracy among the open-weight models, outperforming several frontier closed-source configurations.


\paragraph{Staged decision results.}
\textbf{Stage-1} (real-vs-artifact) exhibits consistent high performance across all model scales, with an accuracy spread of $77.15$--$87.07\%$. \textbf{Stage-2} (physical origin) begins to separate the closed-source frontier and reasoning-enabled open-weight models from standard configurations. Leading results from \texttt{Claude Opus 4.7 think} ($81.60\%$) and \texttt{GPT-5.4 high-think} ($80.00\%$) are joined by strong open-weight performances, such as \texttt{Kimi K2.5 think} ($71.94\%$). \textbf{Stage-3 conditional accuracy} reveals the most significant performance gap. Its spread of $32.12$--$60.78\%$ is among the largest in Table~\ref{tab:partc-cascade}, indicating that while most models effectively clear the initial artifact filter, few maintain high accuracy in fine-grained astrophysical categorization.

\paragraph{Class-level failure modes.}
Figure~\ref{fig:perclass-heatmap} reveals three universal patterns: (i) \textbf{AGN-VS Confusion}: AGN accuracy remains below $8\%$ for all models, with sources overwhelmingly misclassified as variable stars. (ii) \textbf{Gemini's asteroid collapse}: The \texttt{Gemini 2.5} family exhibits a localized blind spot on asteroids ($\leq 1.0\%$), despite \texttt{2.5~Pro} otherwise being competitive on the supernova class at $80.0\%$. (iii) \textbf{Qwen bogus collapse}: Most smaller \texttt{Qwen} variants fail to identify the \textit{bogus} class ($<4\%$), driven by a strong majority-class bias in Stage-1 that predicts \textit{real\_object} for nearly all inputs.

\paragraph{Impact of reasoning.}
To quantify the impact of reasoning, we compare model performance across standard and thinking-enabled configurations, measuring the accuracy improvement provided by internal chain-of-thought. For the three strongest models, enabling adaptive or high-thought reasoning yields significant accuracy gains: \texttt{Claude Opus 4.7} ($+11.73\%$), \texttt{Qwen3.5-397B-A17B} ($+9.29 \%$), and \texttt{GPT-5.4} ($+7.40 \%$). This "think-win" margin scales with overall model capacity, whereas the mid-sized \texttt{Qwen3.5-35B-A3B} result is a statistical tie ($+1.23 \%$). Notably, the smallest \texttt{Qwen3.5-4B} exhibits a performance regression when forced into a thinking mode. Detailed statistical comparisons, including $z$-scores and standard errors for all pairs, are provided in Appendix~\ref{app:reasoning-ablation}.



\section{Evaluating Model Honesty and Scientific Reliability}
\label{sec:honesty}

While high accuracy is a primary benchmark objective, the practical utility of an LLM as a scientific assistant is fundamentally bounded by its calibration. In high-volume fields like time-domain astronomy, where follow-up observations are a scarce resource, a system that cannot accurately signal its own uncertainty poses a significant risk to efficient triage. We therefore evaluate the ``honesty'' of these models—defined as the alignment between a model's internal self-evaluation and its objective correctness—through three distinct lenses: the \textbf{macro} view (\S\ref{sec:honesty-macro}) investigating population-level modesty, the \textbf{inner} view (\S\ref{sec:honesty-inner}) assessing per-alert calibration, and the \textbf{behavioral} view (\S\ref{sec:honesty-retry}) testing iterative error recovery. These findings are supplemented by qualitative analysis and expert reviews case studies in Appendix~\ref{app:ensemble-deepdive}.

\subsection{Population-level trends: Performance versus modesty}
\label{sec:honesty-macro}

\begin{figure}[h]
    \centering
    \IfFileExists{figure/msrs_vs_accuracy.png}{\includegraphics[width=0.7\linewidth]{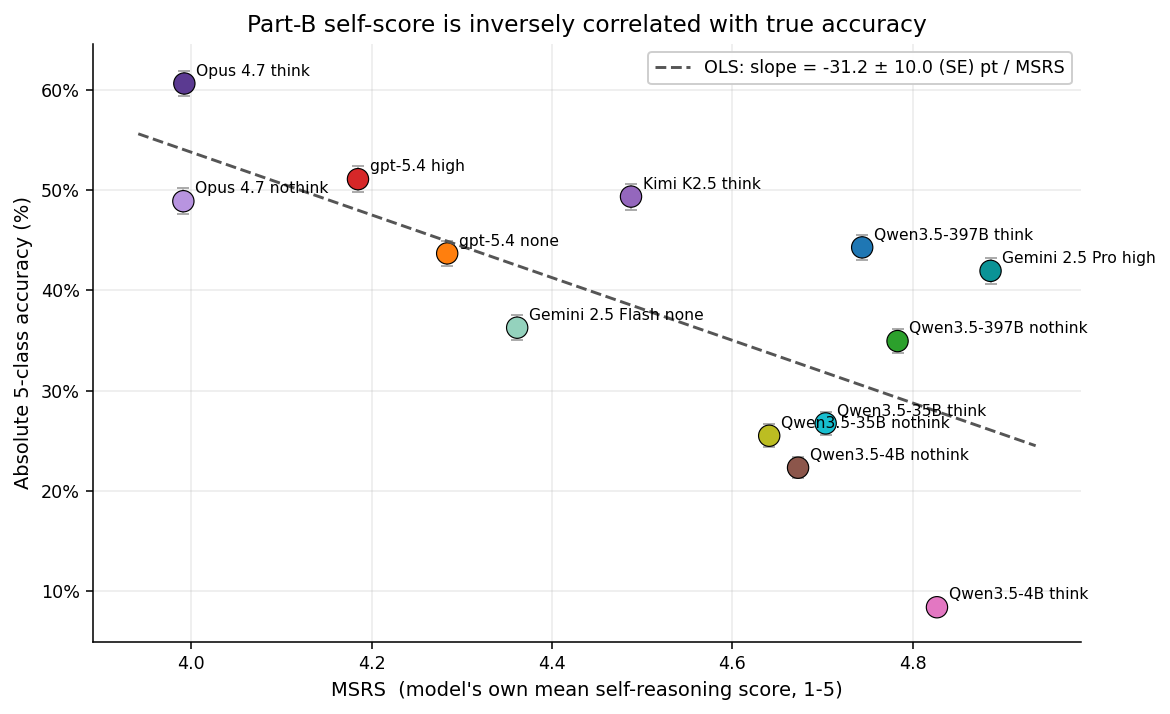}}{\fbox{\parbox[c][5cm][c]{0.7\linewidth}{\centering\itshape [Figure placeholder: cross-model scatter of mean self-reasoning score versus accuracy]}}}
    \caption{\textbf{Inverse correlation between self-assessment and accuracy.} Each point represents one of the 13 evaluated runs ($n=1,500$). The dashed line indicates an OLS fit with a slope of $-31.2 \pm 10.0$ percentage points of accuracy per unit on the 0--5 self-reasoning scale. Models with higher empirical accuracy tend to be more modest in their self-evaluations.} 
    \label{fig:msrs-vs-accuracy}
\end{figure}

Figure~\ref{fig:msrs-vs-accuracy} reveals a counterintuitive population-level trend: models with higher classification accuracy are consistently more modest in their self-evaluations. An ordinary least-squares fit through the 13 models yields a negative slope of $-31.2 \pm 10.0$ percentage points of accuracy per unit on the 0--5 self-reasoning scale. The most capable models are also the most self-critical. The two \texttt{Claude Opus 4.7} runs occupy the upper-left of the plane, pairing the highest accuracies in the batch with the lowest mean self-ratings ($\approx 4.0$). Conversely, the smaller \texttt{Qwen3.5} model configurations cluster in the lower-right, reporting near-perfect self-scores ($\geq 4.6$) despite having mid-to-bottom tier empirical performance. 

\subsection{Instance-level analysis: Calibration of individual models}
\label{sec:honesty-inner}

While stronger models are more modest on average, a self-assessment score is only useful in discovery workflows if it can reliably distinguish correct classifications from errors within the same run. We evaluate this per-alert reliability using the Pearson correlation between confidence and correctness, where values near zero indicate uninformative self-confidence and larger values indicate that scores meaningfully separate correct from incorrect predictions. Detailed rankings and auxiliary metrics are provided in Appendix~\ref{app:calibration}.

Only a handful of models exhibit a meaningful instance-level link between confidence and performance. \texttt{Claude Opus 4.7 nothink} leads the cohort because its mistakes are primarily concentrated in the alerts where it reported lower self-scores. Both \texttt{GPT-5.4} configurations and \texttt{Gemini 2.5 Flash no-think} show a similar pattern, as their self-ratings vary enough to distinguish likely successes from potential errors. In contrast, the remaining runs—including the \texttt{Qwen3.5} think variants and \texttt{Gemini 2.5 Pro high-think}—assign high confidence scores to nearly all outputs regardless of actual correctness, rendering their ratings uninformative. Most strikingly, while adaptive thinking on \texttt{Opus 4.7} significantly improves absolute accuracy, it effectively silences this self-evaluation signal by producing uniformly high confidence scores across both easy and difficult alerts.

\subsection{Behavioral: second-rollout retry on low-confidence rows}
\label{sec:honesty-retry}

A within-model confidence dial is only useful if the model can act on it. We test this directly: for each closed-source run, we draw a stratified subset of low-confidence alerts (instances the model itself flagged as uncertain on the first pass), re-prompt the model with its own first-pass answer attached, and ask for a second pass without revealing the ground-truth classification. Full prompts and results are in Appendix~\ref{app:second roll out prompt} and Appendix~\ref{app:second-rollout}. An honest, well-calibrated model should fix some of the classifications it was wrong about without breaking the others it was right about.

\begin{figure}[h]
    \centering
    \includegraphics[width=0.95\linewidth]{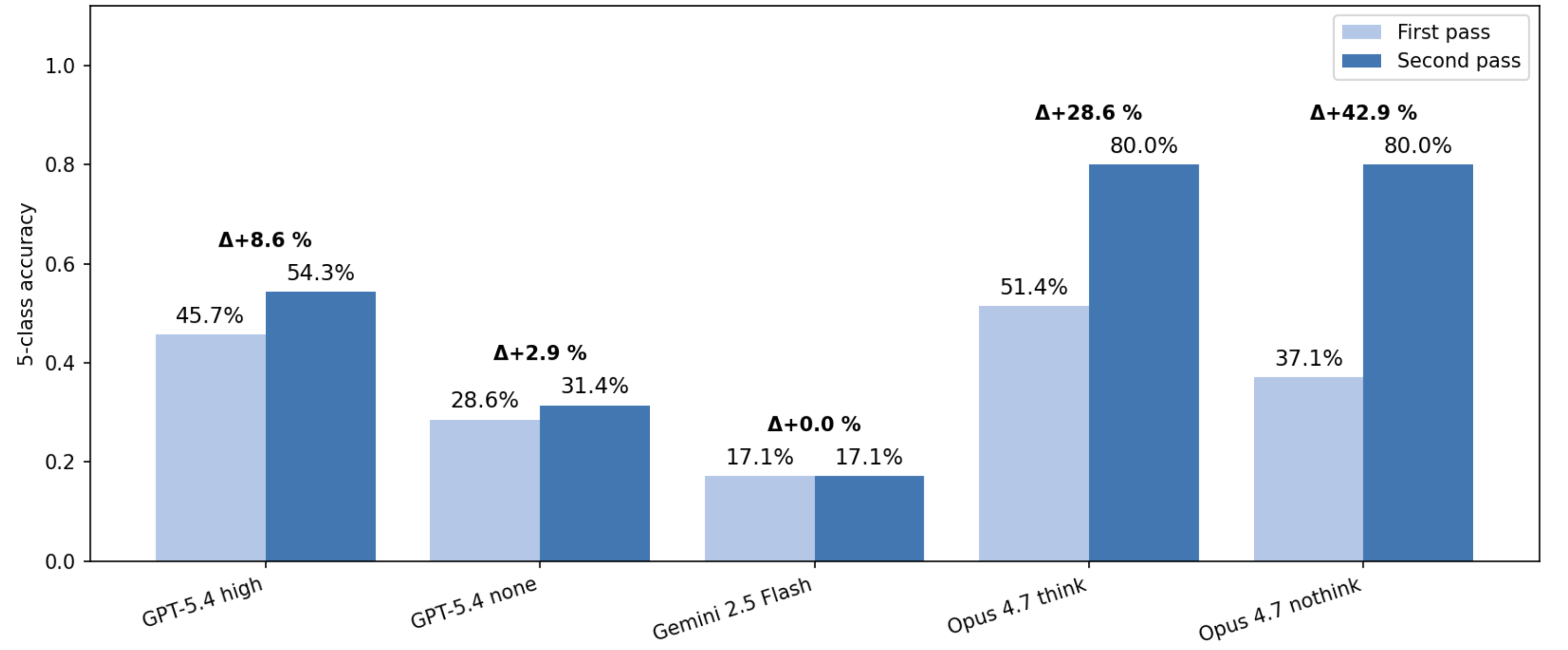}
    \caption{\textbf{Second-rollout retry on the low-confidence subset.} Comparison of first-pass and second-pass accuracy over the same 35 alerts per model, drawn from each model's original low-confidence alert pool. Gains ($\Delta$) are reported in absolute percentage points. 
    }
    \label{fig:second-rollout-acc}
\end{figure}

Figure~\ref{fig:second-rollout-acc} summarizes the results of providing models with a second opportunity to classify the alerts for which they originally reported low confidence. Three distinct behavioral regimes emerge from this test. \texttt{Claude Opus 4.7} (in both reasoning modes) delivers a highly effective repair process, achieving substantial accuracy gains on the retry without breaking a single answer that was already correct on the first pass. In contrast, \texttt{GPT-5.4 high-think} exhibits a trade-off where repairs are balanced against regressions; while it corrects some initial errors, it simultaneously breaks nearly one in five of the answers it previously got right. The \texttt{GPT-5.4 no-think} configuration fails to achieve any meaningful improvement. Finally, \texttt{Gemini 2.5 Flash} demonstrates a complete mismatch between self-rating and performance. Although the model assigns itself significantly higher self-ratings during the second pass, it does not change a single classification outcome, leaving its accuracy entirely static. This serves as direct evidence in this benchmark that \textbf{a model sounding more sure of itself does not always lead to improved scientific results}.



\section{Discussion and Conclusion}
\label{sec:conclusion}

In this work, we introduced \textit{AstroAlertBench}, a multimodal benchmark for evaluating Large Language Models in the specialized domain of time-domain astronomy. Our three-stage evaluation shows that while frontier models excel at metadata perception and can exceed human classification accuracy, they frequently fail to recognize their own errors. We identified systematic blind spots, such as a uniform inability to distinguish active galactic nuclei (AGN) from variable stars and family-specific failures like the Gemini family's collapse on asteroids. These results emphasize that high accuracy alone does not guarantee scientific reliability; a model that sounds confident while making mistakes is difficult to trust in operational workflows. \textit{AstroAlertBench} establishes a comprehensive benchmark for LLMs in astronomy and provides a standardized framework for evaluating model honesty.

Our findings suggest several promising directions for developing autonomous agents. The difficulty in classifying AGNs indicates that future research should prioritize incorporating time-series data, such as light curves, to provide necessary temporal context. Besides, the successful self-correction observed in frontier models suggests that the most capable systems are ready to transition from passive observers to active participants in discovery loops. By serving as autonomous coordinators, such agents could synthesize multi-survey data to prioritize high-value transients and automate the drafting of follow-up requests. Finally, integrating tool-use to query external catalogs may eventually facilitate autonomous hypothesis verification against known astronomical records.

\section{Acknowledgment}
Matthew Graham acknowledges support from NSF grant PHY-2117997. This work was supported in part by compute resources from the Thinking Machines Lab Tinker Research Grant awarded to Claire Chen. The authors thank Shrinivas Kulkarni, Dale Frail, and Xiaoxi Zhou for their valuable discussions and insights.

\newpage
\bibliographystyle{apalike}
\bibliography{bibliography}

\clearpage
\appendix
\tableofcontents
\section{Implementation Details and Resource Analysis}
\label{app:implementation}

This appendix provides the technical specifics for the 13 evaluated model configurations. We detail the hyperparameters used and a comprehensive breakdown of the computational time associated with the 1,500-alert benchmark rollout.
\subsection{Benchmark Implementations}
We set temperature to \texttt{0.2} when the API accepts it. This is because we believe that our tasks should have fixed answers and thus require a more deterministic and consistent output. OpenAI's GPT-5.4 and Claude Opus 4.7 omit \texttt{temperature}; Google and Tinker use \texttt{0.2}. No top-$p$ or frequency-penalty tuning on this path. The maximum allowed output tokens is set to 2048, and the limit is raised to 20\,000 for reasoning models so traces can finish before the JSON. 
\begin{table}[H]
  \centering
  \begin{tabular}{@{}ll@{}}
    \toprule
    Setting & Value \\
    \midrule
    \texttt{temperature} & 0.2 \\
    \texttt{max\_tokens} (non-reasoning) & 2\,048 \\
    \texttt{max\_tokens} (reasoning) & 20\,000 \\
    \texttt{num\_samples} & 1 \\
    Image input & 1 \\
    \bottomrule
  \end{tabular}
    \caption{Primary decoding parameters (released inference code).}
  \label{tab:inference-hparams}
\end{table}

\subsection{Model Configurations}
\label{app:model-configs}
Table~\ref{tab:model_sources_full} and Table~\ref{tab:models-fullbench} provides the technical specifications and primary references for the models evaluated in our benchmarks.

\begin{table}[ht]
  \centering
  \scriptsize
  \begin{tabular}{clll}
    \toprule
    \# & Configuration Name & Params & Reference \\
    \midrule
    \rowcolor{gray!5} 1 & Claude Opus 4.7 think & Proprietary & \citep{anthropic2026claudeopus47} \\
    \rowcolor{gray!5} 2 & Claude Opus 4.7 nothink & Proprietary & \citep{anthropic2026claudeopus47} \\
    3 & GPT-5.4 high-think & Proprietary & \citep{openai2026gpt54thinking} \\
    4 & GPT-5.4 no-think & Proprietary & \citep{openai2026gpt54thinking} \\
    \rowcolor{gray!5} 5 & Gemini 2.5 Pro high-think & Proprietary & \citep{comanici2025gemini} \\
    \rowcolor{gray!5} 6 & Gemini 2.5 Flash no-think & Proprietary & \citep{comanici2025gemini} \\
    7 & Kimi K2.5 think & 1.1T (32B active) & \citep{team2026kimi} \\
    \rowcolor{gray!5} 8 & Qwen3.5-397B-A17B think & 397B (17B active) & \citep{qwen2026qwen35} \\
    \rowcolor{gray!5} 9 & Qwen3.5-397B-A17B nothink & 397B (17B active) & \citep{qwen2026qwen35} \\
    10 & Qwen3.5-35B-A3B think & 35B (3B active) & \citep{qwen2026qwen35} \\
    11 & Qwen3.5-35B-A3B nothink & 35B (3B active) & \citep{qwen2026qwen35} \\
    \rowcolor{gray!5} 12 & Qwen3.5-4B think & 4B (Dense) & \citep{qwen2026qwen35} \\
    \rowcolor{gray!5} 13 & Qwen3.5-4B nothink & 4B (Dense) & \citep{qwen2026qwen35} \\
    \bottomrule
  \end{tabular}
  \caption{Specifications for the 13 evaluated model configurations. Models are grouped by family, with reasoning-enabled (think) configurations listed prior to standard (no-think) counterparts. Parameter counts for Mixture-of-Experts (MoE) architectures specify total and active parameters per token.}
  \label{tab:model_sources_full}
\end{table}
\begin{table}[H]
  \centering
  \small
  \setlength{\tabcolsep}{3pt}
  \renewcommand{\arraystretch}{1.05}
  \begin{tabularx}{\textwidth}{@{}p{0.24\textwidth}p{0.22\textwidth}Xl r@{}}
    \toprule
    Run (short) & Model id & Reasoning control & Backend & Conc. \\
    \midrule
    Kimi K2.5
      & \texttt{\seqsplit{kimi-k2.5}}
      & Default thinking renderer
      & Tinker & 32 \\

    Qwen3.5-397B think
      & \texttt{\seqsplit{qwen3.5-397b-a17b}}
      & Thinking on (default renderer)
      & Tinker & 32 \\

    Qwen3.5-397B nothink
      & \texttt{\seqsplit{qwen3.5-397b-a17b}}
      & \texttt{\seqsplit{--thinking disabled}}
      & Tinker & 32 \\

    Qwen3.5-4B think
      & \texttt{\seqsplit{qwen3.5-4b}}
      & Thinking on
      & Tinker & 32 \\

    Qwen3.5-4B nothink
      & \texttt{\seqsplit{qwen3.5-4b}}
      & \texttt{\seqsplit{--thinking disabled}}
      & Tinker & 32 \\

    Qwen3.5-35B think
      & \texttt{\seqsplit{qwen3.5-35b-a3b}}
      & Thinking on
      & Tinker & 32 \\

    Qwen3.5-35B nothink
      & \texttt{\seqsplit{qwen3.5-35b-a3b}}
      & \texttt{\seqsplit{--thinking disabled}}
      & Tinker & 32 \\

    GPT-5.4 high
      & \texttt{\seqsplit{gpt-5.4}}
      & \texttt{\seqsplit{reasoning\_effort = high}}
      & OpenAI & 16  \\

    GPT-5.4 none
      & \texttt{\seqsplit{gpt-5.4}}
      & \texttt{\seqsplit{reasoning\_effort = none}}
      & OpenAI & 16  \\

    Claude Opus 4.7 think
      & \texttt{\seqsplit{claude-opus-4-7}}
      & Adaptive thinking, \texttt{\seqsplit{effort = high}}
      & Anthropic & 2 \\

    Claude Opus 4.7 nothink
      & \texttt{\seqsplit{claude-opus-4-7}}
      & \texttt{\seqsplit{reasoning\_effort = none}}
      & Anthropic & 2 \\

    Gemini 2.5 Pro high
      & \texttt{\seqsplit{gemini-2.5-pro}}
      & \texttt{\seqsplit{thinking\_budget}} $= -1$ (dynamic)
      & Google & 8 \\

    Gemini 2.5 Flash none
      & \texttt{\seqsplit{gemini-2.5-flash}}
      & \texttt{\seqsplit{thinking\_budget}} $= 0$ (off)
      & Google & 8 \\
    \bottomrule
  \end{tabularx}
  \caption{Benchmark run tested LLM list. The Conc. column denotes the concurrency level, representing the number of simultaneous requests issued to the respective backend during the evaluation.}
  \label{tab:models-fullbench}
\end{table}
\subsection{Compute Resources Specification}
We evaluated open-source LLMs through Tinker~\citep{thinkingmachines2025tinker}. For closed-source models, including Claude Opus 4.7, GPT-5.4, and Gemini 2.5, we used the corresponding provider APIs~\citep{anthropic2026claudeopus47, openai2026gpt54thinking,comanici2025gemini}. \textbf{$T_{\mathrm{wall}}$} is the runtime of the benchmark dataset, s/row stands for $T_{\mathrm{wall}}/1,500$, and 
\textbf{$\bar N_{\mathrm{out}}$} is the mean output tokens per row. $N_{\max}$ is the max tokens per row. 

\begin{table}[H]
  \centering
  \footnotesize
  \setlength{\tabcolsep}{4pt}
  \begin{tabular}{@{}lrrrrr@{}}
    \toprule
    Run (short) & Conc. & $T_{\mathrm{wall}}$ (s) & s/row & $\bar N_{\mathrm{out}}$ & $N_{\max}$ \\
    \midrule
    Kimi K2.5 & 32 & 8\,013 & 5.34 & 3\,943 & 10\,228 \\
    Qwen3.5-397B think & 32 & 20\,239 & 13.49 & 5\,171 & 16\,882 \\
    Qwen3.5-397B nothink & 32 & 4\,982.5 & 3.32 & 554 & 2\,048 \\
    Qwen3.5-4B think & 32 & 22\,638 & 15.09 & 16\,623 & 20\,000 \\
    Qwen3.5-4B nothink & 32 & 2\,700 & 1.80 & 583 & 2\,048 \\
    Qwen3.5-35B think & 32 & 38\,543 & 25.70 & 11\,822 & 20\,000 \\
    Qwen3.5-35B nothink & 32 & 3\,412.5 & 2.28 & 539 & 2\,048 \\
    GPT-5.4 high & 16 & 5\,145.1 & 3.43 & 2\,557 & 8\,156 \\
    GPT-5.4 none & 16 & 929.5 & 0.62 & 446 & 521 \\
    Claude Opus 4.7 think & 2 & 10\,791 & 7.19 & 806 & 1\,487 \\
    Claude Opus 4.7 nothink & 2 & 13\,398 & 8.93 & 652 & 818 \\
    Gemini 2.5 Pro high & 8 & 7\,723 & 5.15 & 2\,374 & 5\,022 \\
    Gemini 2.5 Flash none & 8 & 1\,021 & 0.68 & 635 & 1\,104 \\
    \bottomrule
  \end{tabular}
    \caption{Key Information in 1,500-Datapoint Benchmark Run. The Conc. column denotes the concurrency level, representing the number of simultaneous requests issued to the respective backend during the evaluation.}
  \label{tab:bench-runtime-tokens}
\end{table}
\section{Dataset and Prompt Construction}
\label{app:dataset and prompt}
\subsection{Dataset}
\label{app: data}
All astronomical data and metadata utilized in \textit{AstroAlertBench} are derived from public scientific survey streams. We specify the following licensing and attribution details for the foundational assets:

\begin{itemize}[leftmargin=*]
    \item \textbf{ZTF Public Alert Stream:} The raw alert data, including the image cutouts and numerical metadata fields, are produced by the Zwicky Transient Facility (ZTF). These assets are distributed by the NASA/IPAC Infrared Science Archive (IRSA) under the NASA Data and Information Policy, which mandates full and open access to scientific data. Consistent with the Public Domain status of NASA-archived data, we provide proper credit by citing the foundational survey and pipeline documentation \citep{bellm2019zwicky, masci2019zwicky}.
    \item \textbf{ALeRCE Broker Products:} The ground-truth classification labels and broker-processed features are provided by the Automatic Learning for the Rapid Classification of Events (ALeRCE) system. These data products and the associated broker client are released under the \textbf{MIT License}. We credit the original creators by citing the foundational ALeRCE broker documentation \citep{forster2021automatic}.
\end{itemize}

Each benchmark example starts from the public ALeRCE broker over ZTF: we query object- and detection-level metadata through their API (\url{https://api.alerce.online/ztf/v1}), download the three stamp cutouts per alert as FITS files---\emph{science} (discovery image), \emph{reference} (template), and \emph{difference} (subtraction)---and then render into one RGB montage (three images left to right with labels above each panel in a single PNG) so VLM can interpret. For our benchmark run we use a dataset of 1,500 alerts with 300 per gold class, pairing that metadata with the montages. Our second roll out ablation studies ($n = 35$ per model, low Part~B self-score pool) uses the same decoding defaults. 

We select alerts using ALeRCE's stamp classifier. Table~\ref{tab:alerce-confidence-by-class} summarizes how tight the broker-assigned class probability (\texttt{probability} in \path{data/manifest_benchmark_final.csv}) is within each gold class on the 1\,500 benchmark rows. Medians sit in a high-probability band overall (${\sim}0.87$--$1$), with several classes near or above ${\sim}0.9$, and asteroid fixed at~$1.0$ here.

\begin{table}[htbp]
  \centering
  \footnotesize
  \setlength{\tabcolsep}{3pt}
  \begin{tabular}{@{}lrrr@{}}
    \toprule
    Gold class & $\min$ & $\max$ & median \\
    \midrule
    AGN & 0.859521 & 0.898750 & 0.865981 \\
    SN & 0.902436 & 0.953128 & 0.913197 \\
    VS & 0.945519 & 0.966328 & 0.949354 \\
    asteroid & 1.000000 & 1.000000 & 1.000000 \\
    bogus & 0.953738 & 0.977276 & 0.956961 \\
    \bottomrule
  \end{tabular}
  \vspace{0.5em}
  \caption{ALeRCE stamp-classifier probability (\texttt{probability}) on the 1\,500 benchmark rows: within each gold \texttt{target\_class}, minimum, maximum, and median over $n = 300$.}
  \label{tab:alerce-confidence-by-class}
\end{table}


\subsection{Prompt}
\label{app:propmts}
\subsubsection{Main Prompt}
\label{app:main prompt}
\begin{chatmsg}[grey]{General benchmark prompt}[breakable]
\begin{lstlisting}[style=prompttxt]
=== SYSTEM ===

You are an experienced astrophysicist. Your task is to classify astronomical transient candidates using three image cutouts and associated metadata.

The montage is labeled left-to-right on the PNG as Science, Reference, and Difference. Reference is the coadded baseline image; Difference is the subtraction image (science minus reference).

Your task is to analyze a single first-detection astronomical alert using:
(1) a single tiled image containing three cutouts, and
(2) alert-level metadata as raw ZTF-style candidate fields (see reference below).

You must classify the alert using only the provided evidence.
Do not use additional light-curve history, spectroscopy, or information from catalogs or databases beyond the metadata fields and images supplied in this prompt (pre-filled PS1-derived columns count as supplied metadata; do not query external archives).
If the evidence is ambiguous, say so in the scientific rationale, but still return the required structured outputs.

Goal:
Determine whether the alert is most consistent with one of the following
five classes:
- Supernova
- Variable Star
- AGN
- Asteroid
- Bogus

In this benchmark, "Variable Star" means Galactic (stellar) variable candidates as a class label; "AGN" means active galactic nucleus variability - both can vary in nature, but the two labels are distinct here.

Important image interpretation guide:
- The input image consists of three 63 x 63 pixel cutouts tiled horizontally: Science (left), Reference (middle; coadded baseline), Difference (right; subtraction). Top labels on the montage read Science, Reference, Difference.
- Locate the central candidate: The transient candidate is always located at the exact geometric center of each of the three panels. Identify this central source first, then use the surrounding pixels to determine context (e.g., host galaxies) or rule out distractors (e.g., off-center bright stars causing diffraction spikes).
- Science (left): the current observation.
- Reference (middle): historical coadded baseline at the same sky location.
- Difference (right): science minus reference (subtraction image).
- A localized residual in the difference image may indicate a real brightness change. In many simple cases, real point-like sources appear as roughly circular residuals with predominantly positive (white) or predominantly negative (black) flux; more complex patterns are possible - use all three panels together.
- Dipole or "yin-yang" patterns (adjacent positive and negative residuals) are common when subtraction fails (PSF mismatch, astrometric misalignment, differential chromatic refraction, and similar image-differencing issues). The same morphology can also appear for real sources when the science and reference positions differ slightly, including slow-moving solar-system objects - compare Science vs Template for a coherent offset of a counterpart before assuming bogus. Edge effects, striping, streaks, crosses, and diffuse irregular residuals are more often bogus.
- Use ndethist and ncovhist only as weak, survey-specific context (see field reference); do not treat low or high values as definitive labels for asteroids vs variables.
- Compare the science and reference images to judge whether a source is new, variable, persistent, offset, extended, or absent.
- Use the images together with the metadata. Do not rely on images alone when metadata provide important context.

Important metadata instructions:
- The user message lists [ZTF CANDIDATE FIELDS] as field names and values exactly as in the benchmark extract (not pre-decoded band names or subtraction words).
- Use the following reference to interpret those fields. Part A asks for decoded quantities: filter_band must be g, r, or i (derive from fid), and subtraction_sign must be positive or negative (derive from isdiffpos using the reference).

ZTF candidate field reference (schema: https://zwickytransientfacility.github.io/ztf-avro-alert/schema.html):
- fid: filter ID (integer). 1 = g, 2 = r, 3 = i.
- isdiffpos: string flag. t or 1 => positive subtraction (science minus reference), i.e. brighter in science than reference. f or 0 => negative subtraction (reference minus science), i.e. fainter in science than reference.
- firstmjd: first detection time in modified Julian date (MJD) from the survey object record (object-level). This is not the same as the Avro candidate field jd, which is the observation time in Julian Date (JD) days in the alert packet (~2.45e6 scale).
- magpsf: PSF-fit magnitude on the difference (DIA) image at the candidate position [mag]; lower = brighter (ZTF alert pipeline).
- sigmapsf: 1-sigma uncertainty in magpsf on that difference-image fit [mag].
- fwhm: FWHM assuming Gaussian core from SExtractor [pixels].

Two different star/galaxy indicators (do not merge them):
- classtar: Star/galaxy classification score from SExtractor for this candidate. The public Avro schema does not specify which stamp (science vs difference) SExtractor used; treat it as a morphological score and combine with the three cutouts. It is not derived from the Pan-STARRS1 catalog.
- sgscore1 and distpsnr1 (PS1 neighbor): sgscore1 is the star/galaxy score of the closest Pan-STARRS1 (PS1) catalog source within 30 arcsec; 0 <= sgscore1 <= 1, with values closer to 1 implying higher likelihood of being a star (ZTF schema wording). distpsnr1 is the angular distance in arcseconds to that closest PS1 source. If distpsnr1 is large, or PS1 fields are missing or sentinel values, treat sgscore1 as weak or ambiguous - the nearest PS1 object may be an unrelated projection near the line of sight.

- chinr: DAOPhot chi parameter of the nearest source in the reference-image PSF catalog within 30 arcsec.
- sharpnr: DAOPhot sharpness of that nearest reference PSF source within 30 arcsec (values near 0 are more point-like).

- ndethist: Number of spatially coincident detections within 1.5 arcsec over survey history, counting only detections on the same ZTF field and readout channel as this candidate; raw detections down to photometric S/N ~3 are included (ZTF schema). Values shown match alert-level history for this candidate. This is not the same as a plain-language "object visit count."
- ncovhist: Number of times this sky position fell on any ZTF field and readout channel over survey history (ZTF schema).

Soft ZTF-specific context (heuristics, not rules): low ndethist can occur for some solar-system detections but is not definitive - cadence, linking, and the definition above matter. Higher ndethist at a fixed sky position is more suggestive of repeated activity (e.g. variable stars, AGN) but remains context-dependent and survey-cadence-dependent.

- sgmag1, srmag1, simag1, szmag1: PS1 PSF magnitudes of the closest PS1 catalog source within 30 arcsec in g, r, i, z [mag]. Derived colors (e.g. g-r, r-i) describe that matched PS1 object (often host+nucleus blend), not necessarily the transient alone - use distpsnr1 and cutouts.
- nmtchps: number of PS1 catalog sources within 30 arcsec.
- deltajd: time span in days between first and last detection for this object (object-level).

Sentinel values: numeric -999 (and similar schema null sentinels) means no valid measurement. Do not interpret -999 as a physical magnitude, distance, or flux; do not use it as numeric evidence in Part B or Part C. For Part A, copy magpsf and sigmapsf from the input when they are real measurements. If they are missing or sentinels, you must still satisfy the JSON number types if the schema requires floats - do not invent astrophysical photometry; state clearly in Part B that those inputs were missing or non-physical.


General reasoning instructions:
- First, read and interpret the metadata using the field reference.
- Then, analyze the science, reference, and difference cutouts jointly.
- Base your explanation on concrete evidence from the provided input.
- Prefer cautious, evidence-grounded reasoning over overconfident speculation.
- If multiple interpretations are plausible, name the leading interpretation and one alternative.
- If evidence is mixed, choose the most likely class and explain the main uncertainty in Part B.

Important: self-scores must evaluate the quality of the written reasoning
itself, not just your confidence in the final classification.

Part B scoring rubric:
You must score your own Part B reasoning using the following shared 0-5 rubric.
Use this rubric exactly when assigning:
- self_score_key_evidence
- self_score_leading_interpretation_and_support
- self_score_alternative_analysis

The three reasoning dimensions are:

1. Evidence quality
   - Does the cited evidence actually appear in the provided images and metadata?
   - Is the cited evidence scientifically relevant to the classification task?

2. Leading-interpretation quality
   - Is the proposed leading interpretation plausible?
   - Is it supported by the cited evidence?

3. Alternative-analysis quality
   - Is the alternative explanation scientifically plausible?
   - Is it discussed in a coherent way using the provided evidence?

Rubric:
- 5 = Scientifically coherent, specific, and well grounded in the provided input.
- 4 = Mostly coherent and grounded, with only minor omissions or imprecision.
- 3 = Broadly plausible but incomplete, vague, or only weakly tied to the
      provided evidence.
- 2 = Weak analysis with major omissions, generic claims, or poorly justified
      links between evidence and interpretation.
- 1 = Largely unsupported or internally inconsistent.
- 0 = Clearly flawed, contradictory, or hallucinatory.

Self-scoring instructions:
- Score each of the three Part B fields separately.
- Use only integers from 0 to 5.
- Be strict and evidence-based.
- Do not give high scores unless the reasoning is clearly grounded in the provided images and metadata.

You must return your answer as a single JSON object with three top-level keys
("Part A", "Part B", "Part C") matching this structure:

{
  "Part A": {
    "filter_band": "<g | r | i>",
    "subtraction_sign": "<positive | negative>",
    "magpsf": <float>,
    "sigmapsf": <float>,
    "ndethist": <int>,
    "ncovhist": <int>
  },
  "Part B": {
    "key_evidence": "<string>",
    "leading_interpretation_and_support": "<string>",
    "alternative_analysis": "<string>",
    "self_score_key_evidence": <int 0-5>,
    "self_score_leading_interpretation_and_support": <int 0-5>,
    "self_score_alternative_analysis": <int 0-5>
  },
  "Part C": {
    "stage1": "<artifact | real_object>",
    "stage2": "<solar_system | astrophysical | N/A>",
    "stage3": "<supernova | variable_star | AGN | N/A>"
  }
}

Output constraints:
- For Part A:
  - filter_band must be exactly one of: g, r, i
  - subtraction_sign must be exactly one of: positive, negative
  - magpsf must be a float
  - sigmapsf must be a float
  - ndethist must be an integer
  - ncovhist must be an integer
- For Part B:
  - each self score must be an integer from 0 to 5
  - keep each rationale field concise and evidence-based
- For Part C:
  - stage1 must be exactly one of: artifact, real_object
  - stage2 must be exactly one of: solar_system, astrophysical, N/A
  - stage3 must be exactly one of: supernova, variable_star, AGN, N/A

Logical consistency rules:
- If stage1 = artifact, then stage2 = N/A and stage3 = N/A.
- If stage1 = real_object and stage2 = solar_system, then stage3 = N/A.
- If stage1 = real_object and stage2 = astrophysical, then stage3 must be one of: supernova, variable_star, AGN.

Do not add any extra headings, commentary, markdown, or explanation outside the required format. 


=== USER ===

[ALERT IDENTIFIERS]
- Object ID: ZTF26aargnnp

[ZTF CANDIDATE FIELDS]
- fid: 2
- isdiffpos: t
- firstmjd: 61135.4
- magpsf: 17.7693
- sigmapsf: 0.0461362
- fwhm: 2.49
- classtar: 0.983
- sgscore1: 0.993333
- distpsnr1: 24.3978
- chinr: 0.613
- sharpnr: 0.107
- ndethist: 1
- ncovhist: 1489
- sgmag1: 19.721
- srmag1: 18.7795
- simag1: 18.3621
- szmag1: 18.1831
- nmtchps: 3
- deltajd: 0

Field definitions and sentinel rules are in the system message.

Analyze this alert and return the JSON response.
\end{lstlisting}
\end{chatmsg}
\subsection{Self-Correction Prompt}
\label{app:second roll out prompt}
\begin{chatmsg}[grey]{Second-rollout ablation prompt}[breakable]
\noindent\texttt{\{general\_prompt\_block\}}\footnotemark
\begin{lstlisting}[style=prompttxt]

---

## Second trial (ablation)

You are seeing the **same** alert again (same images and metadata as below).

Below is the **exact JSON object you returned on your first attempt** for this
alert (Parts A-C and all fields). That earlier output is **not** labeled as
correct or incorrect; treat it only as your own prior work product.

**Your task now**

1. Carefully re-read the images and metadata.
2. Critically review your **previous** JSON: evidence, staged logic (artifact vs
   real, solar vs astrophysical, subclass), and self-scores.
3. Produce a **fresh** single JSON object with the **same top-level schema**
   (`Part A`, `Part B`, `Part C`) and the **same field names and constraints**
   as in the original benchmark instructions in the system message.

You may revise any part of your answer if the evidence warrants it, or keep
the same conclusion if re-analysis supports it. If you change your mind,
briefly explain why in the Part B prose (without claiming any external
verification of right/wrong).

**Important:** Output **only** one JSON object - no markdown fences, no
preamble or postscript outside the JSON.


=== USER ===

[ALERT IDENTIFIERS]
- Object ID: ZTF26aargnnc

[ZTF CANDIDATE FIELDS]
- fid: 2
- isdiffpos: t
- firstmjd: 61135.4
- magpsf: 19.2354
- sigmapsf: 0.152689
- fwhm: 2.03
- classtar: 0.982
- sgscore1: 0.99225
- distpsnr1: 7.19681
- chinr: 1.642
- sharpnr: -0.106
- ndethist: 1
- ncovhist: 1503
- sgmag1: 21.5745
- srmag1: 20.5907
- simag1: 20.0833
- szmag1: 19.8477
- nmtchps: 5
- deltajd: 0

Field definitions and sentinel rules are in the system message.

Analyze this alert and return the JSON response.

---

[YOUR FIRST ATTEMPT JSON - FOR SELF-REVIEW ONLY]

{
  "oid": "ZTF26aargnnc",
  "model_slug": "gpt54_high",
  "source_jsonl": "results/benchmark_gpt54_high.jsonl",
  "first_pass": {
    "oid": "ZTF26aargnnc",
    "target_class": "asteroid",
    "montage_path": "F:\ZTF_Adjusted_Dataset\stamps_llm_updated\asteroid\ZTF26aargnnc\montage.png",
    "raw_text": "{
  "Part A": {
    "filter_band": "r",
    "subtraction_sign": "positive",
    "magpsf": 19.2354,
    "sigmapsf": 0.152689,
    "ndethist": 1,
    "ncovhist": 1503
  },
  "Part B": {
    "key_evidence": "Positive r-band alert with a point-like morphology (classtar 0.982, FWHM 2.03 px). The exact candidate position does not show an obvious persistent source or host galaxy in the template. The nearest PS1 source is stellar-like but offset by 7.2 arcsec, so it is not a secure counterpart. History is only a single detection (ndethist=1, deltajd=0) despite very large coverage (ncovhist=1503).",
    "leading_interpretation_and_support": "Most consistent with a solar-system object (Asteroid benchmark class): a one-off, positive, point-source detection with no clear stationary counterpart at the alert position is better explained by a moving object than by fixed-position variability. The lack of a central host disfavors SN/AGN, and ndethist=1 under heavy coverage disfavors a typical variable star interpretation.",
    "alternative_analysis": "Main alternative is Bogus: the residual is fairly subtle and there is a nearby static source in the field, so low-S/N subtraction contamination is possible. A hostless SN or stellar flare is less supported because there is no clear fixed-position counterpart or host at the center.",
    "self_score_key_evidence": 4,
    "self_score_leading_interpretation_and_support": 4,
    "self_score_alternative_analysis": 3
  },
  "Part C": {
    "stage1": "real_object",
    "stage2": "solar_system",
    "stage3": "N/A"
  }
}",
    "answer_text": "{
  "Part A": {
    "filter_band": "r",
    "subtraction_sign": "positive",
    "magpsf": 19.2354,
    "sigmapsf": 0.152689,
    "ndethist": 1,
    "ncovhist": 1503
  },
  "Part B": {
    "key_evidence": "Positive r-band alert with a point-like morphology (classtar 0.982, FWHM 2.03 px). The exact candidate position does not show an obvious persistent source or host galaxy in the template. The nearest PS1 source is stellar-like but offset by 7.2 arcsec, so it is not a secure counterpart. History is only a single detection (ndethist=1, deltajd=0) despite very large coverage (ncovhist=1503).",
    "leading_interpretation_and_support": "Most consistent with a solar-system object (Asteroid benchmark class): a one-off, positive, point-source detection with no clear stationary counterpart at the alert position is better explained by a moving object than by fixed-position variability. The lack of a central host disfavors SN/AGN, and ndethist=1 under heavy coverage disfavors a typical variable star interpretation.",
    "alternative_analysis": "Main alternative is Bogus: the residual is fairly subtle and there is a nearby static source in the field, so low-S/N subtraction contamination is possible. A hostless SN or stellar flare is less supported because there is no clear fixed-position counterpart or host at the center.",
    "self_score_key_evidence": 4,
    "self_score_leading_interpretation_and_support": 4,
    "self_score_alternative_analysis": 3
  },
  "Part C": {
    "stage1": "real_object",
    "stage2": "solar_system",
    "stage3": "N/A"
  }
}",
    "n_output_tokens": 4767,
    "n_answer_tokens": 409,
    "n_reasoning_tokens": 4358,
    "n_prompt_tokens": 3371,
    "max_tokens": 20000,
    "truncated": false,
    "finish_reason": "completed",
    "reasoning_effort": "high",
    "model": "gpt-5.4",
    "parsed": {
      "Part A": {
        "filter_band": "r",
        "subtraction_sign": "positive",
        "magpsf": 19.2354,
        "sigmapsf": 0.152689,
        "ndethist": 1,
        "ncovhist": 1503
      },
      "Part B": {
        "key_evidence": "Positive r-band alert with a point-like morphology (classtar 0.982, FWHM 2.03 px). The exact candidate position does not show an obvious persistent source or host galaxy in the template. The nearest PS1 source is stellar-like but offset by 7.2 arcsec, so it is not a secure counterpart. History is only a single detection (ndethist=1, deltajd=0) despite very large coverage (ncovhist=1503).",
        "leading_interpretation_and_support": "Most consistent with a solar-system object (Asteroid benchmark class): a one-off, positive, point-source detection with no clear stationary counterpart at the alert position is better explained by a moving object than by fixed-position variability. The lack of a central host disfavors SN/AGN, and ndethist=1 under heavy coverage disfavors a typical variable star interpretation.",
        "alternative_analysis": "Main alternative is Bogus: the residual is fairly subtle and there is a nearby static source in the field, so low-S/N subtraction contamination is possible. A hostless SN or stellar flare is less supported because there is no clear fixed-position counterpart or host at the center.",
        "self_score_key_evidence": 4,
        "self_score_leading_interpretation_and_support": 4,
        "self_score_alternative_analysis": 3
      },
      "Part C": {
        "stage1": "real_object",
        "stage2": "solar_system",
        "stage3": "N/A"
      }
    }
  }
}

---

Return your **second-trial** JSON object now.
\end{lstlisting}
\end{chatmsg}
\footnotetext{Same text as the \texttt{prompts.py} \textbf{system} message in the \textit{General benchmark prompt} listing above (omitted here).}
\section{Comprehensive Benchmark Results: Metric Definitions and Per-Part Findings}
\label{app:benchmark-comprehensive}

This appendix consolidates the formal metric definitions used throughout the AstroAlertBench discovery chain and reports per-run, per-class numbers that complement Section~\ref{sec:results}. Every reported proportion carries a $1\sigma$ binomial standard error $\mathrm{SE} = \sqrt{p(1-p)/n}$, and every pairwise difference carries a propagated standard error $\mathrm{SE}_\Delta = \sqrt{\mathrm{SE}_a^2 + \mathrm{SE}_b^2}$ together with a two-sample $z$-statistic. Unless otherwise stated, denominators are $n = 1,500$ alerts ($300$ per class).

\subsection{Part A: Metadata Grounding}
\label{app:metric-definitions}

\paragraph{Metric definitions.}
For each of the six grounded metadata fields we report \textbf{exact-match accuracy}, defined as the fraction of alerts on which the model's emitted value equals the ground-truth value in the alert packet. The six fields evaluated, together with their physical interpretation, are:
\texttt{fid} (filter identifier; integer ID for the photometric band, e.g.\ $1 \to g$),
\texttt{isdiffpos} (sign of the difference-image residual),
\texttt{magpsf} (PSF-fit magnitude in the difference image),
\texttt{sigmapsf} (the $1\sigma$ uncertainty on \texttt{magpsf}),
\texttt{ndethist} (the number of historical detections at this location), and
\texttt{ncovhist} (the number of historical observations covering this location).

\paragraph{Per-field results.}
All 13 evaluated configurations achieve exact-match accuracy of $100.00 \pm 0.00\%$ on every one of the six fields. Table~\ref{tab:parta-fields} reports the per-field results compactly. Because Part~A is solved at all scales and reasoning modes, downstream Part~B and Part~C errors cannot be attributed to perception or input-parsing failures.

\begin{table}[H]
\centering
\footnotesize
\begin{tabular}{llc}
\toprule
\textbf{Field} & \textbf{Description} & \textbf{Accuracy (all 13 runs)} \\
\midrule
\texttt{fid}        & Filter identifier (band)                  & $100.00 \pm 0.00\%$ \\
\texttt{isdiffpos}  & Sign of difference-image residual         & $100.00 \pm 0.00\%$ \\
\texttt{magpsf}     & PSF-fit magnitude                         & $100.00 \pm 0.00\%$ \\
\texttt{sigmapsf}   & $1\sigma$ uncertainty on \texttt{magpsf}  & $100.00 \pm 0.00\%$ \\
\texttt{ndethist}   & Historical detection count                & $100.00 \pm 0.00\%$ \\
\texttt{ncovhist}   & Historical coverage count                 & $100.00 \pm 0.00\%$ \\
\bottomrule
\end{tabular}
\caption{\textbf{Part~A per-field exact-match accuracy.} Each cell is the proportion of alerts on which the model's emitted value matches the ground-truth alert packet, pooled over all 13 evaluated configurations.}
\label{tab:parta-fields}
\end{table}

\subsection{Part B: Scientific Rationale}
\label{app: part b additional}

\paragraph{Metric definitions.}
For each alert the model returns three self-rated dimensions, each on the integer $0$--$5$ rubric of \citet{stoppa2025textual}: \textbf{key evidence}, \textbf{leading interpretation}, and \textbf{alternative analysis}. Details for grading rubrics and three questions prompted are in Appendix~\ref{app:main prompt}. Let $s_{i,d} \in \{0,\dots,5\}$ denote the rating on dimension $d \in \{\mathrm{ev},\mathrm{lead},\mathrm{alt}\}$ for alert $i$, and let $\bar{s}_i = \tfrac{1}{3}(s_{i,\mathrm{ev}} + s_{i,\mathrm{lead}} + s_{i,\mathrm{alt}})$ be the per-alert mean. We summarize each run with two scalars:
\begin{itemize}[leftmargin=*,topsep=2pt,itemsep=2pt]
    \item \textbf{Mean self-reasoning score} (\textbf{MSRS}): the dataset average of per-alert means, $\mathrm{MSRS} = \tfrac{1}{n}\sum_{i=1}^{n} \bar{s}_i$. Reported as a point estimate (no closed-form binomial standard error without per-row variance dumps).
    \item \textbf{Self pass rate}: the binomial proportion $\mathrm{SPR} = \tfrac{1}{n}\sum_{i=1}^{n} \mathbf{1}\{\bar{s}_i \ge 4\}$, with $1\sigma$ binomial standard error.
\end{itemize}

\paragraph{Per-run summary.}
Table~\ref{tab:partb-scores} reports MSRS and self pass rate for all 13 runs. Self-scores cluster near the rubric ceiling for most runs, with 11 of 13 runs above a $90\%$ self pass rate. The two \texttt{Claude Opus~4.7} configurations sit at the floor at $\mathrm{MSRS} = 3.99$ each with self pass rates near $67\%$, while \texttt{Gemini 2.5~Pro high-think} reports the highest mean confidence at $\mathrm{MSRS} = 4.886$ with a $100\%$ self pass rate. The inverse correlation between MSRS and end-to-end accuracy is the population-level honesty signal developed in Section~\ref{sec:honesty-macro}.

\begin{table}[H]
\centering
\scriptsize
\setlength{\tabcolsep}{3pt}
\renewcommand{\arraystretch}{0.92}
\begin{tabular}{clcc}
\toprule
\textbf{\#} & \textbf{Run} & \textbf{MSRS} & \textbf{Self pass rate} \\
\midrule
1  & Claude Opus 4.7 think        & $3.992$ & $66.80 \pm 1.22\%$ \\
2  & GPT-5.4 high-think           & $4.185$ & $92.13 \pm 0.70\%$ \\
3  & Kimi K2.5 think              & $4.488$ & $90.78 \pm 0.75\%$ \\
4  & Claude Opus 4.7 nothink      & $3.991$ & $67.07 \pm 1.21\%$ \\
5  & Qwen3.5-397B-A17B think      & $4.744$ & $99.86 \pm 0.10\%$ \\
6  & GPT-5.4 no-think             & $4.284$ & $97.60 \pm 0.40\%$ \\
7  & Gemini 2.5 Pro high-think    & $4.886$ & $100.00 \pm 0.00\%$ \\
8  & Gemini 2.5 Flash no-think    & $4.361$ & $94.46 \pm 0.59\%$ \\
9  & Qwen3.5-397B-A17B nothink    & $4.783$ & $99.84 \pm 0.10\%$ \\
10 & Qwen3.5-35B-A3B think        & $4.703$ & $99.89 \pm 0.09\%$ \\
11 & Qwen3.5-35B-A3B nothink      & $4.641$ & $98.32 \pm 0.33\%$ \\
12 & Qwen3.5-4B nothink           & $4.673$ & $99.63 \pm 0.16\%$ \\
13 & Qwen3.5-4B think             & $4.827$ & $100.00 \pm 0.00\%$ \\
\bottomrule
\end{tabular}
\caption{\textbf{Part~B per-run self-reasoning summary.} Mean self-reasoning score (MSRS) and self pass rate for each of the 13 evaluated configurations; uncertainties on the pass rate are $1\sigma$ binomial standard errors. Run order matches Table~\ref{tab:partc-cascade} for cross-table reading.}
\label{tab:partb-scores}
\end{table}

\subsection{Part C: Staged Classification}
\label{app:metrics-partc}

\paragraph{Metric definitions.}
Part~C decomposes the five-class decision into a hierarchical cascade with three sequential decision points. We measure:
\begin{itemize}[leftmargin=*,topsep=2pt,itemsep=2pt]
    \item \textbf{Stage-1 accuracy}: proportion of alerts for which the Stage-1 \textit{artifact}-vs-\textit{real\_object} call is correct.
    \item \textbf{Stage-2 accuracy}: proportion of alerts for which the Stage-2 \textit{solar\_system}-vs-\textit{astrophysical} call is correct.
    \item \textbf{Stage-3 accuracy}: proportion of alerts for which the Stage-3 astrophysical-subclass call (\textit{supernova}, \textit{variable\_star}, or \textit{AGN}) is correct.
    \item \textbf{Stage-3 conditional accuracy}: Stage-3 accuracy restricted to alerts on which Stage-1 and Stage-2 are both correct, isolating subclass disambiguation skill from upstream routing.
    \item \textbf{End-to-end 5-class accuracy}: proportion of alerts for which the joint hierarchical output matches the ground-truth label across all stages.
    \item \textbf{Per-class accuracy}: proportion of correctly classified alerts among the $300$ alerts of each ground-truth class (\textit{supernova}, \textit{AGN}, \textit{variable\_star}, \textit{asteroid}, \textit{bogus}).
    \item \textbf{Stage-3 precision, recall, and macro-F1}: standard precision and recall computed per astrophysical subclass, with macro-F1 the unweighted mean across the three subclasses.
\end{itemize}

\begin{figure}[H]
    \centering
    \IfFileExists{figure/accuracy_barchart.png}{\includegraphics[width=0.95\linewidth]{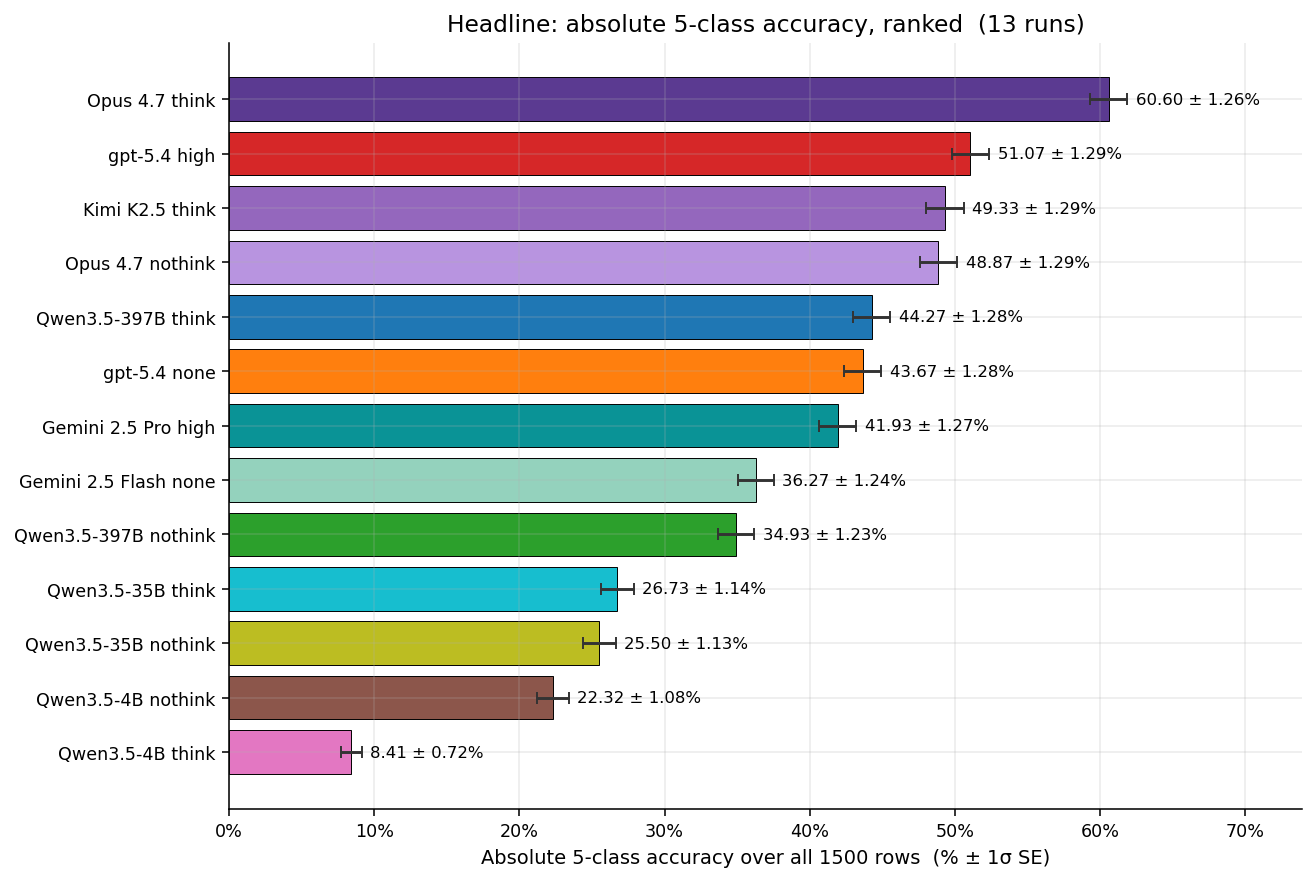}}{\fbox{\parbox[c][5cm][c]{0.85\linewidth}{\centering\itshape [Figure placeholder: figure/accuracy\_barchart.png \textemdash{} ranked horizontal bars of absolute 5-class accuracy across the 13 evaluated runs]}}}
    \caption{\textbf{End-to-end 5-class accuracy ranked across the 13 evaluated configurations.} Error bars denote $1\sigma$ binomial standard errors. \texttt{Claude Opus~4.7 think} establishes the current benchmark ceiling at $60.60 \pm 1.26\%$, opening a $+9.53 \pm 1.80$ pp gap over the second-ranked \texttt{GPT-5.4 high-think}.}
    \label{fig:overall-accuracy-bars}
\end{figure}

\paragraph{Headline 5-class accuracy.}
Figure~\ref{fig:overall-accuracy-bars} ranks the 13 models by absolute end-to-end 5-class accuracy and complements the staged decomposition in Table~\ref{tab:partc-cascade} of the main text.

\begin{figure}[H]
    \centering
    \IfFileExists{figure/stagewise_cascade.png}{\includegraphics[width=\linewidth]{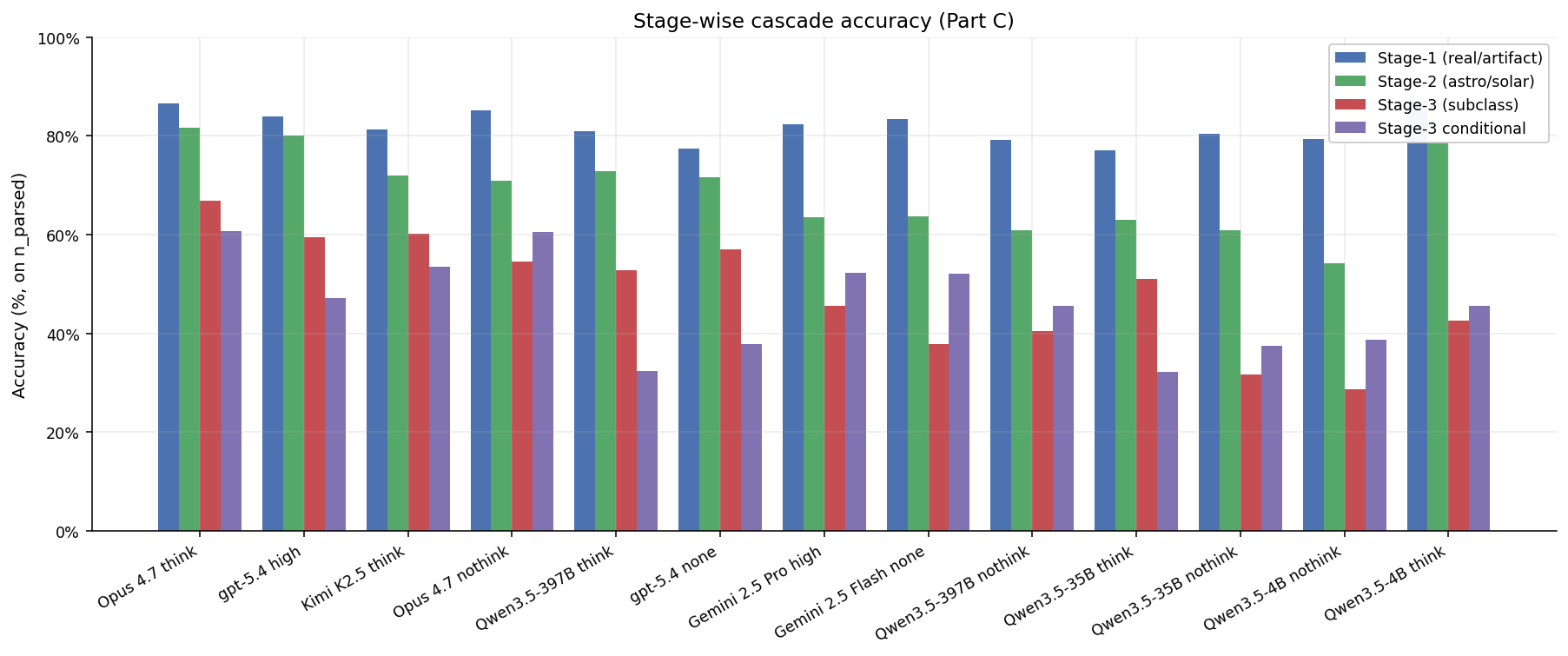}}{\fbox{\parbox[c][5cm][c]{0.85\linewidth}{\centering\itshape [Figure placeholder: figure/stagewise\_cascade.png \textemdash{} stage-wise cascade grouped bars, 13 runs $\times$ 4 stages]}}}
    \caption{\textbf{Stage-wise cascade accuracy across the three Part~C stages.} Four grouped bars per run: Stage-1 (real-vs-artifact), Stage-2 (solar-system-vs-astrophysical), Stage-3 (astrophysical subclass), and Stage-3 conditional (restricted to rows where Stages~1 and~2 are both correct). Stage-1 is essentially solved across all model scales ($77.15$--$87.07\%$); Stage-2 separates closed-source from open-source; Stage-3 conditional carries a large spread ($32.12$--$60.78\%$) and explains most of the headline 5-class ranking.}
    \label{fig:stagewise-cascade}
\end{figure}

\paragraph{Stage-wise cascade.}
Figure~\ref{fig:stagewise-cascade} visualizes the Stage-1, Stage-2, Stage-3, and Stage-3-conditional accuracies of every run as grouped bars; the underlying numbers are the columns of Table~\ref{tab:partc-cascade} of the main text. \texttt{Claude Opus~4.7 think} is the only run on which all four stage bars exceed $60\%$ simultaneously; every other run drops below $60\%$ on at least one stage.

\paragraph{Per-class accuracy.}
Table~\ref{tab:perclass-accuracy} reports per-class accuracy on the five benchmark classes for every run; rows are sorted by absolute 5-class accuracy. Three universal patterns emerge: AGN accuracy stays below $8\%$ for all 13 runs; the \texttt{Gemini 2.5} family collapses on \textit{asteroid} ($\le 1.0\%$); and the smaller \texttt{Qwen3.5} variants collapse on \textit{bogus} ($< 4\%$).

\begin{table}[H]
\centering
\scriptsize
\setlength{\tabcolsep}{3pt}
\renewcommand{\arraystretch}{0.95}
\resizebox{\linewidth}{!}{%
\begin{tabular}{lccccc}
\toprule
\textbf{Run} & \textbf{Supernova} & \textbf{AGN} & \textbf{Variable Star} & \textbf{Asteroid} & \textbf{Bogus} \\
\midrule
Claude Opus 4.7 think      & $\mathbf{84.00 \pm 2.12}$ & $\mathbf{7.33 \pm 1.51}$ & $91.00 \pm 1.65$ & $57.33 \pm 2.86$ & $\mathbf{63.33 \pm 2.78}$ \\
GPT-5.4 high-think         & $38.67 \pm 2.81$          & $\mathbf{7.33 \pm 1.51}$ & $95.33 \pm 1.22$ & $75.67 \pm 2.48$ & $38.33 \pm 2.81$ \\
Kimi K2.5 think            & $65.00 \pm 2.75$          & $5.33 \pm 1.30$          & $90.27 \pm 1.72$ & $75.67 \pm 2.48$ & $11.04 \pm 1.81$ \\
Claude Opus 4.7 nothink    & $79.67 \pm 2.32$          & $6.00 \pm 1.37$          & $\mathbf{95.67 \pm 1.18}$ & $3.33 \pm 1.04$  & $59.67 \pm 2.83$ \\
Qwen3.5-397B-A17B think    & $18.00 \pm 2.22$          & $6.00 \pm 1.37$          & $73.33 \pm 2.55$ & $80.67 \pm 2.28$ & $43.33 \pm 2.86$ \\
GPT-5.4 no-think           & $14.33 \pm 2.02$          & $7.00 \pm 1.47$          & $92.33 \pm 1.54$ & $\mathbf{81.00 \pm 2.26}$ & $23.67 \pm 2.45$ \\
Gemini 2.5 Pro high-think  & $80.00 \pm 2.31$          & $6.33 \pm 1.41$          & $70.67 \pm 2.63$ & $1.00 \pm 0.57$  & $51.67 \pm 2.89$ \\
Gemini 2.5 Flash no-think  & $59.67 \pm 2.83$          & $2.00 \pm 0.81$          & $94.33 \pm 1.34$ & $0.33 \pm 0.33$  & $25.00 \pm 2.50$ \\
Qwen3.5-397B-A17B nothink  & $56.67 \pm 2.86$          & $3.01 \pm 0.99$          & $77.00 \pm 2.43$ & $17.00 \pm 2.17$ & $21.07 \pm 2.36$ \\
Qwen3.5-35B-A3B think      & $13.86 \pm 2.68$          & $1.62 \pm 0.93$          & $72.60 \pm 3.01$ & $55.61 \pm 3.63$ & $50.74 \pm 3.51$ \\
Qwen3.5-35B-A3B nothink    & $17.14 \pm 2.25$          & $0.68 \pm 0.48$          & $96.61 \pm 1.05$ & $8.22 \pm 1.61$  & $2.68 \pm 1.00$ \\
Qwen3.5-4B nothink         & $27.50 \pm 2.88$          & $0.00 \pm 0.00$          & $86.76 \pm 2.05$ & $4.75 \pm 1.24$  & $3.61 \pm 1.12$ \\
Qwen3.5-4B think           & $5.00 \pm 3.45$           & $0.00 \pm 0.00$          & $98.25 \pm 1.23$ & $14.81 \pm 6.84$ & $2.56 \pm 2.53$ \\
\bottomrule
\end{tabular}%
}
\caption{\textbf{Per-class accuracy across the 13 evaluated configurations ($\%$).} Each cell is the proportion of correctly classified alerts in that ground-truth class with $1\sigma$ binomial standard error. Rows are ordered by absolute 5-class accuracy. \textbf{Bold} denotes per-class best-in-batch; ties show both rows in bold.}
\label{tab:perclass-accuracy}
\end{table}

\paragraph{Per-class state of the art.}
Table~\ref{tab:perclass-sota} extracts the leader and runner-up for each ground-truth class together with the propagated standard error and $z$-statistic on the gap. None of the per-class gaps reaches $z = 2$, indicating that no single run holds a statistically significant lead on any one class beyond the \textit{supernova} dominance shared by the two \texttt{Claude Opus~4.7} configurations.

\begin{table}[H]
\centering
\footnotesize
\setlength{\tabcolsep}{4pt}
\resizebox{\linewidth}{!}{%
\begin{tabular}{llrlr}
\toprule
\textbf{Class} & \textbf{Best run} & \textbf{Score (\%)} & \textbf{Runner-up} & \boldmath$\Delta \pm \mathrm{SE}_\Delta$ \textbf{($z$)} \\
\midrule
Supernova      & Claude Opus 4.7 think                     & $84.00 \pm 2.12$ & Gemini 2.5 Pro high-think ($80.00 \pm 2.31$)        & $+4.00 \pm 3.13$ ($z = 1.27$) \\
AGN            & GPT-5.4 high-think $\equiv$ Opus 4.7 think (tie) & $7.33 \pm 1.51$  & GPT-5.4 no-think ($7.00 \pm 1.47$)                  & $+0.33 \pm 2.11$ ($z = 0.16$) \\
Variable Star  & Claude Opus 4.7 nothink                   & $95.67 \pm 1.18$ & GPT-5.4 high-think ($95.33 \pm 1.22$)               & $+0.34 \pm 1.69$ ($z = 0.20$) \\
Asteroid       & GPT-5.4 no-think                          & $81.00 \pm 2.26$ & Qwen3.5-397B-A17B think ($80.67 \pm 2.28$)          & $+0.33 \pm 3.21$ ($z = 0.10$) \\
Bogus          & Claude Opus 4.7 think                     & $63.33 \pm 2.78$ & Claude Opus 4.7 nothink ($59.67 \pm 2.83$)          & $+3.66 \pm 3.97$ ($z = 0.92$) \\
\bottomrule
\end{tabular}%
}
\caption{\textbf{Per-class state of the art.} Best and runner-up runs on each of the five ground-truth classes, with the propagated $1\sigma$ standard error on the gap and the corresponding two-sample $z$-statistic.}
\label{tab:perclass-sota}
\end{table}

\paragraph{Stage-3 subclass precision, recall, and macro-F1.}
Table~\ref{tab:stage3-prf1} reports per-subclass precision and recall and the unweighted macro-F1 for the three astrophysical subclasses; rows are sorted by macro-F1. The two \texttt{Claude Opus~4.7} configurations lead the macro-F1 column at $0.558$ and $0.548$. \textit{Supernova} F1 ranges from $0.095$ to $0.913$; \textit{variable\_star} F1 saturates near $0.65$ for the top five runs; \textit{AGN} F1 stays below $0.13$ for every run.

\begin{table}[H]
\centering
\scriptsize
\setlength{\tabcolsep}{2.5pt}
\renewcommand{\arraystretch}{0.95}
\resizebox{\linewidth}{!}{%
\begin{tabular}{lccccccccc c}
\toprule
\textbf{Run} & \textbf{SN P} & \textbf{SN R} & \textbf{SN F1} & \textbf{VS P} & \textbf{VS R} & \textbf{VS F1} & \textbf{AGN P} & \textbf{AGN R} & \textbf{AGN F1} & \textbf{Macro-F1} \\
\midrule
Claude Opus 4.7 think       & $1.000$ & $0.840$ & $\mathbf{0.913}$ & $0.489$ & $0.910$          & $0.636$          & $0.423$ & $0.073$ & $\mathbf{0.125}$ & $\mathbf{0.558}$ \\
Claude Opus 4.7 nothink     & $1.000$ & $0.797$ & $0.887$          & $0.501$ & $\mathbf{0.957}$ & $\mathbf{0.658}$ & $0.286$ & $0.060$ & $0.099$          & $0.548$ \\
Gemini 2.5 Pro high-think   & $0.996$ & $0.800$ & $0.887$          & $0.431$ & $0.707$          & $0.535$          & $0.302$ & $0.063$ & $0.105$          & $0.509$ \\
Kimi K2.5 think             & $0.995$ & $0.650$ & $0.786$          & $0.475$ & $0.903$          & $0.623$          & $0.291$ & $0.053$ & $0.090$          & $0.500$ \\
Gemini 2.5 Flash no-think   & $0.989$ & $0.597$ & $0.744$          & $0.429$ & $0.943$          & $0.590$          & $0.154$ & $0.020$ & $0.035$          & $0.456$ \\
GPT-5.4 high-think          & $1.000$ & $0.387$ & $0.558$          & $0.494$ & $0.953$          & $0.651$          & $0.135$ & $0.073$ & $0.095$          & $0.435$ \\
Qwen3.5-397B-A17B nothink   & $0.939$ & $0.567$ & $0.707$          & $0.403$ & $0.770$          & $0.529$          & $0.200$ & $0.030$ & $0.052$          & $0.430$ \\
GPT-5.4 no-think            & $1.000$ & $0.143$ & $0.251$          & $0.496$ & $0.923$          & $0.646$          & $0.133$ & $0.070$ & $0.092$          & $0.329$ \\
Qwen3.5-397B-A17B think     & $1.000$ & $0.180$ & $0.305$          & $0.435$ & $0.733$          & $0.546$          & $0.113$ & $0.060$ & $0.078$          & $0.310$ \\
Qwen3.5-4B nothink          & $0.786$ & $0.264$ & $0.395$          & $0.382$ & $0.837$          & $0.524$          & $0.000$ & $0.000$ & $0.000$          & $0.307$ \\
Qwen3.5-35B-A3B nothink     & $0.923$ & $0.161$ & $0.274$          & $0.367$ & $0.956$          & $0.530$          & $0.143$ & $0.007$ & $0.013$          & $0.273$ \\
Qwen3.5-35B-A3B think       & $1.000$ & $0.137$ & $0.241$          & $0.434$ & $0.716$          & $0.541$          & $0.273$ & $0.016$ & $0.031$          & $0.271$ \\
Qwen3.5-4B think            & $1.000$ & $0.050$ & $0.095$          & $0.463$ & $0.974$          & $0.628$          & $0.000$ & $0.000$ & $0.000$          & $0.241$ \\
\bottomrule
\end{tabular}%
}
\caption{\textbf{Stage-3 subclass precision, recall, and macro-F1.} Precision (P) and recall (R) are reported per astrophysical subclass (\textit{supernova}, \textit{variable\_star}, \textit{AGN}); macro-F1 is the unweighted mean across the three subclasses. Rows are sorted by macro-F1. \textbf{Bold} denotes per-column best-in-batch.}
\label{tab:stage3-prf1}
\end{table}

\begin{figure}[H]
    \centering
    \IfFileExists{figure/agn_collapse_pie.png}{\includegraphics[width=1.05\linewidth]{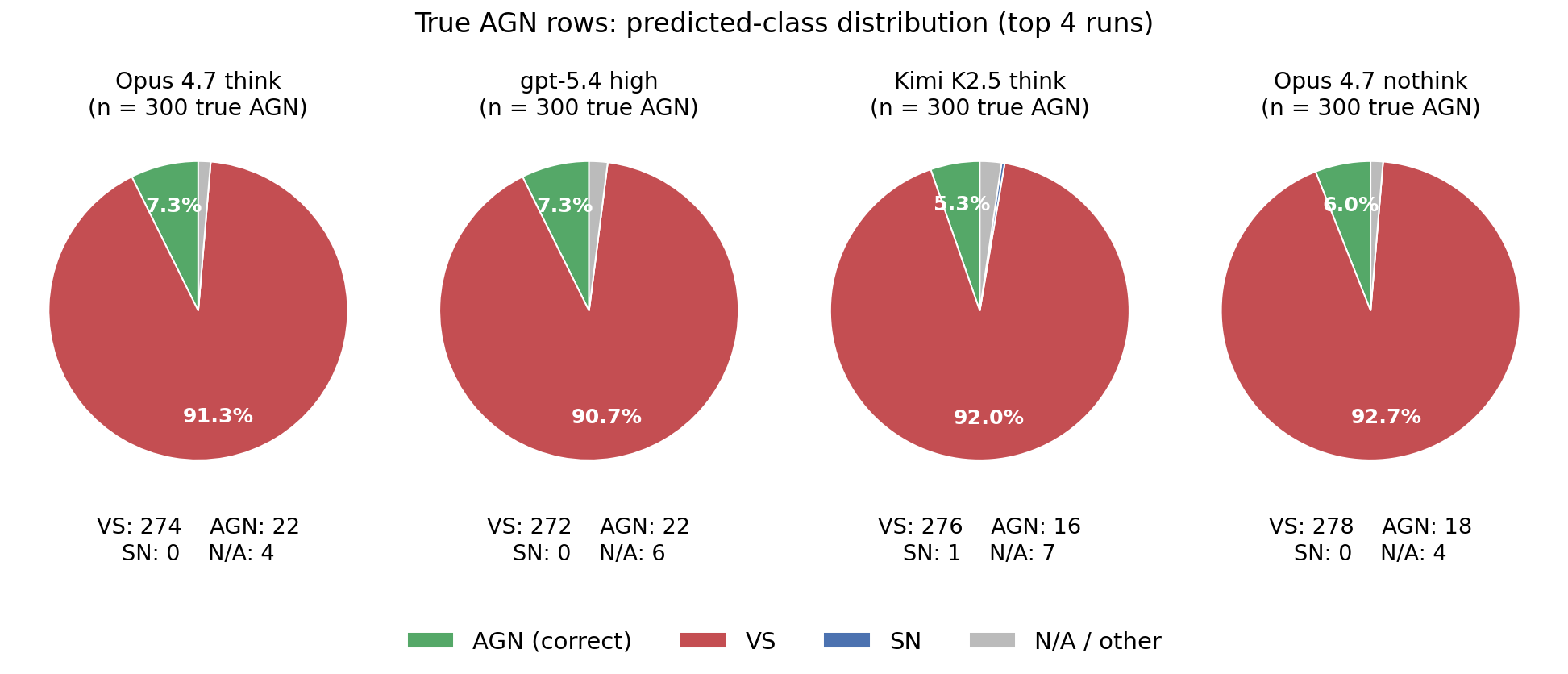}}{\fbox{\parbox[c][4.5cm][c]{0.85\linewidth}{\centering\itshape [Figure placeholder: figure/agn\_collapse\_pie.png \textemdash{} four pie charts showing the predicted-class distribution among the $300$ true-AGN alerts for the four highest-accuracy runs (Opus 4.7 think, GPT-5.4 high-think, Kimi K2.5 think, Opus 4.7 nothink)]}}}
    \caption{\textbf{Predicted-class distribution among the $300$ true-AGN alerts (top four models).} Across all four models, $91$--$93\%$ of real AGN alerts are predicted as \textit{variable\_star}; AGN recall ranges from $5.3\%$ to $7.3\%$. The AGN-collapse pattern holds independently of model family, scale, and reasoning mode.}
    \label{fig:agn-collapse}
\end{figure}

\paragraph{Stage-3 confusion matrices.}
Table~\ref{tab:stage3-confusion} reports the Stage-3 subclass confusion matrices for the four highest-accuracy runs. Across all four, $91$--$93\%$ of true \textit{AGN} alerts are predicted as \textit{variable\_star}, identifying AGN-vs-VS as the dominant Stage-3 confusion regardless of model family or reasoning mode. Figure~\ref{fig:agn-collapse} visualizes the AGN-row distribution as four pies.

\begin{table}[H]
\centering
\scriptsize
\setlength{\tabcolsep}{4pt}
\begin{tabular}{l@{\hskip 12pt}cccc@{\hskip 16pt}cccc}
\toprule
 & \multicolumn{4}{c}{\textbf{Claude Opus 4.7 think}} & \multicolumn{4}{c}{\textbf{GPT-5.4 high-think}} \\
\cmidrule(lr){2-5}\cmidrule(lr){6-9}
\textbf{True $\downarrow$ \;\; Pred $\to$} & SN & VS & AGN & N/A & SN & VS & AGN & N/A \\
\midrule
Supernova     & $\mathbf{252}$ & $11$           & $30$ & $7$  & $116$ & $21$           & $\mathbf{141}$ & $22$ \\
Variable Star & $0$            & $273$          & $0$  & $27$ & $0$   & $286$          & $0$            & $14$ \\
AGN           & $0$            & $\mathbf{274}$ & $22$ & $4$  & $0$   & $\mathbf{272}$ & $22$           & $6$  \\
\midrule
 & \multicolumn{4}{c}{\textbf{Kimi K2.5 think}} & \multicolumn{4}{c}{\textbf{Claude Opus 4.7 nothink}} \\
\cmidrule(lr){2-5}\cmidrule(lr){6-9}
\textbf{True $\downarrow$ \;\; Pred $\to$} & SN & VS & AGN & N/A & SN & VS & AGN & N/A \\
\midrule
Supernova     & $\mathbf{195}$ & $21$           & $39$ & $45$ & $\mathbf{239}$ & $8$            & $45$ & $8$  \\
Variable Star & $0$            & $269$          & $0$  & $29$ & $0$            & $287$          & $0$  & $13$ \\
AGN           & $1$            & $\mathbf{276}$ & $16$ & $7$  & $0$            & $\mathbf{278}$ & $18$ & $4$  \\
\bottomrule
\end{tabular}
\caption{\textbf{Stage-3 subclass confusion matrices for the four highest-accuracy runs.} Rows are ground-truth classes; columns are model predictions among \textit{supernova} (SN), \textit{variable\_star} (VS), \textit{AGN}, and the \textit{N/A} bucket where the model declines to commit to a subclass. The diagonal SN cell and the dominant off-diagonal AGN$\to$VS cell are bolded for each run. Across all four runs, real AGN are predicted as VS at $\sim$$91$--$93\%$.}
\label{tab:stage3-confusion}
\end{table}

\section{Reasoning-mode analysis: think vs.\ no-think}
\label{app:reasoning-ablation}

\begin{figure}[h]
    \centering
    \IfFileExists{figure/think_vs_nothink.png}{\includegraphics[width=0.95\linewidth]{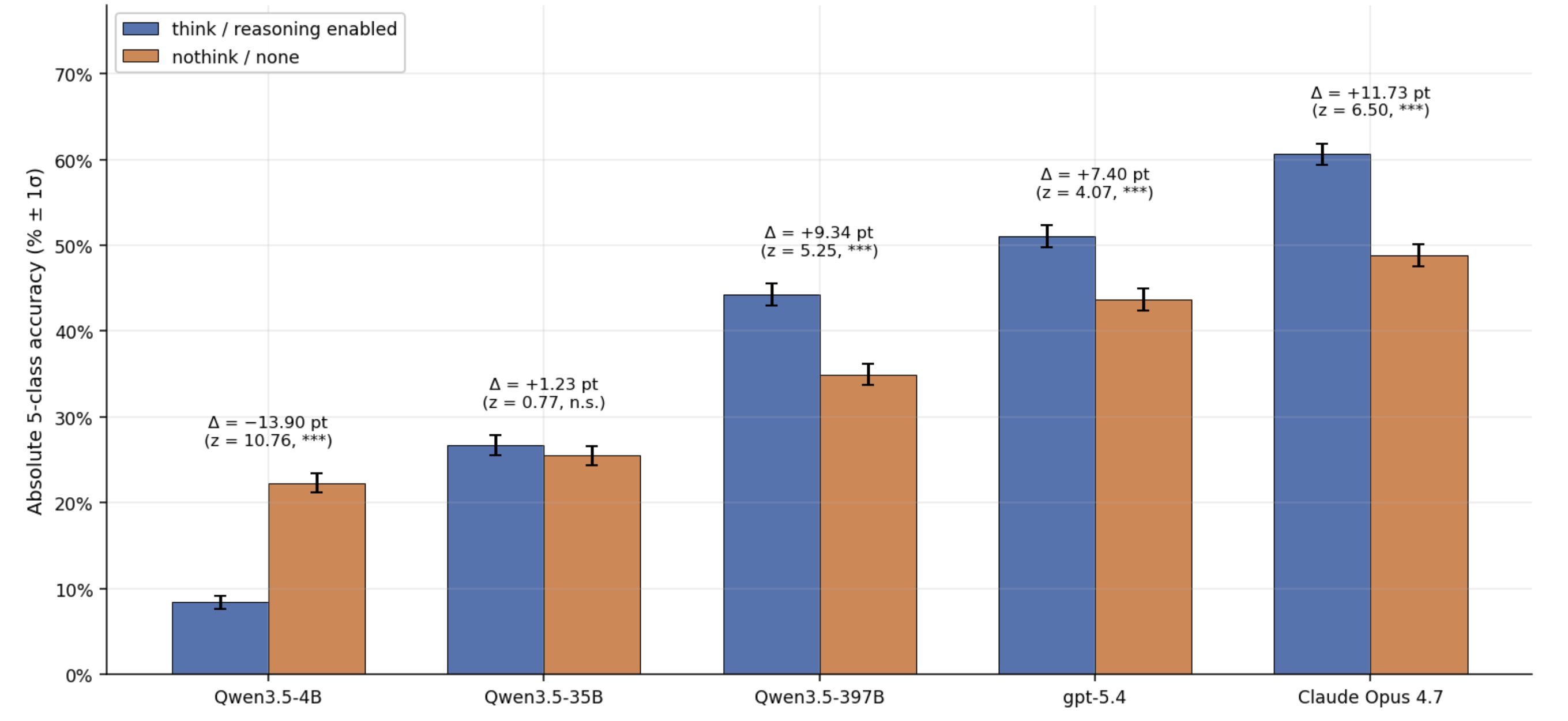}}{\fbox{\parbox[c][5cm][c]{0.8\linewidth}{\centering\itshape [Figure placeholder: figure/think\_vs\_nothink.png \textemdash{} paired bars of think vs.\ no-think absolute 5-class accuracy across the five model families, with $\Delta$ and $z$ annotations per pair]}}}
    \caption{\textbf{Reasoning-mode dial across five model families ($n = 1,500$ per run).} For each family, blue bars show absolute 5-class accuracy with thinking enabled and orange bars show the same with thinking disabled; $\Delta$ annotations report the gap and the two-sample $z$-statistic. The think-win margin grows with backbone capacity: Claude Opus~4.7 shows the largest gap ($+11.73$ pp), Qwen3.5-397B-A17B and GPT-5.4 follow with significant gains, Qwen3.5-35B-A3B is a statistical tie, and Qwen3.5-4B inverts.
    }
    \label{fig:reasoning-ablation}
\end{figure}

To quantify the impact of reasoning on the end-to-end 5-class accuracy reported in Section~\ref{sec:results-partc}, we compare the same model architecture across its two reasoning modes (thinking enabled and direct answer). Five families admit a same-architecture pair at $n = 1,500$: Claude Opus~4.7 \{think, nothink\}, GPT-5.4 \{high-think, no-think\}, and the three Qwen3.5 sizes \{4B, 35B-A3B, 397B-A17B\} $\times$ \{think, nothink\}. Gemini 2.5 Pro versus 2.5 Flash is not a clean A/B because the two configurations are different models rather than the same model with the reasoning dial flipped, and we therefore omit it from the paired comparison. For each pair we report the absolute accuracy gap $\Delta = \text{Acc}_{\text{think}} - \text{Acc}_{\text{nothink}}$, its propagated $1\sigma$ standard error $\text{SE}_\Delta = \sqrt{\text{SE}_{\text{think}}^2 + \text{SE}_{\text{nothink}}^2}$, and the corresponding two-sample $z$-statistic. Table~\ref{tab:reasoning-ablation} reports the comparison and Figure~\ref{fig:reasoning-ablation} visualizes the paired bars across the five families.

\begin{table}[h]
\centering
\footnotesize
\setlength{\tabcolsep}{4pt}
\caption{\textbf{Think vs.\ no-think pairs across five model families.} $\Delta$ is the absolute 5-class accuracy gap (think minus no-think) with propagated $1\sigma$ standard error; $z$ is the two-sample $z$-statistic. Significance markers: $\ast\ast\ast$ denotes $|z| > 4$. Rows are ordered from most negative $\Delta$ to most positive.}
\label{tab:reasoning-ablation}
\resizebox{\linewidth}{!}{%
\begin{tabular}{lcccrl}
\toprule
\textbf{Family} & \textbf{Think (\%)} & \textbf{No-think (\%)} & \boldmath$\Delta \pm \text{SE}_\Delta$ \textbf{(pt)} & \boldmath$z$ & \textbf{Verdict} \\
\midrule
Qwen3.5-4B           & $8.41 \pm 0.72$  & $22.32 \pm 1.08$ & $-13.90 \pm 1.29$ & $-10.76$ & no-think wins ($\ast\ast\ast$); think truncated \\
Qwen3.5-35B-A3B      & $26.73 \pm 1.14$ & $25.50 \pm 1.13$ & $+1.23 \pm 1.60$  & $0.77$   & essentially tied (n.s.) \\
Qwen3.5-397B-A17B    & $44.27 \pm 1.28$ & $34.98 \pm 1.23$ & $+9.29 \pm 1.78$  & $5.25$   & think wins ($\ast\ast\ast$) \\
GPT-5.4              & $51.07 \pm 1.29$ & $43.67 \pm 1.28$ & $+7.40 \pm 1.82$  & $4.07$   & think wins ($\ast\ast\ast$) \\
Claude Opus~4.7      & $60.60 \pm 1.26$ & $48.87 \pm 1.29$ & $+11.73 \pm 1.80$ & $6.52$   & think wins, largest margin ($\ast\ast\ast$) \\
\bottomrule
\end{tabular}%
}
\end{table}

\paragraph{Interpretation.}
Three frontier-scale model pairs (\texttt{Opus 4.7}, \texttt{Qwen3.5-397B-A17B}, and \texttt{GPT-5.4}) show statistically significant accuracy gains of $+7$ to $+12$ percentage points ($z > 4$) when extended reasoning is enabled. The mid-sized \texttt{Qwen3.5-35B-A3B} pair results in a statistical tie, while the smallest \texttt{Qwen3.5-4B} configuration exhibits a decisive performance inversion, suggesting that the efficacy of internal chain-of-thought is highly sensitive to base model scale on this specialized task. Among the models capable of successfully leveraging reasoning-enabled modes, the accuracy uplift scales monotonically with backbone capacity, indicating that the benefits of extended thinking are most pronounced in the strongest models.


\section{Alignment and Calibration of Model Self-Grading}
\label{app:calibration}

This appendix supplies the per-run numerical detail behind the within-model calibration analysis in Section~\ref{sec:honesty-inner}.

\subsection{Definitions and standard errors}
\label{app:calibration-definitions}

Let $c_i \in \{0, 1, 2, 3, 4, 5\}$ denote the per-instance Part~B mean self-score (averaged over the three rubric dimensions: key evidence, leading interpretation, alternative analysis) and $y_i \in \{0, 1\}$ denote the per-row Part~C correctness indicator under end-to-end 5-class scoring. We report three calibration scalars per run, computed over the $n_{\mathrm{linked}}$ parsed rows that admit both a self-score and an evaluable Part~C label.

\begin{itemize}
    \item \textbf{Calibration gap}: $\Delta\mu_{\mathrm{conf}} = \overline{c}\mid_{y=1} - \overline{c}\mid_{y=0}$. We report a $1\sigma$ standard error under the difference-of-means approximation $\mathrm{SE} = \sqrt{\sigma^2/n_{\mathrm{correct}} + \sigma^2/n_{\mathrm{incorrect}}}$ with $\sigma = 1$ (range/4 on the 0--5 ordinal) as a conservative-side estimate; without per-row score variances, this overstates the SE relative to the true pooled SD.
    \item \textbf{Pearson $r$}: the standard product-moment correlation between $c_i$ and $y_i$, reported with Fisher $1/\sqrt{n_{\mathrm{linked}} - 3}$ as a $1\sigma$ standard error.
    \item \textbf{Per-bin accuracy}: with cutoff $\tau$, partition rows into $\mathrm{high} = \{i : c_i \ge \tau\}$ and $\mathrm{low} = \{i : c_i < \tau\}$ and report $\mathrm{acc}\mid_{\mathrm{high}}$, $\mathrm{acc}\mid_{\mathrm{low}}$, the bin sizes $n_{\mathrm{high}}, n_{\mathrm{low}}$, and the bin-difference $\Delta_{\mathrm{bin}} = \mathrm{acc}\mid_{\mathrm{high}} - \mathrm{acc}\mid_{\mathrm{low}}$ in percentage points. The threshold is $\tau = 4$ (mean of three integer 0--5 scores at least 4); we additionally report midpoint $\tau = 3.5$ and strict $\tau = 4.5$.
\end{itemize}

\subsection{Per-model calibration scalars}
\label{app:calibration-scalars}

Table~\ref{tab:calibration-scalars} reports the calibration gap and Pearson~$r$ for all 13 models alongside the absolute 5-class accuracy. Rows are sorted in ascending calibration gap. Standard errors are conservative (see \S\ref{app:calibration-definitions}); for the gap column, a value within $\pm 2\sigma$ of zero should be read as ``no measurable calibration signal'' rather than as ``small calibration error.''

\begin{table}[h]
\centering
\footnotesize
\setlength{\tabcolsep}{4pt}
\caption{\textbf{Per-run calibration scalars (sorted by calibration gap).} $n_{\mathrm{linked}}$ is the number of parsed rows admitting both a self-score and an evaluable Part~C label. Calibration gap is $\overline{c}\mid_{\mathrm{correct}} - \overline{c}\mid_{\mathrm{incorrect}}$ in points on the 0--5 self-score scale; Pearson $r$ is between per-row self-score and per-row 0/1 correctness.}
\label{tab:calibration-scalars}
\begin{tabular}{clccccc}
\toprule
\textbf{\#} & \textbf{Run} & \textbf{$n_{\mathrm{linked}}$} & \textbf{Abs.\ 5-class (\%)} & \textbf{Calibration gap} & \textbf{$1\sigma$ SE} & \textbf{Pearson $r$} \\
\midrule
1  & Qwen3.5-397B-A17B think      & 1,500 & 44.27 & $0.0014$ & $\pm 0.052$ & $+0.0047$ \\
2  & Qwen3.5-4B think             & 317  & 8.41  & $0.0068$ & $\pm 0.116$ & $+0.0195$ \\
3  & Gemini 2.5 Pro high-think    & 1,500 & 41.93 & $0.0098$ & $\pm 0.052$ & $+0.0302$ \\
4  & Qwen3.5-35B-A3B think        & 967  & 26.73 & $0.0147$ & $\pm 0.069$ & $+0.0617$ \\
5  & Qwen3.5-397B-A17B nothink    & 1498 & 34.98 & $0.0214$ & $\pm 0.058$ & $+0.0592$ \\
6  & Qwen3.5-35B-A3B nothink      & 1485 & 25.50 & $0.0298$ & $\pm 0.061$ & $+0.0946$ \\
7  & Kimi K2.5 think              & 1497 & 49.34 & $0.0325$ & $\pm 0.052$ & $+0.0486$ \\
8  & Claude Opus 4.7 think        & 1,500 & 60.60 & $\mathbf{0.0350}$ & $\pm 0.053$ & $+0.0470$ \\
9  & Qwen3.5-4B nothink           & 1367 & 22.32 & $0.0381$ & $\pm 0.064$ & $+0.1064$ \\
10 & Gemini 2.5 Flash no-think    & 1,500 & 36.27 & $0.1263$ & $\pm 0.054$ & $+0.1847$ \\
11 & GPT-5.4 high-think           & 1,500 & 51.07 & $0.1299$ & $\pm 0.052$ & $+0.2028$ \\
12 & GPT-5.4 no-think             & 1,500 & 43.67 & $0.1413$ & $\pm 0.052$ & $+0.2185$ \\
13 & Claude Opus 4.7 nothink      & 1,500 & 48.87 & $\mathbf{0.1886}$ & $\pm 0.052$ & $\mathbf{+0.2517}$ \\
\bottomrule
\end{tabular}
\end{table}

\begin{figure}[h]
    \centering
    \IfFileExists{figure/calibration_pearson_bar.png}{\includegraphics[width=0.95\linewidth]{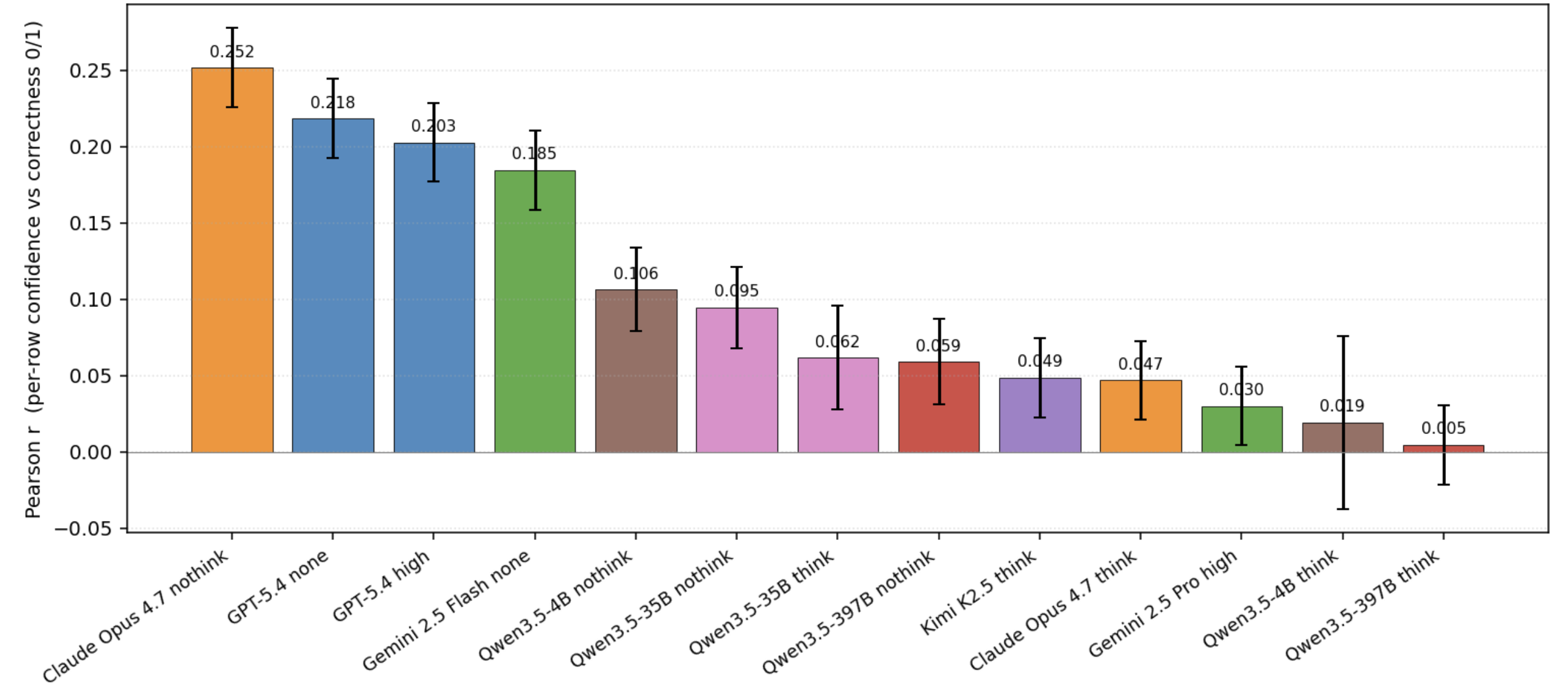}}{\fbox{\parbox[c][5cm][c]{0.8\linewidth}{\centering\itshape [Figure placeholder: figure/calibration\_pearson\_bar.png \textemdash{} sorted bar chart of per-run Pearson correlation between per-row self-confidence and per-row 0/1 correctness with $1\sigma$ Fisher error bars (mirrors the calibration report's Fig C2)]}}}
    \caption{\textbf{Per-run Pearson correlation between self-confidence and correctness, sorted descending.} Larger values mean the model's self-confidence rank-orders its right answers from its wrong ones; near-zero values mean the confidence dial is uninformative. Error bars are $1\sigma$ standard error.}
    \label{fig:calibration-pearson}
\end{figure}



\subsection{Per-bin accuracy at the $\tau = 4$ threshold}
\label{app:calibration-bin}

Table~\ref{tab:calibration-bin-default} reports the high-/low-confidence bin accuracies and the operational triage gap at the default cutoff. The most actionable view of within-run calibration: if a downstream pipeline wants to use the model's self-score as a per-row triage signal, the high-confidence bin should be both well-populated and meaningfully more accurate than the low-confidence bin.

\begin{table}[h]
\centering
\footnotesize
\setlength{\tabcolsep}{4pt}
\caption{\textbf{Per-bin accuracy at $\tau = 4$.} High bin: rows with self-mean $\geq 4$. Low bin: rows with self-mean $< 4$. $\Delta_{\mathrm{bin}}$ is the high-minus-low accuracy gap in percentage points. Rows sorted by $\Delta_{\mathrm{bin}}$ descending. Runs whose low-bin denominator is too small to support a stable estimate ($n_{\mathrm{low}} \le 5$) are flagged.}
\label{tab:calibration-bin-default}
\begin{tabular}{lcccccr}
\toprule
\textbf{Run} & \textbf{acc\,$|$\,high (\%)} & \textbf{$n_{\mathrm{high}}$} & \textbf{acc\,$|$\,low (\%)} & \textbf{$n_{\mathrm{low}}$} & \textbf{$\Delta_{\mathrm{bin}}$ (pp)} & \\
\midrule
GPT-5.4 no-think           & 44.06 & 1464 & 27.78 & 36   & $+16.28$ & \\
\textbf{Claude Opus 4.7 nothink} & \textbf{53.48} & \textbf{1006} & \textbf{39.47} & \textbf{494}  & $\mathbf{+14.01}$ & \\
GPT-5.4 high-think         & 51.88 & 1382 & 41.53 & 118  & $+10.35$ & \\
Gemini 2.5 Flash no-think  & 36.79 & 1416 & 27.71 & 83   & $+9.08$ & \\
Kimi K2.5 think            & 49.96 & 1359 & 44.20 & 138  & $+5.76$ & \\
Claude Opus 4.7 think      & 60.68 & 1002 & 60.44 & 498  & $+0.24$ & \\
Qwen3.5-35B-A3B nothink    & 26.16 & 1399 & 0.00  & 21   & $+26.16$ & noise ($n_{\mathrm{low}} \le 5$) \\
Qwen3.5-4B nothink         & 24.59 & 1326 & 0.00  & 5    & $+24.59$ & noise ($n_{\mathrm{low}} \le 5$) \\
\multicolumn{7}{l}{(Qwen3.5-397B-A17B think and Gemini 2.5 Pro high-think: $n_{\mathrm{low}} \le 2$, bin accuracy undefined.)} \\
\bottomrule
\end{tabular}
\end{table}

\begin{figure}[h]
    \centering
    \IfFileExists{figure/calib_perbin_default.png}{\includegraphics[width=\linewidth]{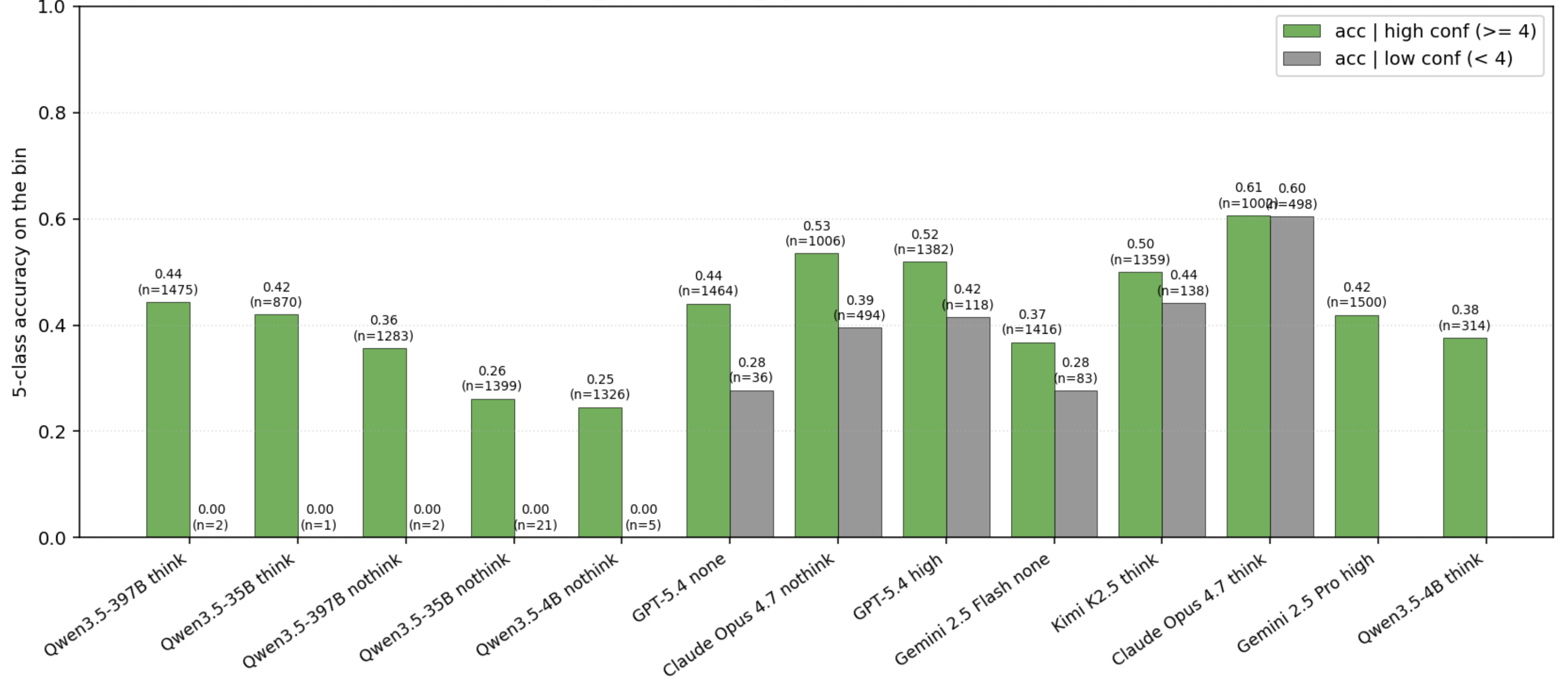}}{\fbox{\parbox[c][4.5cm][c]{0.8\linewidth}{\centering\itshape [Figure placeholder: figure/calib\_perbin\_default.png \textemdash{} per-run grouped bars of accuracy on the high-confidence bin (green) versus the low-confidence bin (grey) at the $\tau = 4$ cutoff (report Fig~C5)]}}}
    \caption{\textbf{5-class accuracy on rows the model labels high-confidence (self-mean $\geq 4$) versus low-confidence (self-mean $< 4$).} 
    \texttt{Claude Opus 4.7 nothink} has both a populated low-confidence bin 
    and a meaningful accuracy gap; \texttt{GPT-5.4 high-think} has a strong gap on a much narrower low-confidence population.
    }
    \label{fig:calib-perbin-default}
\end{figure}

\subsection{Joint views and confidence-mass distribution}
\label{app:calibration-joint}

\begin{figure}[h]
    \centering
    \IfFileExists{figure/calib_joint_gap_r.png}{\includegraphics[width=\linewidth]{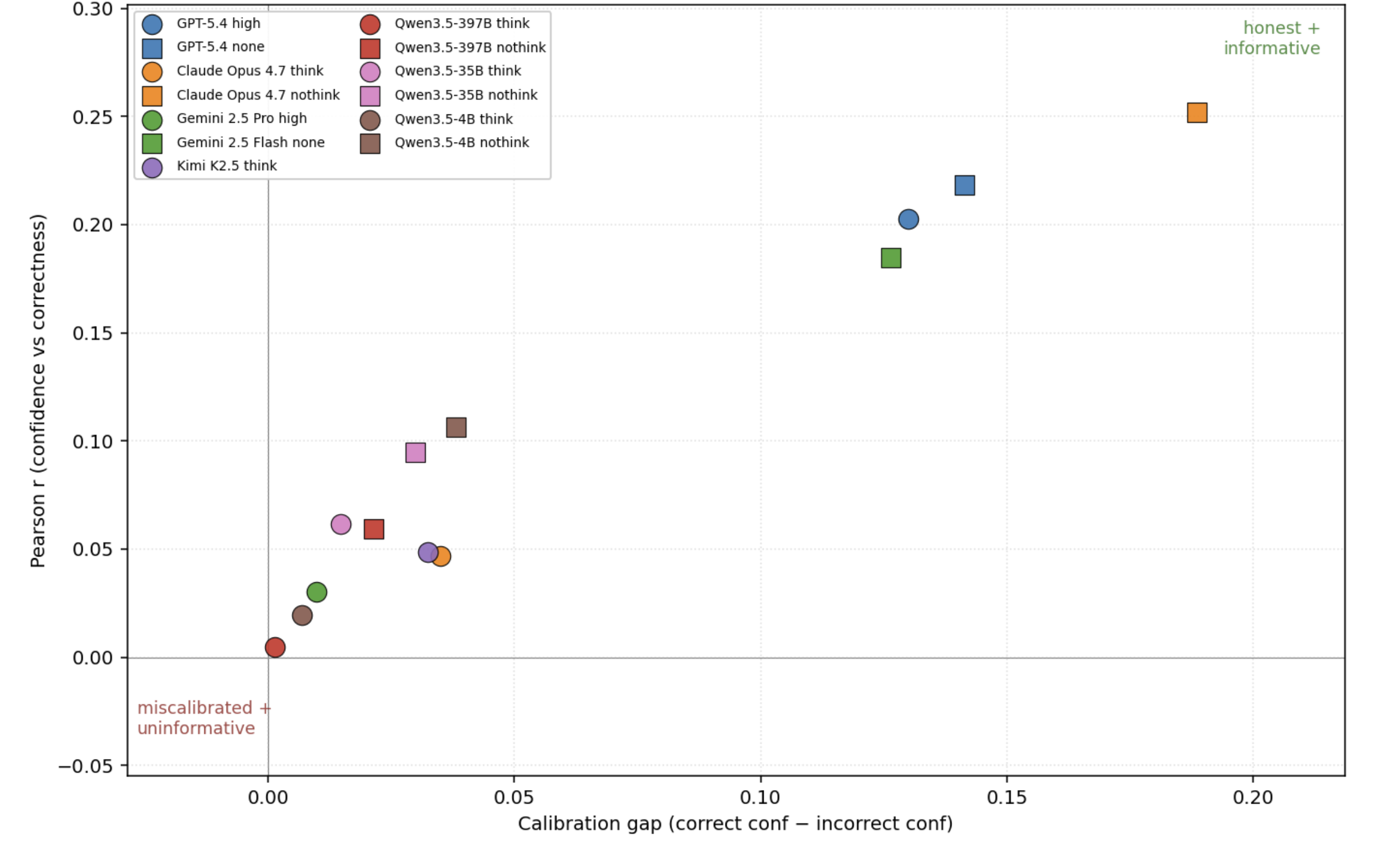}}{\fbox{\parbox[c][5cm][c]{0.6\linewidth}{\centering\itshape [Figure placeholder: figure/calib\_joint\_gap\_r.png \textemdash{} cross-run scatter, calibration gap (x) versus Pearson $r$ (y), 13 dots, with the upper-right ``honest + informative'' quadrant labeled (report Fig~C3)]}}}
    \caption{\textbf{Calibration gap vs.\ Pearson $r$ across 13 runs.} Each marker is one run. The horizontal axis is the calibration gap defined in \S\ref{app:calibration-definitions}; the vertical axis is the per-row Pearson correlation also shown as a sorted bar chart in main-text Figure~\ref{fig:calibration-pearson}. The two views are alternative summaries of the same correlation; this scatter additionally reveals two regimes: an upper-right honest-and-informative cluster (\texttt{Opus 4.7 nothink}, both \texttt{GPT-5.4} variants, \texttt{Gemini 2.5 Flash no-think}) and a low-gap-low-$r$ corner near the origin populated by runs whose dial is essentially constant.}
    \label{fig:calib-joint-app}
\end{figure}

The calibration gap and Pearson~$r$ are alternative summaries of the same correlation; Figure~\ref{fig:calib-joint-app} plots them against each other. The diagonal alignment is mechanical (Pearson $r$ is bounded by the gap divided by the SD of the confidence distribution), but the position of each run reveals two distinct regimes: the upper-right honest-and-informative quadrant (\texttt{Opus 4.7 nothink}, both \texttt{GPT-5.4} variants, \texttt{Gemini 2.5 Flash no-think}) versus a low-gap-low-$r$ corner near the origin populated by runs whose dial is essentially constant.

\begin{figure}[h]
    \centering
    \IfFileExists{figure/calib_bin_distribution.png}{\includegraphics[width=\linewidth]{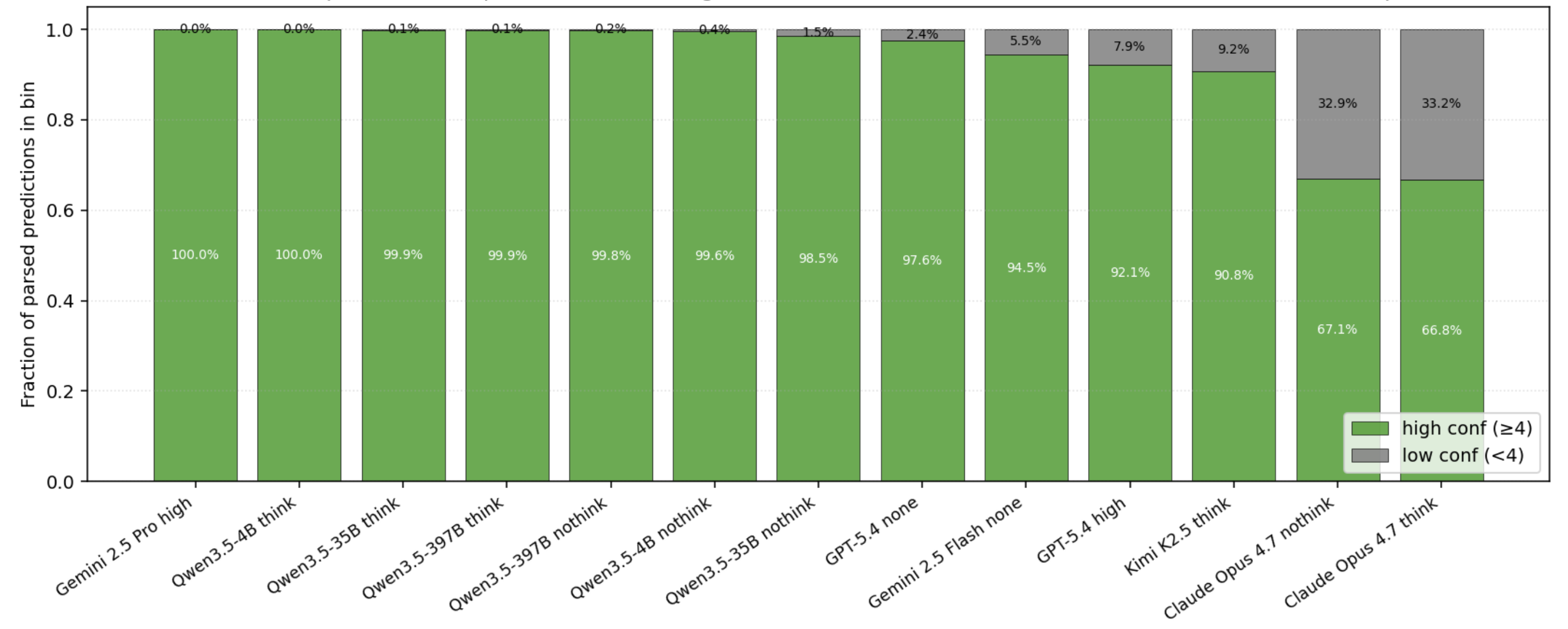}}{\fbox{\parbox[c][4.5cm][c]{0.8\linewidth}{\centering\itshape [Figure placeholder: figure/calib\_bin\_distribution.png \textemdash{} per-run stacked bars of high-confidence vs.\ low-confidence mass at $\tau = 4$ (report Fig~C6)]}}}
    \caption{\textbf{Confidence-bin distribution per run.} Most runs park more than $90\%$ of predictions in the high-confidence bin; only the two \texttt{Opus 4.7} runs come anywhere near a balanced split (think $67/33$, nothink $67/33$). For the high-mass-only runs there is no usable low-confidence subset to calibrate against.}
    \label{fig:calib-bin-distribution}
\end{figure}

\section{Self-Correction and Behavioral Reliability}
\label{app:second-rollout}

This appendix supplies the full numerical detail behind the behavioral honesty test in Section~\ref{sec:honesty-retry}.

\subsection{Methodology}
\label{app:second-rollout-method}

For each closed-source run we constructed a low-confidence subset by selecting first-pass benchmark rows where the mean Part~B self-score (averaged over the three rubric dimensions) was strictly below $4$ and the Part~C output parsed to a valid 5-class label. From this pool we drew $n = 35$ alerts per model under Hamilton-quota stratification on the gold class (proportional to the low-confidence pool's class histogram with random tie-breaking; the $n = 35$ alert IDs differ across models because each model's low-confidence set differs). Pool sizes were $\{118, 36, 83, 109, 121\}$ for the five tested runs respectively.

Each model was re-prompted with the original images and metadata plus its own first-pass JSON appended verbatim to the user message; the gold class was \emph{not} revealed and the second-pass instruction asked only for a fresh classification under the same Part~A/B/C schema as the benchmark. We define $C/W$ as first-pass correct/wrong and $C'/W'$ as second-pass correct/wrong, and report the four cell counts of the $2\!\times\!2$ paired-outcome table along with derived rates: \textbf{correction} $W \to C'$ as a fraction of the wrong stratum, \textbf{damage} $C \to W'$ as a fraction of the correct stratum, and \textbf{persistence} $W \to W'$ as a fraction of the wrong stratum. McNemar's exact two-sided test on the two off-diagonal cells gives a significance reading on the asymmetry of discordant flips.

Cross-model McNemar $p$-values are not directly comparable because the alert IDs differ across models; we treat them as a within-model significance reading rather than a pooled A/B.


\begin{table}[h!]
\centering
\scriptsize
\setlength{\tabcolsep}{3pt}
\resizebox{\linewidth}{!}{%
\begin{tabular}{lcccccccccccc}
\toprule
\textbf{Model} & \textbf{Acc$_1$} & \textbf{Acc$_2$} & \textbf{$\Delta$Acc} & \textbf{Correction} & \textbf{Damage} & \textbf{Persistence} & \textbf{McNemar $p$} & \textbf{$\Delta$MSRS} & \textbf{Out 2/1} & \textbf{Prior ref} & \textbf{S3 flip} & \textbf{2nd MSRS\,$\geq\!4$} \\
\midrule
GPT-5.4 high-think         & $45.71\%$ & $54.29\%$ & $+8.57$  & $31.58\%$ & $18.75\%$ & $68.42\%$ & $0.508$ & $+0.276$ & $1.10$ & $34.29\%$ & $17.14\%$ & $74.29\%$ \\
GPT-5.4 no-think           & $28.57\%$ & $31.43\%$ & $+2.86$  & $16.00\%$ & $30.00\%$ & $84.00\%$ & $1.000$ & $+0.371$ & $1.04$ & $\mathbf{74.29\%}$ & $11.43\%$ & $82.86\%$ \\
Gemini 2.5 Flash no-think  & $17.14\%$ & $17.14\%$ & $\phantom{+}0.00$ & $0.00\%$ & $0.00\%$ & $100.00\%$ & $1.000$ & $+0.429$ & $1.03$ & $40.00\%$ & $0.00\%$ & $48.57\%$ \\
\textbf{Claude Opus 4.7 think}    & $51.43\%$ & $\mathbf{80.00\%}$ & $\mathbf{+28.57}$ & $58.82\%$ & $\mathbf{0.00\%}$ & $41.18\%$ & $\mathbf{0.00195}$ & $+0.048$ & $0.97$ & $57.14\%$ & $20.00\%$ & $11.43\%$ \\
\textbf{Claude Opus 4.7 nothink}  & $37.14\%$ & $\mathbf{80.00\%}$ & $\mathbf{+42.86}$ & $\mathbf{68.18\%}$ & $\mathbf{0.00\%}$ & $31.82\%$ & $\mathbf{6.1\!\times\!10^{-5}}$ & $+0.038$ & $1.12$ & $54.29\%$ & $\mathbf{34.29\%}$ & $14.29\%$ \\
\bottomrule
\end{tabular}%
}
\caption{\textbf{Second-rollout retry headline metrics ($n = 35$ per model).} Acc$_1$ and Acc$_2$ are first- and second-pass 5-class accuracy on the low-confidence subset. Correction and Damage are the per-stratum flip rates as defined above. McNemar $p$ is the exact two-sided test on the discordant cells. $\Delta$MSRS is the change in mean self-rating from first to second pass. Out tok 2/1 is the ratio of mean second-pass to first-pass output tokens. Prior-ref rate is the fraction of second-pass rows whose text matches a regex for explicit reference to the model's prior answer. Stage3-flip is the fraction of rows whose Stage-3 subclass label changed between passes. 2nd MSRS$\,\geq\!4$ is the fraction of second-pass rows assigned mean self-score $\geq 4$.}
\label{tab:second-rollout-headline}
\end{table}


\begin{table}[h]
\centering
\footnotesize
\begin{tabular}{lcc}
\toprule
\textbf{Model} & \textbf{$W \to C'$} & \textbf{$C \to W'$} \\
\midrule
GPT-5.4 high-think         & 6  & 3 \\
GPT-5.4 no-think           & 4  & 3 \\
Gemini 2.5 Flash no-think  & 0  & 0 \\
Claude Opus 4.7 think      & 10 & 0 \\
Claude Opus 4.7 nothink    & 15 & 0 \\
\bottomrule
\end{tabular}
\caption{\textbf{Discordant pair counts $(W \to C', C \to W)$.} McNemar uses only these two cells; the symmetric (concordant) cells $C \to C'$ and $W \to W'$ do not enter the test.}
\label{tab:second-rollout-discordants}
\end{table}

\subsection{Auxiliary figures}
\label{app:second-rollout-figures}

\begin{figure}[H]
    \centering
    \IfFileExists{figure/second_rollout_correction_flow.png}{\includegraphics[width=\linewidth]{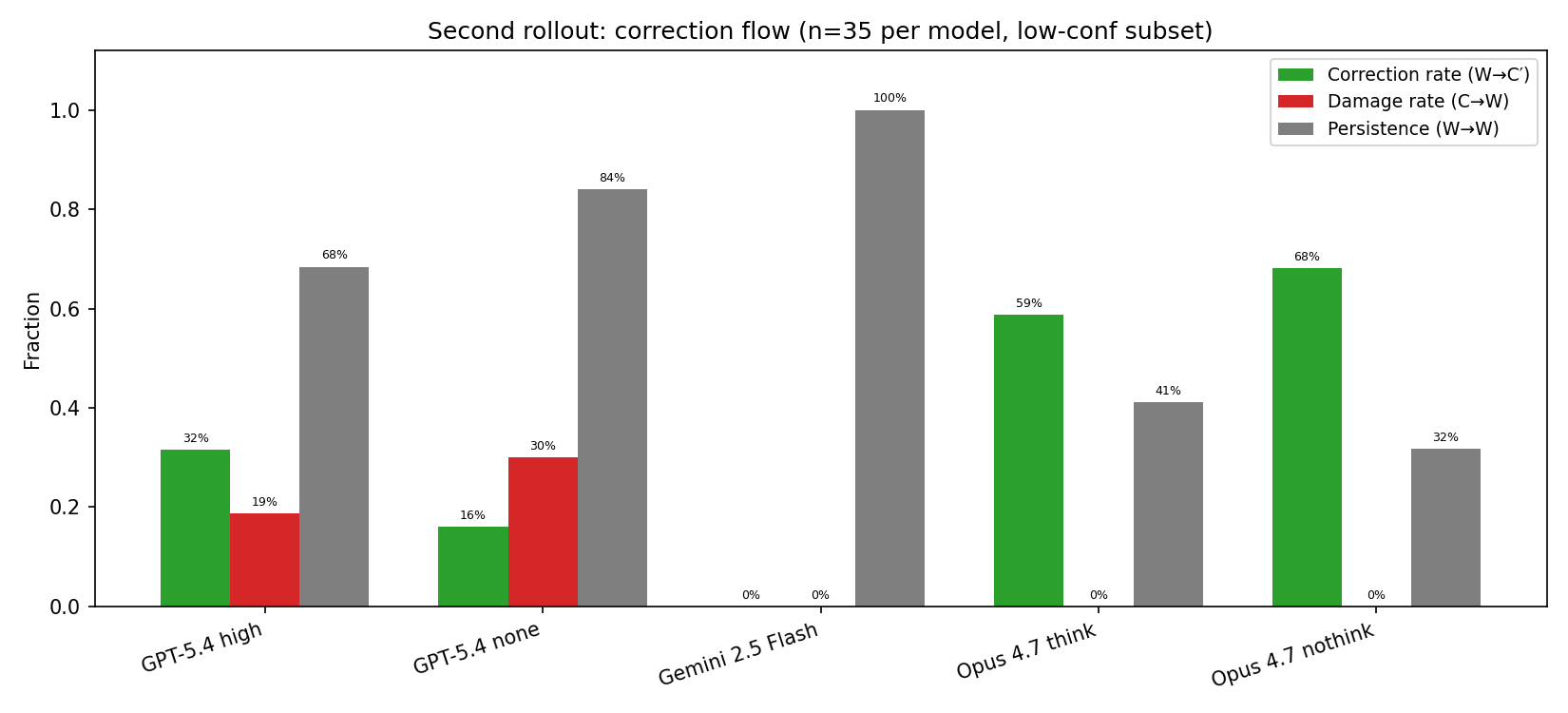}}{\fbox{\parbox[c][4.5cm][c]{0.8\linewidth}{\centering\itshape [Figure placeholder: figure/second\_rollout\_correction\_flow.png \textemdash{} grouped bars of correction / damage / persistence rates per model (report Fig~1)]}}}
    \caption{\textbf{Correction, damage, and persistence rates per model.} Tall green (correction) plus zero red (damage) on both \texttt{Claude Opus 4.7} variants; \texttt{GPT-5.4 high-think} trades correction against a visible damage segment.}
    \label{fig:second-rollout-correction}
\end{figure}

\begin{figure}[H]
    \centering
    \IfFileExists{figure/second_rollout_contingency.png}{\includegraphics[width=0.85\linewidth]{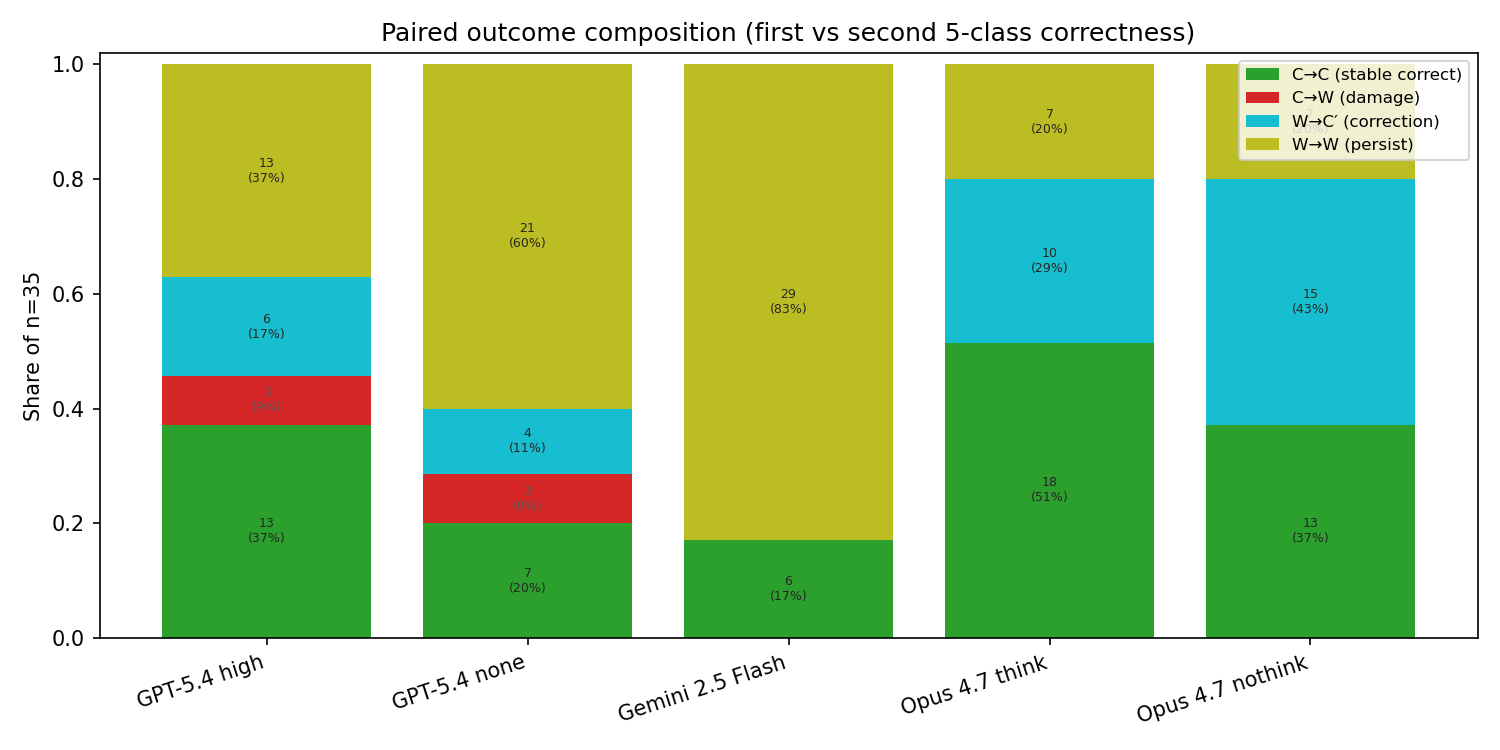}}{\fbox{\parbox[c][4.5cm][c]{0.8\linewidth}{\centering\itshape [Figure placeholder: figure/second\_rollout\_contingency.png \textemdash{} normalized stacked bars of paired outcomes $C \to C'$, $C \to W'$, $W \to C'$, $W \to W'$ as a fraction of all 35 rows per model (report Fig~4)]}}}
    \caption{\textbf{Normalized $2\!\times\!2$ paired-outcome composition.} Stacked bars sum to $100\%$ within each model. \texttt{Opus 4.7} runs devote a large slice to $W \to C'$ with no visible $C \to W'$ band; \texttt{Gemini 2.5 Flash no-think} is dominated by $W \to W'$ plus the rows that stay correct.}
    \label{fig:second-rollout-contingency}
\end{figure}

\begin{figure}[H]
    \centering
    \IfFileExists{figure/second_rollout_mcnemar.png}{\includegraphics[width=0.9\linewidth]{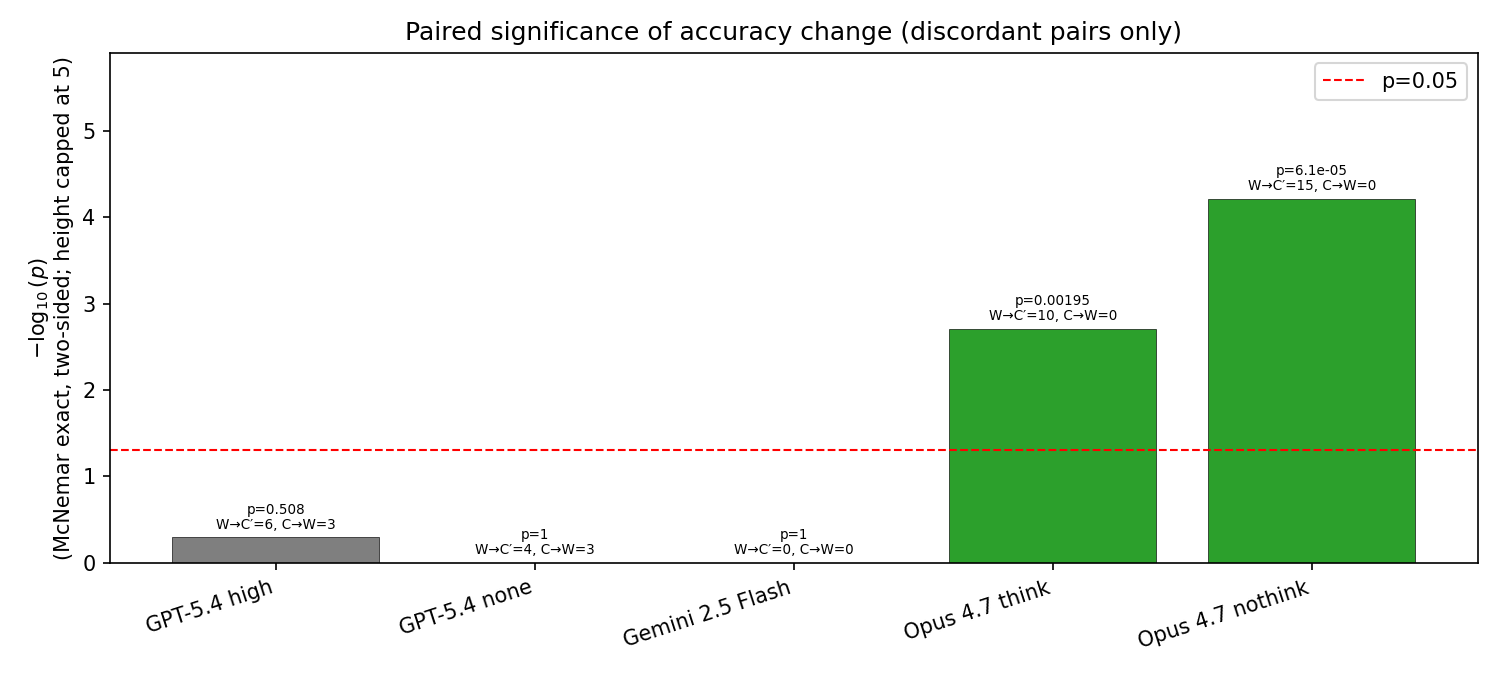}}{\fbox{\parbox[c][4cm][c]{0.65\linewidth}{\centering\itshape [Figure placeholder: figure/second\_rollout\_mcnemar.png \textemdash{} bar chart of $-\log_{10} p$ for McNemar's exact two-sided test per model with $\alpha = 0.05$ reference line (report Fig~3)]}}}
    \caption{\textbf{McNemar two-sided $p$-values on $-\log_{10}$ scale.} The test evaluates the asymmetry of accuracy changes using only discordant pairs ($W \rightarrow C'$ vs. $C \rightarrow W$); the dashed line marks $p = 0.05$. Both \texttt{Claude Opus 4.7} variants sit above this threshold, indicating statistically significant improvements. \texttt{GPT-5.4} configurations remain below the line, while \texttt{Gemini 2.5 Flash} has zero discordant pairs, rendering the test degenerate ($p = 1$).}
    \label{fig:second-rollout-mcnemar}
\end{figure}

\subsection{Per-family interpretation}
\label{app:second-rollout-interpretation}

\paragraph{\texttt{Claude Opus 4.7} (think and nothink).}
Largest $\Delta$Acc on the cohort, second-pass accuracy $80\%$ in both modes, and \emph{zero damage} ($C \to W' = 0$ for both); McNemar $p = 2.0\!\times\!10^{-3}$ (think) and $6.1\!\times\!10^{-5}$ (nothink). Mean $\Delta$MSRS is small and positive ($+0.05$ / $+0.04$) and second-pass MSRS$\,\geq\!4$ stays low ($11$--$14\%$): many rows remain self-critical even after a correct second answer, consistent with honest self-grading under revision rather than indiscriminate score inflation. Stage-3 subclass flip rate is higher for nothink ($34.29\%$) than think ($20.00\%$), suggesting that direct-answer mode more often re-reads its first response and revises the fine subclass.

\paragraph{\texttt{GPT-5.4} (high-think and no-think).}
\texttt{High-think} delivers $+8.57$~pp net but with $18.75\%$ damage---roughly one in five initially-correct low-confidence rows regresses; McNemar $p = 0.51$, so the asymmetry is not significant despite a positive point estimate. \texttt{No-think} adds only $+2.86$~pp, with $30\%$ damage and the highest prior-reference rate in the cohort ($74.29\%$): explicit talk about the prior answer does not translate into reliable repair.

\paragraph{\texttt{Gemini 2.5 Flash no-think}.}
$\Delta$Acc $= 0$; correction and damage are both $0$; every row stays in the same correctness class on retry ($6$ stayed correct, $29$ stayed wrong; zero discordant pairs, McNemar $p = 1$ uninformative). Mean $\Delta$MSRS is the largest in the cohort ($+0.43$) and second-pass MSRS$\,\geq\!4$ is near coin-flip ($48.57\%$): the model re-grades more confidently without changing any staged classification---the clearest calibration--behavior decoupling in this ablation.

\subsection{Qualitative Case Studies of Second-Rollout Behavior}
To supplement the quantitative findings, we provide two representative case studies from the second-rollout cohort. These examples illustrate the contrast between "calibrated self-correction" (Section~\ref{subsec:self_correction}) and "static re-grading" (Section~\ref{subsec:static_persistence}). In both cases, models were provided with their original first-pass response and asked to perform a second review of the visual and metadata inputs to reanswer the questions without being informed of their prior correctness.

\subsubsection{Evidence Re-weighting and Successful Self-Correction}
\label{subsec:self_correction}
This example, featuring \texttt{claude-opus-4.7-nothink}, demonstrates the highly effective repair process observed in frontier reasoning models. In the first pass, the model over-weights the temporal baseline (9 detections over 19 days) to predict \textit{AGN}. During the second trial, the model exhibits genuine behavioral gain by re-evaluating the spatial metadata it previously under-weighted—specifically the 1.52~arcsec offset (\texttt{distpsnr1}) and the extended host indicators (\texttt{chinr}). This transition from \textit{AGN} to \textit{supernova} represents a successful revision driven by a re-reading of the scientific context rather than indiscriminate score inflation.

\noindent\textit{Gold: SN. Paired correctness: wrong$\to$correct.}

Below are the inputs fed to the model for its first attempt:

\begin{figure}[htbp]
\centering
\includegraphics[width=0.88\textwidth]{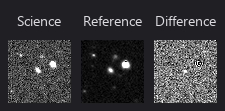}
\caption{Science--Reference--Difference montage for \texttt{ZTF21abculew}.}
\end{figure}

\begin{chatmsg}[grey]{System — attempt 1}[breakable]
\noindent\texttt{\{general\_prompt\_block\}}\footnote{The general system prompt is provided in Section~\ref{app:main prompt}.}
\end{chatmsg}
\begin{chatmsg}[blue]{User — attempt 1}[breakable]
\begin{lstlisting}[style=prompttxt]
[ALERT IDENTIFIERS]
- Object ID: ZTF21abculew

[ZTF CANDIDATE FIELDS]
- fid: 2
- isdiffpos: t
- firstmjd: 59361.3
- magpsf: 19.9901
- sigmapsf: 0.184607
- fwhm: 1.93
- classtar: 0.963
- sgscore1: 0.5
- distpsnr1: 1.5198
- chinr: 6.499
- sharpnr: 0.708
- ndethist: 9
- ncovhist: 2664
- sgmag1: 20.1168
- srmag1: 19.0823
- simag1: -999
- szmag1: 19.3916
- nmtchps: 13
- deltajd: 19.0387

Field definitions and sentinel rules are in the system message.

Analyze this alert and return the JSON response.
\end{lstlisting}
\end{chatmsg}
The corresponding model response for the first attempt follows:
\begin{chatmsg}[green]{Output 1}[breakable]
\begin{lstlisting}[style=rawouttxt]
{
  "Part A": {
    "filter_band": "r",
    "subtraction_sign": "positive",
    "magpsf": 19.9901,
    "sigmapsf": 0.184607,
    "ndethist": 9,
    "ncovhist": 2664
  },
  "Part B": {
    "key_evidence": "Positive r-band detection at magpsf ~19.99 coincident with a PS1 source 1.52 arcsec away (sgscore1=0.5, ambiguous star/galaxy). classtar=0.963 suggests point-like morphology in the science image. ndethist=9 over deltajd=19 days indicates repeated detections at the same position. The reference image shows a visible source at the candidate location, and the difference image shows a positive residual.",
    "leading_interpretation_and_support": "Repeated positive-flux detections (ndethist=9) over ~19 days at a location with a persistent PS1 counterpart (sgscore1=0.5, ambiguous) favor AGN variability. The PS1 colors (g-r ~ 1.03) and ambiguous star/galaxy score are consistent with a compact galaxy nucleus, and sustained variability over weeks is characteristic of AGN rather than a single SN rise.",
    "alternative_analysis": "A supernova in the host galaxy at ~1.5 arcsec offset is plausible given the positive difference residual and host presence; however, 9 detections spanning 19 days without clear decline, plus the near-nuclear offset consistent with distpsnr1, tilt the interpretation toward AGN. A variable star is less likely because sgscore1 is not close to 1 and the counterpart color is redder than typical stellar variables at this depth.",
    "self_score_key_evidence": 4,
    "self_score_leading_interpretation_and_support": 3,
    "self_score_alternative_analysis": 4
  },
  "Part C": {
    "stage1": "real_object",
    "stage2": "astrophysical",
    "stage3": "AGN"
  }
}
\end{lstlisting}
\end{chatmsg}
Below are the inputs fed to the model for its second attempt:
\begin{chatmsg}[grey]{System — attempt 2}[breakable]
\noindent\texttt{\{second\_rollout\_ablation\_block\}}\footnote{The general system prompt is provided in Section~\ref{app:main prompt}.}
\end{chatmsg}
\begin{chatmsg}[blue]{User — attempt 2}[breakable]
\begin{lstlisting}[style=prompttxt]
[ALERT IDENTIFIERS]
- Object ID: ZTF21abculew

[ZTF CANDIDATE FIELDS]
- fid: 2
- isdiffpos: t
- firstmjd: 59361.3
- magpsf: 19.9901
- sigmapsf: 0.184607
- fwhm: 1.93
- classtar: 0.963
- sgscore1: 0.5
- distpsnr1: 1.5198
- chinr: 6.499
- sharpnr: 0.708
- ndethist: 9
- ncovhist: 2664
- sgmag1: 20.1168
- srmag1: 19.0823
- simag1: -999
- szmag1: 19.3916
- nmtchps: 13
- deltajd: 19.0387

Field definitions and sentinel rules are in the system message.

Analyze this alert and return the JSON response.

---

Below this metadata block, the live API user message also embeds the model's complete first-trial JSON (verbatim) under [YOUR FIRST ATTEMPT JSON --- FOR SELF-REVIEW ONLY], then asks for the second-trial JSON. Primary-manuscript placeholder for that payload: {embedded_first_attempt_json_block} (constructed by prompts_second_roll_out_ablation.build_user_prompt; archived under data_second_roll_out_ablation/priors/).
\end{lstlisting}
\end{chatmsg}
The corresponding model response for the second attempt follows:
\begin{chatmsg}[green]{Output 2}[breakable]
\begin{lstlisting}[style=rawouttxt]
{
  "Part A": {
    "filter_band": "r",
    "subtraction_sign": "positive",
    "magpsf": 19.9901,
    "sigmapsf": 0.184607,
    "ndethist": 9,
    "ncovhist": 2664
  },
  "Part B": {
    "key_evidence": "Positive r-band detection at magpsf ~19.99 with sigmapsf 0.18 (modest S/N). The science image shows a source near the center with a brighter neighbor visible to the lower-left; the reference shows the same field with a clearly resolved/extended-looking host at the candidate location. The difference image shows mostly noise with a faint positive residual near center. Metadata: distpsnr1=1.52 arcsec to nearest PS1 source (sgscore1=0.5, ambiguous), classtar=0.963 (point-like on science), ndethist=9 over deltajd=19.0 days, chinr=6.5 (high chi suggests non-stellar/extended reference source).",
    "leading_interpretation_and_support": "The 1.5 arcsec offset from the PS1 counterpart, combined with chinr=6.5 (extended reference object) and a positive point-like residual in difference, is most consistent with a supernova offset from its host galaxy nucleus. Nine detections over 19 days is fully compatible with an SN light curve near peak/plateau. The point-like classtar on a faint magpsf~20 source above an extended host favors a transient on a galaxy rather than nuclear AGN activity.",
    "alternative_analysis": "AGN is plausible if the offset is small enough that the candidate sits on the nucleus (sgscore1=0.5 is ambiguous), and sustained detections over weeks could reflect stochastic AGN variability. However, the 1.52 arcsec offset and high chinr (extended host) favor an off-nuclear SN. A variable star is disfavored because the reference source appears extended (high chinr) and sgscore1 is not stellar. On reflection, my first-pass AGN call under-weighted the offset and host extendedness, so I revise to supernova.",
    "self_score_key_evidence": 4,
    "self_score_leading_interpretation_and_support": 4,
    "self_score_alternative_analysis": 4
  },
  "Part C": {
    "stage1": "real_object",
    "stage2": "astrophysical",
    "stage3": "supernova"
  }
}
\end{lstlisting}
\end{chatmsg}

\subsubsection{Static Persistence and Score Inflation}
\label{subsec:static_persistence}

This example illustrates a mismatch between self-rating and performance characteristic of the \texttt{Gemini 2.5 Flash} model. In the first pass, the model correctly identifies a point-like source but settles on a variable star interpretation, ignoring the possibility of a supernova despite acknowledging a low \texttt{sgscore1} for the host. Upon the second rollout, the model explicitly mentions "refining the phrasing" and even increases its self-scores for evidence and logic (from 4/4/3 to 5/4/4). However, it \textbf{fails to revise the final classification}, demonstrating that a model sounding more sure of itself does not always lead to improved scientific results.

\noindent\textit{Gold: SN. Paired correctness: wrong$\to$wrong.}
\begin{figure}[htbp]
\centering
\includegraphics[width=0.88\textwidth]{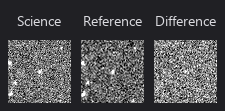}
\caption{Science--Reference--Difference montage for \texttt{ZTF24ablarbs}.}
\end{figure}

Below are the inputs fed to the model for its first attempt:
\begin{chatmsg}[grey]{System — attempt 1}[breakable]
\noindent\texttt{\{general\_prompt\_block\}}\footnote{The general system prompt is provided in Section~\ref{app:main prompt}.}
\end{chatmsg}
\begin{chatmsg}[blue]{User — attempt 1}[breakable]
\begin{lstlisting}[style=prompttxt]
[ALERT IDENTIFIERS]
- Object ID: ZTF24ablarbs

[ZTF CANDIDATE FIELDS]
- fid: 2
- isdiffpos: t
- firstmjd: 60585.2
- magpsf: 21.4729
- sigmapsf: 0.214785
- fwhm: 2.35
- classtar: 0.99
- sgscore1: 0.008
- distpsnr1: 0.0349517
- chinr: 2.314
- sharpnr: 0.366
- ndethist: 4
- ncovhist: 1787
- sgmag1: 22.1523
- srmag1: 21.6716
- simag1: 20.9498
- szmag1: 20.739
- nmtchps: 2
- deltajd: 1.13814

Field definitions and sentinel rules are in the system message.

Analyze this alert and return the JSON response.
\end{lstlisting}
\end{chatmsg}
The corresponding model response for the first attempt follows:
\begin{chatmsg}[green]{Output 1}[breakable]
\begin{lstlisting}[style=rawouttxt]
```json
{
  "Part A": {
    "filter_band": "r",
    "subtraction_sign": "positive",
    "magpsf": 21.4729,
    "sigmapsf": 0.214785,
    "ndethist": 4,
    "ncovhist": 1787
  },
  "Part B": {
    "key_evidence": "The difference image shows a clear, point-like positive residual at the center, indicating a new or brighter source. The science image shows a faint point source at the center that is not present in the reference image. The fwhm of 2.35 pixels is consistent with a point source. The classtar of 0.99 suggests a stellar morphology. The nearest PS1 source (distpsnr1 = 0.03 arcsec) has a very low sgscore1 (0.008), suggesting it is extended or a galaxy, but this is contradicted by the classtar and the image morphology of the transient. The chinr (2.314) and sharpnr (0.366) values for the nearest reference source are somewhat high for a perfect point source, but not definitively indicative of an artifact. The ndethist of 4 and ncovhist of 1787 suggest repeated observations of this field, and the deltajd of 1.13814 indicates multiple detections over a short period.",
    "leading_interpretation_and_support": "The alert is most likely a variable star. The clear point-like detection in the difference image, coupled with the absence of a counterpart in the reference image, confirms it as a real transient. The high classtar (0.99) strongly supports a stellar morphology. While the sgscore1 of the nearest PS1 source is low (0.008), its very small distance (0.03 arcsec) suggests it is likely the same object, and the low sgscore1 might be due to blending or issues with the PS1 classification for faint sources, or it could be a background galaxy with a foreground variable star. However, given the clear point-like nature in the ZTF images, a variable star is more plausible than an AGN, especially with no clear host galaxy visible at this faint magnitude. The ndethist of 4 and deltajd of ~1 day are consistent with a variable star that has been detected a few times.",
    "alternative_analysis": "An alternative interpretation is that this could be an AGN. The low sgscore1 for the nearby PS1 source could indicate a faint host galaxy, and the point-like nature of the transient could be the varying nucleus. However, without a clear extended host visible in the ZTF images, and given the strong stellar classification from classtar, a variable star is a stronger candidate. It is less likely to be a supernova given the lack of a clear host galaxy and the relatively short deltajd, though a very faint, distant supernova cannot be entirely ruled out without more information. It is not an asteroid due to the lack of a dipole signature in the difference image and the stationary appearance between science and reference images. It is also unlikely to be bogus due to the clear, well-defined positive residual.",
    "self_score_key_evidence": 4,
    "self_score_leading_interpretation_and_support": 4,
    "self_score_alternative_analysis": 3
  },
  "Part C": {
    "stage1": "real_object",
    "stage2": "astrophysical",
    "stage3": "variable_star"
  }
}
```
\end{lstlisting}
\end{chatmsg}
Below are the inputs fed to the model for its second attempt:
\begin{chatmsg}[grey]{System — attempt 2}[breakable]
\noindent\texttt{\{second\_rollout\_ablation\_block\}}\footnote{The general system prompt is provided in Section~\ref{app:main prompt}.}
\end{chatmsg}
\begin{chatmsg}[blue]{User — attempt 2}[breakable]
\begin{lstlisting}[style=prompttxt]
[ALERT IDENTIFIERS]
- Object ID: ZTF24ablarbs

[ZTF CANDIDATE FIELDS]
- fid: 2
- isdiffpos: t
- firstmjd: 60585.2
- magpsf: 21.4729
- sigmapsf: 0.214785
- fwhm: 2.35
- classtar: 0.99
- sgscore1: 0.008
- distpsnr1: 0.0349517
- chinr: 2.314
- sharpnr: 0.366
- ndethist: 4
- ncovhist: 1787
- sgmag1: 22.1523
- srmag1: 21.6716
- simag1: 20.9498
- szmag1: 20.739
- nmtchps: 2
- deltajd: 1.13814

Field definitions and sentinel rules are in the system message.

Analyze this alert and return the JSON response.

---

Below this metadata block, the live API user message also embeds the model's complete first-trial JSON (verbatim) under [YOUR FIRST ATTEMPT JSON --- FOR SELF-REVIEW ONLY], then asks for the second-trial JSON. Primary-manuscript placeholder for that payload: {embedded_first_attempt_json_block} (constructed by prompts_second_roll_out_ablation.build_user_prompt; archived under data_second_roll_out_ablation/priors/).
\end{lstlisting}
\end{chatmsg}
The corresponding model response for the second attempt follows:
\begin{chatmsg}[green]{Output 2}[breakable]
\begin{lstlisting}[style=rawouttxt]
```json
{
  "Part A": {
    "filter_band": "r",
    "subtraction_sign": "positive",
    "magpsf": 21.4729,
    "sigmapsf": 0.214785,
    "ndethist": 4,
    "ncovhist": 1787
  },
  "Part B": {
    "key_evidence": "The difference image shows a clear, point-like positive residual at the center, indicating a new or brighter source. The science image shows a faint point source at the center that is not present in the reference image. The fwhm of 2.35 pixels is consistent with a point source. The classtar of 0.99 strongly suggests a stellar morphology. The nearest PS1 source (distpsnr1 = 0.03 arcsec) has a very low sgscore1 (0.008), suggesting it is extended or a galaxy. The chinr (2.314) and sharpnr (0.366) values for the nearest reference source are somewhat high for a perfect point source, but not definitively indicative of an artifact. The ndethist of 4 and ncovhist of 1787 suggest repeated observations of this field, and the deltajd of 1.13814 indicates multiple detections over a short period.",
    "leading_interpretation_and_support": "The alert is most likely a variable star. The clear point-like detection in the difference image, coupled with the absence of a counterpart in the reference image, confirms it as a real transient. The high classtar (0.99) strongly supports a stellar morphology. While the sgscore1 of the nearest PS1 source is low (0.008), its very small distance (0.03 arcsec) suggests it is likely the same object. The discrepancy between classtar and sgscore1 could be due to the PS1 source being a faint, unresolved galaxy with a foreground variable star, or issues with PS1 classification for faint sources. Given the clear point-like nature in the ZTF images and the lack of a clearly extended host, a variable star is more plausible than an AGN. The ndethist of 4 and deltajd of ~1 day are consistent with a variable star that has been detected a few times.",
    "alternative_analysis": "An alternative interpretation is that this could be an AGN. The low sgscore1 for the nearby PS1 source could indicate a faint host galaxy, and the point-like nature of the transient could be the varying nucleus. However, without a clear extended host visible in the ZTF images, and given the strong stellar classification from classtar, a variable star is a stronger candidate. It is less likely to be a supernova given the lack of a clear host galaxy and the relatively short deltajd, though a very faint, distant supernova cannot be entirely ruled out without more information. It is not an asteroid due to the lack of a dipole signature in the difference image and the stationary appearance between science and reference images. It is also unlikely to be bogus due to the clear, well-defined positive residual. My previous analysis was largely consistent, but I've refined the phrasing to better articulate the discrepancy between classtar and sgscore1.",
    "self_score_key_evidence": 5,
    "self_score_leading_interpretation_and_support": 4,
    "self_score_alternative_analysis": 4
  },
  "Part C": {
    "stage1": "real_object",
    "stage2": "astrophysical",
    "stage3": "variable_star"
  }
}
```
\end{lstlisting}
\end{chatmsg}



\section{Human Baseline}
\label{app:human-baseline}

This appendix provides the full methodology, per-expert breakdown, and same-15-alert head-to-head with models for the human baseline reported in Section~\ref{sec:results-partc}.

\subsection{Methodology}
\label{app:human-methodology}

Five astronomers participated as a reference cohort, with experience distributed as follows: two experts with $\ge 20$ years of professional astronomical experience, two with $3$--$10$ years, and one with $3$ years. Each expert independently classified the same 15-alert representative subset. The ground-truth class distribution of the 15-alert subset is AGN: 5, variable star: 3, supernova: 2, bogus: 3, asteroid: 3.

For each alert, an expert assigned one of the five benchmark classes (\textit{supernova}, \textit{AGN}, \textit{variable star}, \textit{asteroid}, \textit{bogus}) or selected ``Don't Know'' (DK) to indicate that the available image and metadata were insufficient for a confident classification. The total number of expert-alert trials is $5 \times 15 = 75$. Whereas models in the main protocol are required to commit to a single class on every alert (forced choice), expert humans were permitted to abstain.

\paragraph{Three accuracy definitions.}
We report three complementary summaries of human performance, capturing different stances on how to score abstention.
\begin{itemize}
    \item \textit{Effective accuracy.} The proportion of the $n = 75$ expert-alert trials whose committed class matches the manifest gold; abstentions count as incorrect. This is the value reported in the main-text Table~\ref{tab:partc-cascade} and provides the most direct comparison with forced-choice models.
    \item \textit{Selective accuracy.} The proportion of the $n_{\text{committed}} = 67$ committed (non-DK) trials that match the manifest gold. This measures expert precision conditional on a willingness to commit.
    \item \textit{Ensemble majority accuracy.} For each of the $n = 15$ alerts, take the modal class among the five non-DK responses; if a strict $3/5$ non-DK majority exists and matches the manifest, the alert is counted as correct. This corresponds to a collective decision-quality reading on this slice.
\end{itemize}

The manifest gold reference is the project's stamp-classifier pipeline. Agreement-with-manifest is intentionally a strict criterion: as in Zooniverse-style annotation, expert disagreement with a project reference on hard cutouts is expected, and the gap between inter-expert agreement and agreement-with-manifest is itself a signal we discuss below.

\subsection{Per-expert and ensemble results}
\label{app:human-results}

\begin{figure}[h]
    \centering
    \IfFileExists{figure/human_accuracy_bars.png}{\includegraphics[width=0.55\linewidth]{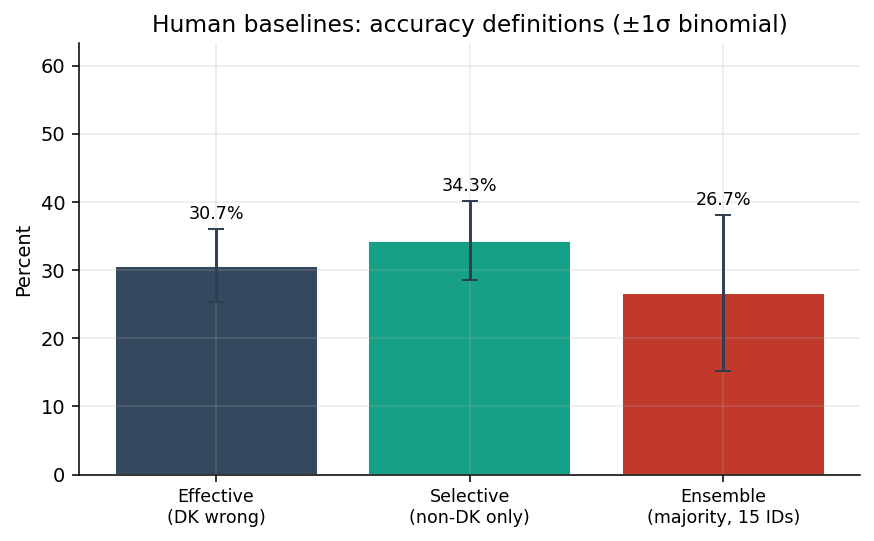}}{\fbox{\parbox[c][4cm][c]{0.5\linewidth}{\centering\itshape [Figure placeholder: figure/human\_accuracy\_bars.png \textemdash{} three-bar chart of effective vs.\ selective vs.\ ensemble human accuracy with $\pm 1\sigma$ binomial bars]}}}
    \caption{\textbf{Three human-accuracy definitions on the 15-alert subset.} \textit{Effective} ($30.67\% \pm 5.32$, abstention scored as incorrect, $n = 75$), \textit{selective} ($34.33\% \pm 5.80$, abstention excluded, $n = 67$), and \textit{ensemble majority} ($26.67\% \pm 11.42$, strict $3/5$ non-DK majority matches gold, $n = 15$). Error bars are $1\sigma$ binomial standard errors on each metric's denominator.}
    \label{fig:human-accuracy-bars}
\end{figure}

Table~\ref{tab:human-per-expert} reports the per-expert accuracy on the 15-alert subset (each expert sees all 15 alerts), and Figure~\ref{fig:human-accuracy-bars} shows the three ensemble-level accuracy definitions side-by-side.

\begin{table}[h]
\centering
\footnotesize
\caption{\textbf{Per-expert accuracy on the 15-alert subset.} Each of the five experts classified the same 15 alerts; uncertainty is the $1\sigma$ binomial standard error.}
\label{tab:human-per-expert}
\begin{tabular}{lccc}
\toprule
\textbf{Expert (alias)} & \textbf{Correct / 15} & \textbf{Accuracy (\%)} & \textbf{$\pm 1\sigma$ (pt)} \\
\midrule
mjgraham    & 7 & $\mathbf{46.67}$ & 12.88 \\
aschig      & 6 & 40.00            & 12.65 \\
libai\_astro & 5 & 33.33            & 12.17 \\
lukehandley & 3 & 20.00            & 10.33 \\
fperezpa    & 2 & 13.33            & 8.78  \\
\bottomrule
\end{tabular}
\end{table}

The expert refusal rate is $10.67\% \pm 3.56$ pt ($8/75$ responses marked DK). Of the 15 alerts, only 4 produced a valid $3/5$ non-DK majority that matched the manifest gold; 7 had no valid $3/5$ majority due to split votes, near-ties, or DK responses; no alert produced a unanimous $5/5$ correct labeling, and the strongest observed agreement was $4/5$ on two cutouts.

\subsection{Same-15-alert head-to-head with models}
\label{app:human-vs-models}

\begin{figure}[H]
    \centering
    \IfFileExists{figure/human_vs_models_15ids.png}{\includegraphics[width=0.85\linewidth]{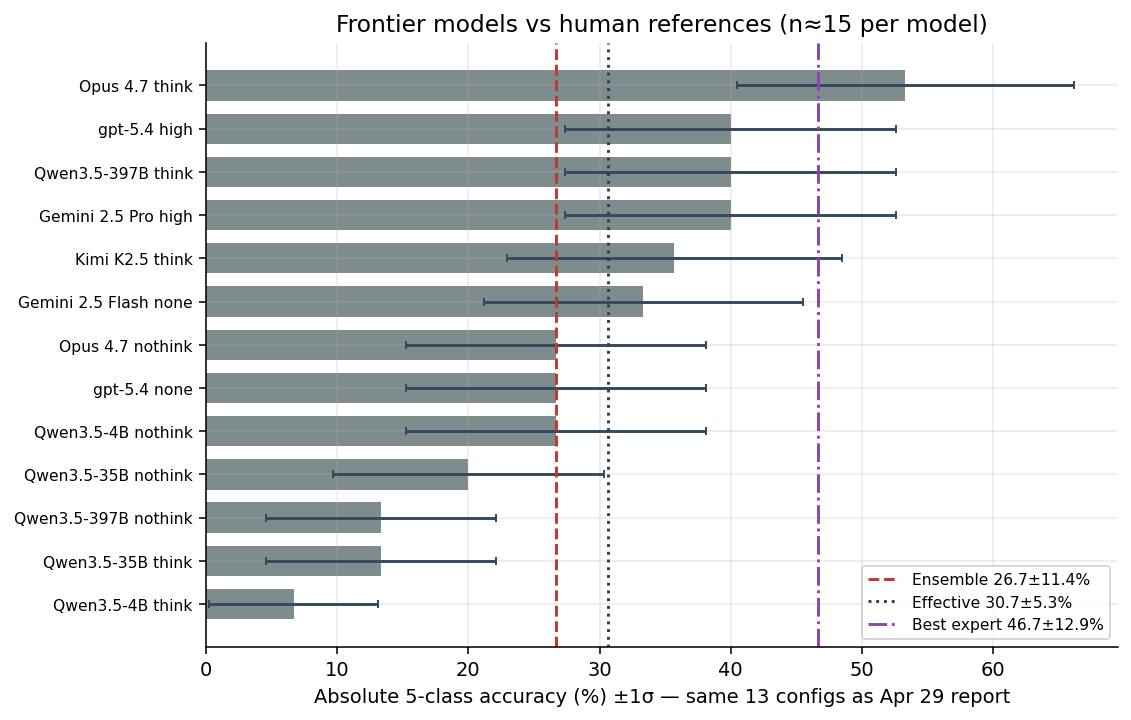}}{\fbox{\parbox[c][5cm][c]{0.8\linewidth}{\centering\itshape [Figure placeholder: figure/human\_vs\_models\_15ids.png \textemdash{} horizontal bars of 13 model runs' absolute 5-class accuracy on the same 15-alert subset, with three vertical dashed reference lines for the human ensemble ($26.67\%$), the effective human rate ($30.67\%$), and the best individual expert ($46.67\%$)]}}}
    \caption{\textbf{Frontier models versus human references on the 15-alert subset.} Each horizontal bar reports a model run's absolute 5-class accuracy on the same 15 alerts the human cohort evaluated (error bars: $\pm 1\sigma$ binomial). Three dashed vertical lines mark the strict ensemble majority ($26.67\%$, $n = 15$), the effective human rate ($30.67\%$, $n = 75$), and the best individual expert ($46.67\%$, $n = 15$). Claude Opus~4.7 think exceeds the best individual expert; mid-pack frontier models near $40\%$ sit between the best individual and the cohort mean and remain above the strict ensemble.}
    \label{fig:human-vs-models-15ids}
\end{figure}

To enable an apples-to-apples comparison with the human cohort, we recompute each model run's accuracy restricted to the same 15 alerts the experts saw. Table~\ref{tab:human-vs-models-15ids} lists each run's absolute accuracy on $n = 15$ alongside the three human reference values; Figure~\ref{fig:human-vs-models-15ids} visualizes the model accuracies as horizontal bars with three vertical reference lines for the human ensemble, effective, and best-individual rates.

\begin{table}[h]
\centering
\footnotesize
\caption{\textbf{Same-15-alert head-to-head: model runs versus human reference values.} Model rows report absolute 5-class accuracy on the same 15 alerts the human cohort evaluated, with $1\sigma$ binomial standard errors and Stage-3 macro-F1 as in Section~\ref{sec:results-partc}; $n$ indicates the number of parseable predictions on the 15-alert slice (Kimi K2.5 think drops one alert due to a parse failure). Reference rows above the rule report the human ensemble (strict $3/5$ majority), the effective rate ($n = 75$ trials, abstention as incorrect), and the best individual expert.}
\label{tab:human-vs-models-15ids}
\begin{tabular}{clcccc}
\toprule
\textbf{Rank} & \textbf{Run} & \textbf{Absolute (\%)} & \textbf{$\pm 1\sigma$ (pt)} & \textbf{Macro-F1} & \textbf{$n$} \\
\midrule
\rowcolor{gray!15}
--- & Best individual expert (mjgraham)   & 46.67 & 12.88 & ---    & 15 \\
\rowcolor{gray!15}
--- & Effective human ensemble ($n = 75$) & 30.67 & 5.32  & ---    & 75 \\
\rowcolor{gray!15}
--- & Strict ensemble majority ($n = 15$) & 26.67 & 11.42 & 0.307  & 15 \\
\midrule
1   & Claude Opus 4.7 think      & $\mathbf{53.33}$ & 12.88 & 0.5333 & 15 \\
2   & Gemini 2.5 Pro high-think  & 40.00            & 12.65 & 0.5333 & 15 \\
3   & Qwen3.5-397B-A17B think    & 40.00            & 12.65 & 0.3056 & 15 \\
4   & GPT-5.4 high-think         & 40.00            & 12.65 & 0.2000 & 15 \\
5   & Kimi K2.5 think            & 35.71            & 12.81 & 0.5000 & 14 \\
6   & Gemini 2.5 Flash no-think  & 33.33            & 12.17 & 0.5333 & 15 \\
7   & Claude Opus 4.7 nothink    & 26.67            & 11.42 & 0.4222 & 15 \\
8   & Qwen3.5-4B nothink         & 26.67            & 11.42 & 0.4148 & 15 \\
9   & GPT-5.4 no-think           & 26.67            & 11.42 & 0.4222 & 15 \\
10  & Qwen3.5-35B-A3B nothink    & 20.00            & 10.33 & 0.1667 & 15 \\
11  & Qwen3.5-397B-A17B nothink  & 13.34            & 8.78  & 0.2963 & 15 \\
12  & Qwen3.5-35B-A3B think      & 13.33            & 8.78  & 0.0952 & 15 \\
13  & Qwen3.5-4B think           & 6.67             & 6.44  & 0.2222 & 15 \\
\bottomrule
\end{tabular}
\end{table}

The best run (Claude Opus~4.7 think, $53.33\%$ on $n = 15$) is above the best single expert ($46.67\%$), the effective human rate ($30.67\%$), and the strict ensemble ($26.67\%$). Mid-pack frontier models near $40\%$---Gemini 2.5~Pro high-think, Qwen3.5-397B-A17B think, GPT-5.4 high-think---sit between the best expert and the cohort mean and remain above the strict ensemble. The strict-ensemble value is low not because individual experts are uniformly weak, but because votes rarely consolidate into a clean $3/5$ majority on hard cutouts.

\subsection{Calibrated refusal versus blind confidence}
\label{app:human-calibration}

The cleanest qualitative contrast between humans and models on this slice is in their stance toward ambiguity.

\textbf{Models are forced to classify; humans abstain.} Section~\ref{sec:results-partc} reports models on a forced-choice protocol; the human ensemble exhibits a $10.67\% \pm 3.56$ refusal rate (DK responses) that is unavailable to models in our schema. The selective accuracy ($34.33\% \pm 5.80$) exceeds the effective accuracy ($30.67\% \pm 5.32$) by approximately $3.7$ pp, which means abstention does carry partial signal---when humans choose to commit, they are slightly more accurate than when they are forced---but the effect is small relative to the cross-model spread reported in Table~\ref{tab:partc-cascade}.

\textbf{Contrast with model self-reasoning scores.} Section~\ref{sec:results-partb} documents that several frontier models, most prominently Gemini 2.5~Pro high-think, reach a mean self-reasoning score of $4.886/5$ with a $100\%$ self pass rate---and yet land at $40\%$ accuracy on this 15-alert slice (Table~\ref{tab:human-vs-models-15ids}), only marginally above the strict human ensemble. The human reference cohort's stance is the inverse: explicit acknowledgment of ambiguity rather than confident commitment. The cross-model self-reasoning-versus-accuracy slope reported in Section~\ref{sec:honesty-macro} places this contrast on a numeric axis: high self-rated reasoning quality tracks negatively with accuracy across this batch of runs.

\textbf{Caveat on ``errors''.} Many strict-majority human ``misses'' on this slice reflect calibrated refusal in the face of genuine scientific ambiguity rather than knowledge failure. The strict ensemble value should be read as a lower bound on collective expert decision quality on hard cutouts; the best-individual rate ($46.67\%$) and the selective accuracy ($34.33\%$) provide complementary upper-bound and committed-precision readings.

\section{Human-Model Grading Comparison: Representative Case Studies}
\label{app:ensemble-deepdive}

To validate the model self-assessment metrics discussed in Section~\ref{sec:honesty}, we performed a multi-expert qualitative audit on a difficult alert: \texttt{ZTF26aargnnp}. This alert was selected because it elicited a wide range of performance levels and reasoning strategies across the 13 model configurations. We established the human reference based on the consensus of five independent graders: two senior experts ($\ge 20$ years experience), two intermediate researchers ($3$--$10$ years), and one early-career astronomer ($3$ years). Furthermore, character-level qualitative markup (distinguishing correct, partially flawed, and hallucinatory claims) was provided by a senior expert with 20+ years of professional experience in time-domain astronomy.

On this alert, we identify a clear “alternative-reasoning advantage'': the share of expert-verified fully correct'' (green) characters in the model's \texttt{alternative analysis} field (Q3) is $44.48\%$, nearly double the $23.19\%$ observed in the \texttt{leading interpretation} field (Q2). This indicates that when the prompt forces a model to give alternative hypotheses, its prose becomes significantly more defensible under expert review. Alongside the findings from our 1,500-alert benchmark, the individual examples in this section show two main flaws in how models grade themselves: their reasoning becomes more accurate only when they are forced to consider alternatives, and their internal confidence often fails to match the judgment of human experts.

\subsection{The alternative-reasoning advantage}
\label{app:ztf26-altreason}

To quantify the impact of alternative-hypothesis prompting on reasoning quality, we first analyzed the character-level expert markup across all 13 model configurations for alert ZTF26aargnnp. When pooling the results, we find that the share of expert-verified ``fully correct'' (green) characters in the models' \texttt{alternative analysis} fields (Q3) is $44.48\%$, nearly double the $23.19\%$ observed in the \texttt{leading interpretation} fields (Q2). Specifically, the composition shifts from $23.19\%$ green, $47.89\%$ yellow, and $20.97\%$ red in Q2 to $44.48\%$ green, $28.90\%$ yellow, and $21.20\%$ red in Q3. Notably, because the red share remains effectively flat across both fields ($\approx 21\%$), this gain represents a yellow-to-green shift rather than a red-to-green correction. This suggests that models do not necessarily change which claims they get wrong, but their successful claims become significantly more grounded and scientifically defensible when they are formally forced to canvas alternative hypotheses.

\begin{figure}[H]
    \centering
    \IfFileExists{figure/ztf26_q2_vs_q3_highlight_mix.png}{\includegraphics[width=0.7\linewidth]{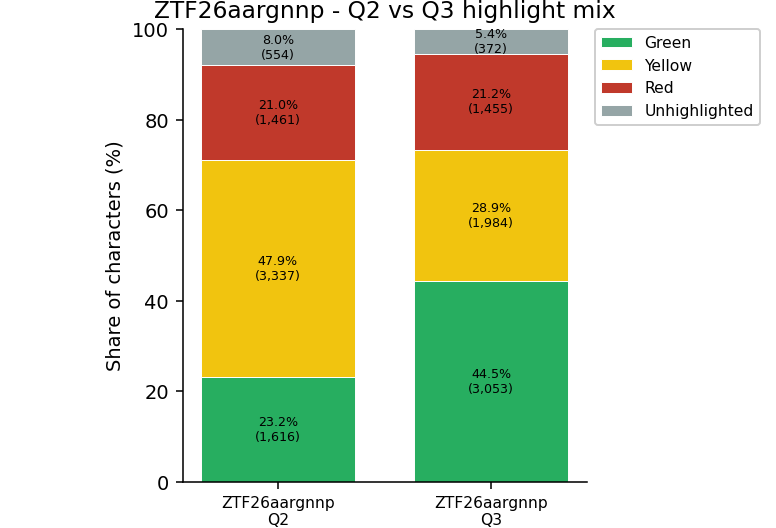}}{\fbox{\parbox[c][5cm][c]{0.7\linewidth}{\centering\itshape [Figure placeholder: stacked bars of green / yellow / red / unhighlighted character shares for the \texttt{leading\_interpretation} (Q2) and \texttt{alternative\_analysis} (Q3) fields on ZTF26aargnnp, pooled over the 13 model blocks.]}}}
    \caption{\textbf{Q2 vs.\ Q3 highlight composition on ZTF26aargnnp.} Stacked bars give the share of expert-marked green (``fully correct''), yellow (``partially correct''), red (``flawed or hallucinatory''), and unhighlighted characters in the model's \texttt{leading\_interpretation} (Q2) and \texttt{alternative\_analysis} (Q3) fields, pooled over the 13 audited model blocks. The green share rises from $23.19\%$ in Q2 to $44.48\%$ in Q3 while the red share stays near $21\%$, isolating the Q2-to-Q3 gain as a yellow-to-green shift.}
    \label{fig:ztf26-q2-vs-q3-mix}
\end{figure}

\subsection{Population-level decoupling and bias}
\label{app:ztf26-population}

\begin{figure}[H]
    \centering
    \IfFileExists{figure/ztf26_calibration_scatter.png}{\includegraphics[width=0.85\linewidth]{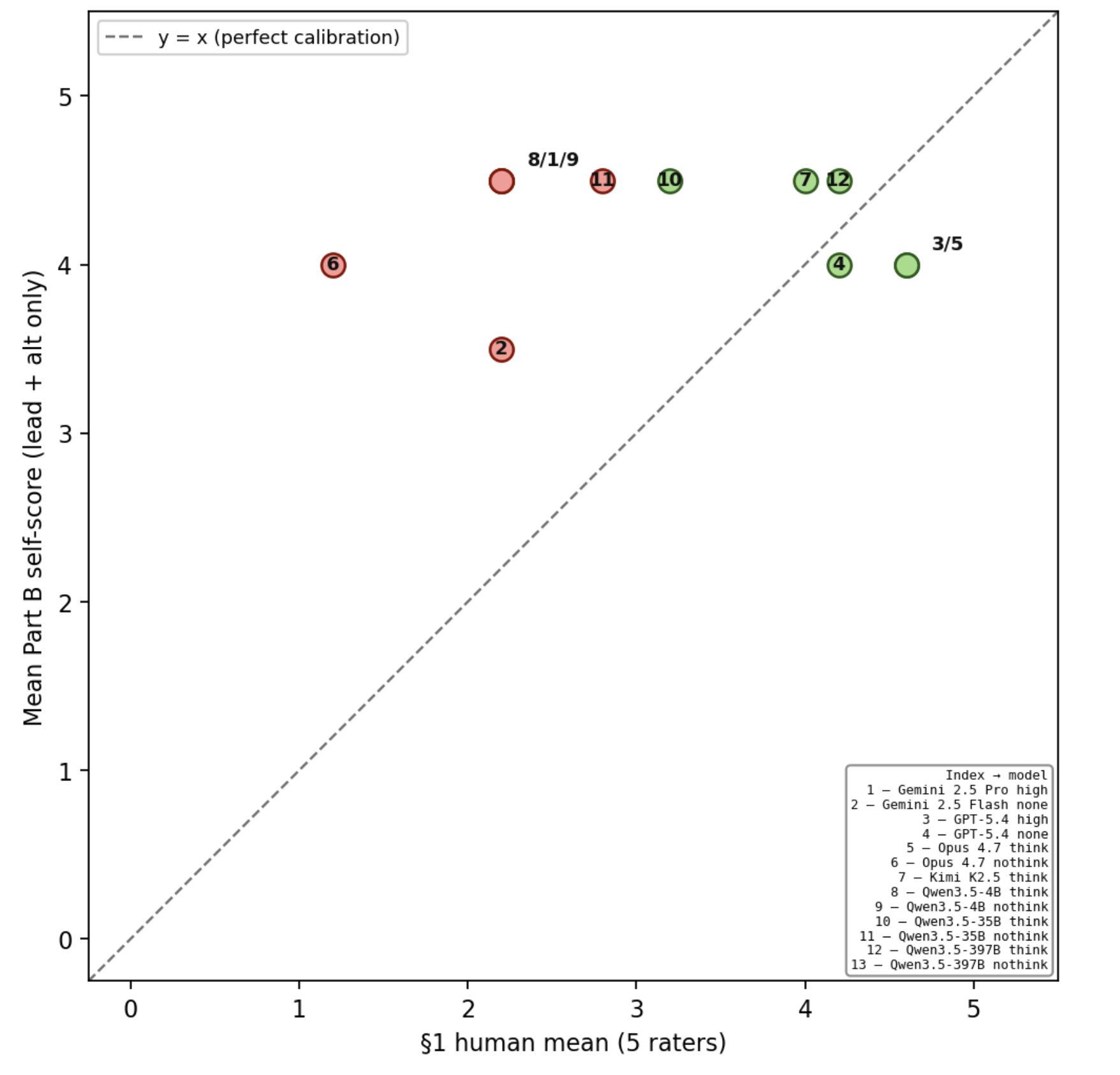}}{\fbox{\parbox[c][5cm][c]{0.7\linewidth}{\centering\itshape [Figure placeholder: per-LLM calibration scatter of mean Part B self-score versus blended human mean on ZTF26aargnnp.]}}}
    \caption{\textbf{Per-LLM calibration on ZTF26aargnnp.} Mean Part B self-score (vertical axis) against the blended human mean (horizontal axis) for the 12 models with complete Part B fields. The dashed $y=x$ line marks perfect calibration; marker fill encodes Part C correctness (green = correct class, red = incorrect class, gray = unparsed).}
    \label{fig:ztf26-calibration}
\end{figure}

\begin{figure}[h]
    \centering
    \IfFileExists{figure/ztf26_bias_bar.png}{\includegraphics[width=0.85\linewidth]{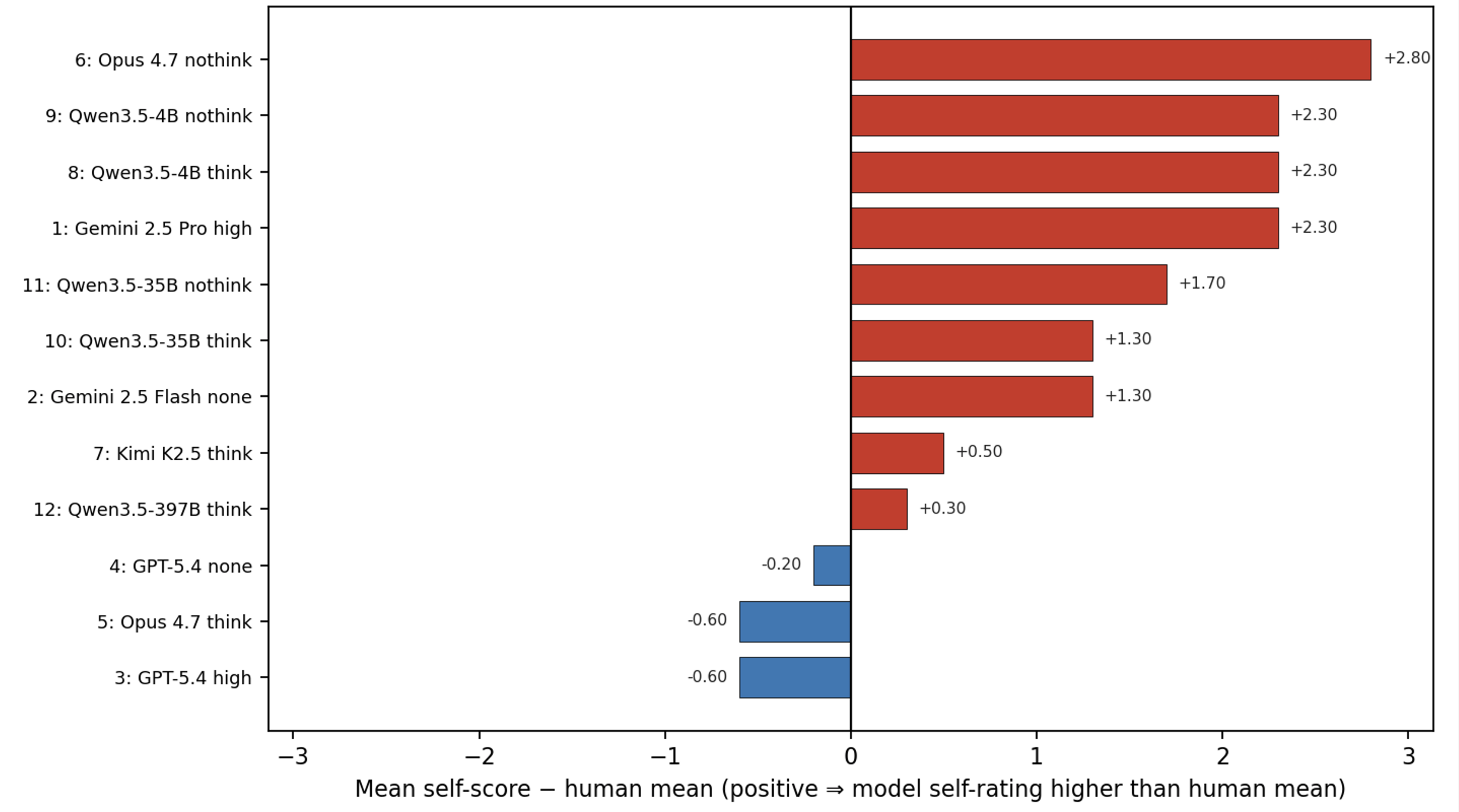}}{\fbox{\parbox[c][5cm][c]{0.85\linewidth}{\centering\itshape [Figure placeholder: per-model self-rating bias (mean self minus blended human grade), sorted, on ZTF26aargnnp.]}}}
    \caption{\textbf{Self-rating bias on ZTF26aargnnp.} Mean Part B self-score minus the blended human grade, sorted. Positive values (red bars) indicate the model rates its own reasoning above the human consensus; negative values (blue bars) indicate the model is stricter than the experts.}
    \label{fig:ztf26-bias}
\end{figure}

Across the 12 model runs with parseable Part B fields\footnote{One model (\texttt{Qwen3.5-397B-A17B nothink}) is excluded from correlation and calibration metrics because it did not output \texttt{Part B} self-score fields in this example; however, its reasoning text was fully audited for character-level highlights.}, the Pearson correlation between the blended human grade and the model's mean Part B self-score is effectively zero ($r = -0.023, p = 0.94$), indicating that on this difficult alert, model self-confidence carries no linear signal regarding actual expert agreement. As shown in Figure~\ref{fig:ztf26-bias}, several models (e.g., \texttt{Opus 4.7 nothink}, \texttt{Gemini 2.5 Pro high}) self-rate two to three scale points above the human consensus. Conversely, GPT-5.4 none sits closest to the $y=x$ line in Figure~\ref{fig:ztf26-calibration}, with self-ratings  closely aligning with a human mean.

\begin{figure}[H]
    \centering
    \IfFileExists{figure/ztf26_self_vs_human_bars.png}{\includegraphics[width=\linewidth]{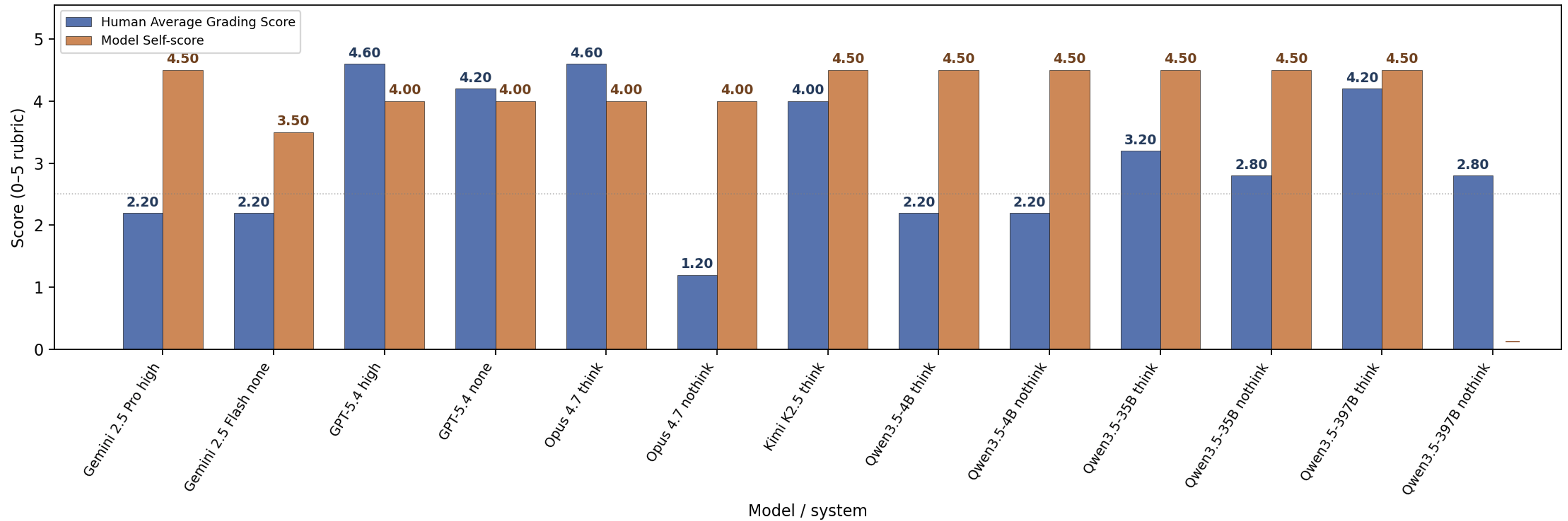}}{\fbox{\parbox[c][5cm][c]{0.85\linewidth}{\centering\itshape [Figure placeholder: per-model paired bars of mean Part B self-score and the blended human grade on ZTF26aargnnp.]}}}
    \caption{\textbf{Self-score vs.\ human mean per model on ZTF26aargnnp.} Paired bars show the blended human grade (five raters) and the model's mean Part B self-score (\texttt{leading\_interpretation} + \texttt{alternative\_analysis}) for each of the 13 audited runs.}
    \label{fig:ztf26-self-vs-human}
\end{figure}

\subsection{Three representative model responses}
\label{app:ztf26-cases}

To provide qualitative grounding for the population-level statistics, we present three model responses that represent the full spectrum of expert grades for alert ZTF26aargnnp. Each case study includes the human mean score, the model’s mean \texttt{Part B} two fields' self-score, and \texttt{Part C} correctness. The reasoning text (\texttt{Q2}+\texttt{Q3}) features character-level markup from a senior expert, where {\color{hlgreen}green text} denotes fully correct'' claims,  {\color{hlyellow}yellow text} denotes partially correct'' analysis, and {\color{red}red text} denotes clearly flawed or hallucinatory'' statements.

\begin{figure}[H]
\centering
\includegraphics[width=0.88\textwidth]{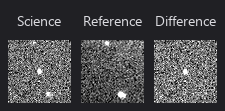}
\caption{Science--Reference--Difference montage for \texttt{ZTF26aargnnp}.}
\end{figure}

\subsubsection{Tier 1 --- \texttt{Claude Opus 4.7 think}}
This example demonstrates high scientific alignment, correctly identifying the ‘mover' signature by reconciling high historical coverage (\texttt{ncovhist=1489}) with a single-epoch detection—logic that is almost entirely validated by expert character-level markup.
\begin{chatmsg}[blue]{Claude Opus 4.7 think --- raw output (Human Grading Score: 4.60 \& Model Self-score: 4.00)}[breakable]
\begin{lstlisting}[style=rawouttxt]
(model=claude-opus-4-7, n_prompt_tokens=4513, n_answer_tokens=1065)

{
  "Part A": {
    "filter_band": "r",
    "subtraction_sign": "positive",
    "magpsf": 17.7693,
    "sigmapsf": 0.0461362,
    "ndethist": 1,
    "ncovhist": 1489
  },
  "Part B": {
    "key_evidence": "Science shows a bright, point-like source at center; the reference/template has no counterpart there; the difference shows a clean positive residual. classtar=0.983 and fwhm=2.49 pix are consistent with a stellar PSF. ndethist=1 with ncovhist=1489 means this field has been imaged many times but the source was detected only once (deltajd=0). The nearest PS1 object is 24.4\" away (distpsnr1), so there is no identified host at the candidate position. isdiffpos=t confirms a brightening, magpsf=17.77 in r.",
    "leading_interpretation_and_support": "(*@\textcolor{hlyellow}{Most consistent with a real solar-system object (Asteroid).}\textcolor{black}{ }\textcolor{hlgreen}{The source is point-like and appears on otherwise blank sky, with no convincing stationary source or galaxy in the template.}\textcolor{black}{ }\textcolor{hlyellow}{The very low same-position history (ndethist=1) despite high coverage (ncovhist=1489) fits a moving object better than a fixed astrophysical variable.}\textcolor{black}{ }\textcolor{hlgreen}{The positive, compact subtraction residual argues against a pure subtraction artifact.}\textcolor{black}{}@*)",
    "alternative_analysis": "(*@\textcolor{hlgreen}{A hostless or very faint-host supernova is the main astrophysical alternative, since a new positive point source on a blank template can also be an extragalactic transient.}\textcolor{black}{ }\textcolor{hlyellow}{However, the lack of any visible host/persistent counterpart and the single same-position detection make that less compelling here than a solar-system interpretation}\textcolor{black}{. }\textcolor{hlgreen}{AGN and ordinary variable star interpretations are weak because there is no close persistent nucleus or stellar counterpart at the alert position.}\textcolor{black}{}@*)",
    "self_score_key_evidence": 4,
    "self_score_leading_interpretation_and_support": 4,
    "self_score_alternative_analysis": 4
  },
  "Part C": {
    "stage1": "real_object",
    "stage2": "solar_system",
    "stage3": "N/A"
  }
}
\end{lstlisting}
\end{chatmsg}
\subsubsection{Tier 2 --- \texttt{Qwen3.5-35B nothink}}
This model typifies the ‘biased middle' behavior, where accurate metadata grounding fails to translate into a correct classification due to a breakdown in positional reasoning, resulting in a high self-score that masks a flawed scientific conclusion.
\begin{chatmsg}[blue]{Qwen3.5-35B nothink --- raw output (Human Grading Score: 2.80 \& Model Self-score: 4.50)}[breakable]
\begin{lstlisting}[style=rawouttxt]
(model=Qwen/Qwen3.5-35B-A3B, n_prompt_tokens=?, n_answer_tokens=526)

{
  "Part A": {
    "filter_band": "r",
    "subtraction_sign": "positive",
    "magpsf": 17.7693,
    "sigmapsf": 0.0461362,
    "ndethist": 1,
    "ncovhist": 1489
  },
  "Part B": {
    "key_evidence": "The difference image shows a clear, point-like positive residual at the center. The source has a high FWHM (2.49) and high sharpness (0.107), consistent with a stellar point source. The PS1 neighbor is at 24.4 arcsec (sgscore1=0.99), indicating a nearby star but not a direct match to the transient position. The history shows ndethist=1 (first detection) despite high coverage (ncovhist=1489).",
    "leading_interpretation_and_support": "(*@\textcolor{hlred}{Variable Star.}\textcolor{black}{ }\textcolor{hlyellow}{The high classtar (0.983) and sharpness indicate a stellar object.}\textcolor{black}{ }\textcolor{hlgreen}{The high ncovhist (1489) confirms the telescope has repeatedly observed this sky position, yet ndethist is only 1.}\textcolor{black}{ }\textcolor{hlyellow}{This implies the object is faint or variable and was only detected above threshold in this specific epoch.}\textcolor{black}{ }\textcolor{hlred}{The Science vs Template comparison shows the source is present in both but significantly brighter in Science, consistent with a variable star (e.g., RR Lyrae, Cepheid) or a flare.}\textcolor{black}{}@*)",
    "alternative_analysis": "(*@\textcolor{hlgreen}{Asteroid.}\textcolor{black}{ }\textcolor{hlgreen}{The low ndethist (1) is a common signature for moving objects that are not linked across epochs.}\textcolor{black}{ }\textcolor{hlred}{However, the Science and Template images show the source at the same geometric center in both frames, with no visible streak or offset relative to the background stars. }\textcolor{hlyellow}{A moving asteroid would typically show a positional shift between Science and Template or a streak in the difference image if the motion was significant during the exposure.}\textcolor{black}{ }\textcolor{hlred}{The lack of motion evidence makes the asteroid hypothesis less likely than a variable star.}\textcolor{black}{}@*)",
    "self_score_key_evidence": 5,
    "self_score_leading_interpretation_and_support": 5,
    "self_score_alternative_analysis": 4
  },
  "Part C": {
    "stage1": "real_object",
    "stage2": "astrophysical",
    "stage3": "variable_star"
  }
}<|im_end|>
\end{lstlisting}
\end{chatmsg}

\subsubsection{Tier 3 --- \texttt{Gemini 2.5 Flash}}
This response highlights a hallucination failure mode, where the model justifies its incorrect classification by explicitly claiming the source is present in the reference cutout despite clear visual evidence of its absence.
\begin{chatmsg}[green]{Gemini 2.5 Flash --- raw output (Human Grading Score: 2.20 \& Model Self-score: 3.50)}[breakable]
\begin{lstlisting}[style=rawouttxt]
(model=gemini-2.5-flash, n_prompt_tokens=3312, n_answer_tokens=642)

```json
{
  "Part A": {
    "filter_band": "r",
    "subtraction_sign": "positive",
    "magpsf": 17.7693,
    "sigmapsf": 0.0461362,
    "ndethist": 1,
    "ncovhist": 1489
  },
  "Part B": {
    "key_evidence": "The science image shows a bright, point-like source at the center that is absent in the reference image, resulting in a clear positive residual in the difference image. The FWHM of 2.49 pixels is consistent with a point source. The classtar (0.983) and sgscore1 (0.993333) values are both very high, strongly indicating a stellar morphology. The ndethist is 1, suggesting this is a first detection. The nearest PS1 source is 24.3978 arcsec away, which is a significant distance, implying the PS1 source is likely unrelated to the transient itself, or at least not a direct counterpart.",
    "leading_interpretation_and_support": "(*@\textcolor{hlred}{Most consistent with a Galactic variable star: the candidate appears as a stellar (point-like) source already present in the reference, brightening enough to produce a positive difference residual}\textcolor{black}{. }\textcolor{hlyellow}{High classtar, low sharpnr, and the lack of any extended host nearby (nearest PS1 source is far at \textasciitilde{}24\textbackslash{}") favor a stellar variable rather than an SN or AGN.}\textcolor{black}{}@*)",
    "alternative_analysis": "(*@\textcolor{hlred}{A supernova is unlikely because the source is clearly present in the reference and shows stellar morphology with no underlying galaxy.}\textcolor{black}{ }\textcolor{hlgreen}{A CV-type outburst or other dwarf nova is plausible given a single new detection after many visits}\textcolor{black}{. }\textcolor{hlred}{AGN is disfavored due to no resolved host.}\textcolor{black}{ }\textcolor{hlred}{An asteroid is ruled out since the source is also in the template at the same position.}\textcolor{black}{}@*)",
    "self_score_key_evidence": 5,
    "self_score_leading_interpretation_and_support": 4,
    "self_score_alternative_analysis": 3
  },
  "Part C": {
    "stage1": "real_object",
    "stage2": "astrophysical",
    "stage3": "variable_star"
  }
}

\end{lstlisting}
\end{chatmsg}
\end{document}